\documentclass[a4paper,fleqn,usenatbib]{mnras}

\usepackage{newtxtext,newtxmath}

\usepackage[T1]{fontenc}
\usepackage{ae,aecompl}

\usepackage{amssymb} 
\usepackage{amsmath} 
\usepackage{hyperref}
\usepackage{booktabs,longtable}
\usepackage{graphicx} 
\usepackage{wasysym} 
\usepackage{lscape} 
\usepackage{caption} 
\usepackage{wrapfig,lipsum,booktabs} 
\usepackage{enumitem} 
\usepackage{mathtools} 
\usepackage{overpic} 
\usepackage{graphbox} 
\usepackage{tabularx} 
\usepackage{caption} 
\usepackage[table]{xcolor} 

\newcommand{\GALFIT}{\textsc{Galfit}}
\newcommand{\Sersic}{S\'{e}rsic}
\newcommand{\Sersicone}{\Sersic{}$^{n=1}$}
\newcommand{\Sersicfour}{\Sersic{}$^{n=4}$}
\newcommand{\Chisq}{$\chi_{\nu}^{2}$}
\newcommand{\sampleAGN}{5}
\newcommand{\sampleUnfittable}{19}
\newcommand{\sample}{345}
\newcommand{\sampleBadMag}{7}
\newcommand{\sampleNotFitted}{3}
\newcommand{\sampleBestFits}{335}
\newcommand{\sampleNoBestFits}{10}
\newcommand{\sampleDecomposed}{274}
\newcommand{\sampleNotDecomposed}{62}
\newcommand{\nthreshold}{1.5}
\newcommand{\PSCz}{PSC$z$}

\newlist{cutenumerate}{enumerate}{1}
\setlist[cutenumerate,1]{
  label={\arabic*},
  leftmargin=*,
  align=left,
  labelsep=0.05cm,
}

\title
[SFRS. IV. Stellar mass distribution] 
{
 The Star Formation Reference Survey. IV.
 \\
 Stellar mass distribution of local star-forming galaxies
} 

\author[P. Bonfini et al.]
{
 \parbox{\textwidth}{
  P. Bonfini$^{1,2,3,4}$\thanks{E-mail: \texttt{paolo@physics.uoc.gr}},
  A. Zezas$^{1,2,5}$,
  M.~L.~N. Ashby$^{5}$,
  S.~P.~Willner$^{5}$,
  A.~Maragkoudakis$^{1,2,6}$,
  K.~Kouroumpatzakis$^{1,2}$,
  P.~H.~Sell$^{1,2,7}$,
  and
  K.~Kovlakas$^{1,2}$
 }
 \vspace{0.4cm}\\
 \parbox{\textwidth}{
  $^{1}$Department of Physics, University of Crete, 70013 Heraklion, Crete, Greece\\
  $^{2}$Foundation for Research and Technology-Hellas, 71110 Heraklion, Crete, Greece\\
  $^{3}$ Institute for Astronomy, Astrophysics, Space Applications \& Remote Sensing, National Observatory of Athens, GR-15236, Penteli, Greece\\
  $^{4}$Computer Science Department, University of Crete, 70013 Heraklion, Crete, Greece\\
  $^{5}$Center for Astrophysics \textbar\ Harvard \& Smithsonian, 60 Garden Street, Cambridge, MA 02138, USA\\
  $^{6}$Western University, 1151 Richmond Street, London, Ontario, N6A 3K7, Canada\\
  $^{7}$Department of Astronomy, Bryant Space Science Center, University of Florida, Gainesville FL 32611, USA\\
 }
}

\date{Published in MNRAS 504, 3831–3861 (2021)}

\pubyear{2021}

\begin{document}
\label{firstpage}
\pagerange{\pageref{firstpage}--\pageref{lastpage}}
\maketitle

\begin{abstract}
 We constrain the mass distribution in nearby, star-forming galaxies with the
 Star Formation Reference Survey (SFRS), a galaxy sample constructed to be
 representative of all known combinations of star formation rate (SFR),
 dust temperature, and specific star formation rate (sSFR) that exist in the
 Local Universe.
 An innovative two-dimensional bulge/disk decomposition of the
 2MASS/$K_{s}$-band images of the SFRS galaxies yields 
 global luminosity and stellar mass functions,
 along with separate mass functions for their bulges and disks.
 These accurate mass functions cover the full range from dwarf galaxies
 to large spirals, and are representative of star-forming galaxies selected
 based on their infra-red luminosity, unbiased by AGN content and environment.
 We measure an integrated luminosity density
 $j$ = 1.72 $\pm$ 0.93 $\times$ 10$^{9}$ L$_{\odot}$ $h^{-1}$ Mpc$^{-3}$
 and a total stellar mass density
 $\rho_{M}$ = 4.61 $\pm$ 2.40 $\times$ 10$^{8}$ M$_{\odot}$ $h^{-1}$ Mpc$^{-3}$.
 While the stellar mass of the \emph{average} star-forming galaxy 
 is equally distributed between its sub-components, disks globally
 dominate the mass density budget by a ratio 4:1 with respect to bulges.  
 In particular, our functions suggest that recent star formation happened
 primarily in massive systems, where they have yielded a disk stellar mass
 density larger than that of bulges by more than 1~dex.
 Our results constitute a reference benchmark for models addressing the assembly
 of stellar mass on the bulges and disks of local ($z = 0$) star-forming galaxies. 
\end{abstract}



\section{Introduction}
\label{Introduction}

\noindent
Although there is strong evidence that galaxies generally build their mass in a 
hierarchical bottom-up sequence, in which smaller units merge to form more massive
systems
\citep[e.g.,][and references therein]{white,springel,hopkins:AGN_feedback,naab:2009,hopkins:2010},
several processes may influence the later stages of the evolution of a galaxy
(since $z\sim2$).
For example, a gas-rich object may lose its gas due to a variety of stripping phenomena
\citep[e.g.,][]{mayer,yagi,brown}, while
an old galaxy can be rejuvenated by the creation of a new disk from infalling gas
\citep[e.g.,][]{steinmetz,marino,serra:2014}.
Different models have been recently proposed in order to explain the
evolution of galaxies after $z \sim 1$ in a comprehensive way, such as the
inside--out growth of ``red nugget'' seeds \citep[][]{red_nuggets} or the
two-stage formation model, in which an early in-situ star-formation via dissipative
processes is followed by accretion of stars formed outside the galaxy
\citep{oser,driver,two_phase}.

Any successful galaxy-formation model must at minimum predict local observed galaxy
mass functions, the separate mass functions of bulges and disks, the colors of these
components, and the dependence of star formation rate (SFR) on stellar mass.
The only solid point is the scaling of SFR with mass \citep[a relation commonly referred to as
``main sequence of star-forming galaxies''; e.g.,][]{elbaz},
for which there is growing evidence against its dependence on the environment,
at least since $z \sim 2$ \citep[e.g.,][]{koyama}.

The present work addresses bulge and disk mass functions and the dependence of star
formation activity on the stellar mass, which can itself be interpretable as the
integrated product of past star formation.
In particular, we constrain the stellar mass function of star-forming
galaxies in the Local Universe.
Our analysis is based on a representative sample of nearby star-forming galaxies
depicting arguably all the star formation
phenomenologies and 'blind' to the environment and to AGN content: the Star
Formation Reference Survey \citep[\href{http://www.cfa.harvard.edu/sfrs/}{SFRS};][]{ashby}.
This paper describes the bulge/disk decomposition of
the SFRS galaxies, which we use to derive their mass functions as well as the
separate mass functions of their disk and bulge sub-components.

This paper is organized as follows.
$\S$\ref{The Star Formation Reference Survey (SFRS)} introduces the SFRS
project and its aims.
$\S$\ref{K-band data} presents the $K$-band data used for the
current study.
$\S$\ref{The 2D fit of SFRS galaxies} through
$\S$\ref{Separation of disk/bulge components} describe the procedure to calculate
the individual bulge and disk masses;
in particular: $\S$\ref{The 2D fit of SFRS galaxies} presents the surface
brightness fitting pipeline; $\S$\ref{Best-fit model selection} presents
the decisional algorithm to identify the best-fitting model, while
$\S$\ref{Separation of disk/bulge components} discusses the bulge/disk
decomposition.
$\S$\ref{Luminosity and Mass Functions} presents the luminosity and stellar
mass function for our representative sample of star-forming galaxies as well as
the mass functions for their separate bulge and disk sub-components.
$\S$\ref{Discussion} discusses our results in the context of the current galaxy
evolution scheme, and, finally, $\S$\ref{Summary and Conclusions} summarizes our
work and conclusions.

Throughout the paper, we assume a flat ($k = 0$) $\Lambda$CDM cosmology with
\mbox{$H_{0}$ = 73~km s$^{-1}$ Mpc$^{-1}$} \mbox{($h$ = $H_{0}$/100 = 0.73}),
\mbox{$\Omega_{m}$ = 0.3}, $\Lambda_{0}$ = 0.70, and $q = -0.55$, consistently
with \cite{ashby}.
The quoted $K$-band magnitudes refer to the Vega system.
The SDSS magnitudes are reported in the native SDSS ``maggy'' system, very close
to the AB system.
All the masses reported in this paper are stellar-only, i.e., they do not account
for gas nor dark matter, and are based on a Salpeter IMF.

\section{The Star Formation Reference Survey (SFRS)}
\label{The Star Formation Reference Survey (SFRS)}

\begin{figure}
 \begin{center}
  \includegraphics[width=0.48\textwidth,angle=0]{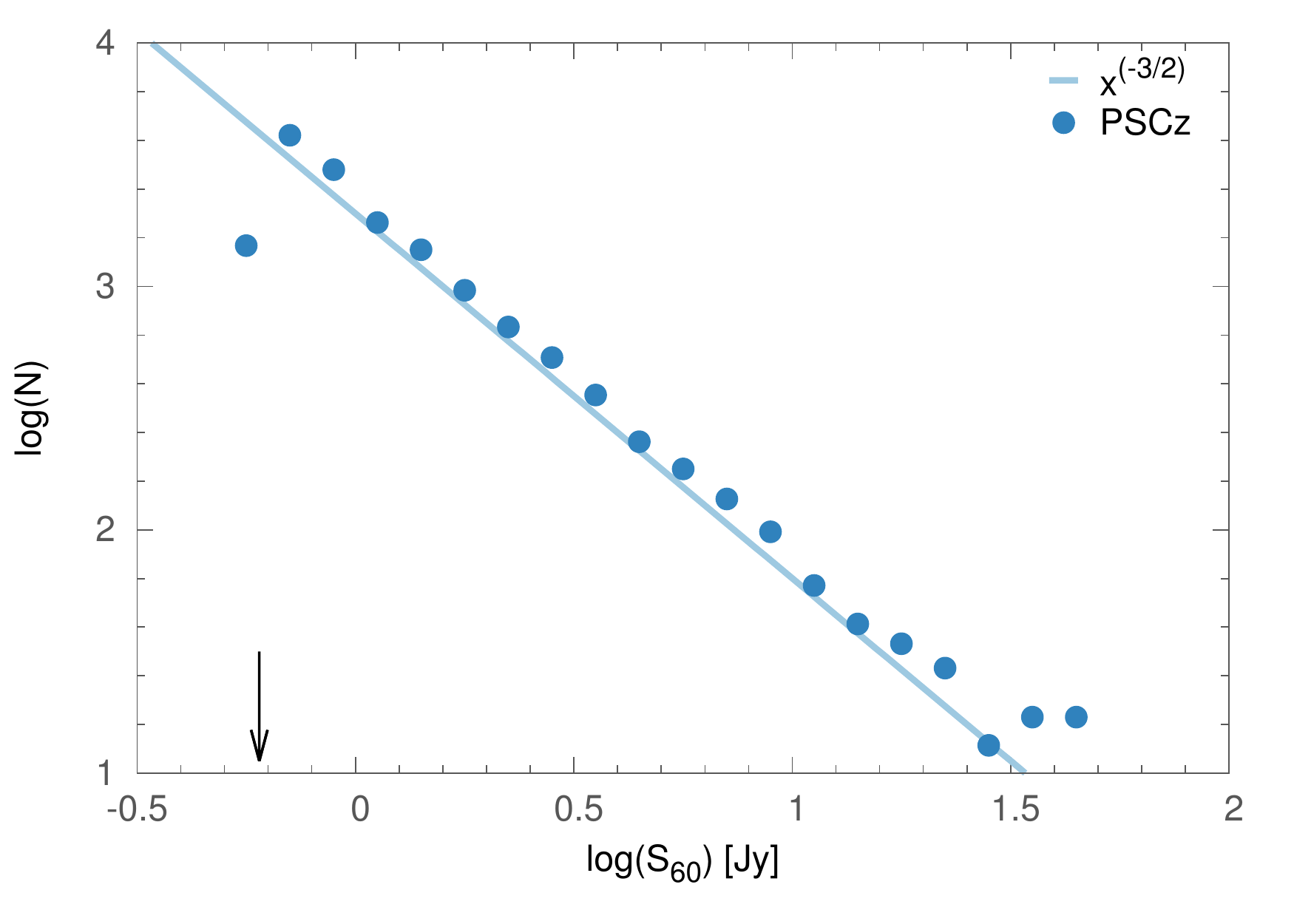}
  \caption[log($N$)-log($S_{60}$) distribution of the \PSCz{} sources]{
   log($N$)-log($S_{60}$) distribution of the \PSCz{} sources (data points).
   The \emph{slope} of the solid line corresponds to what is expected from a uniform
   distribution of sources in an Euclidean space ($N(flux)$ $\sim$ $S^{-(3/2)}$);
   in this figure, its normalization is arbitrary.
   The data follow the slope of the Euclidean distribution down to
   the detection limit of \PSCz{} (0.6~Jy; black arrow), indicating that the catalogue
   is complete to this flux limit.
   \label{figure:LogN-LogS}
  }
 \end{center}
\end{figure}

\noindent
This work is based on the Star Formation Reference Survey (SFRS) galaxy sample.
The SFRS is designed to determine the conditions under which different star
formation estimators are valid.
For this reason, 1) it is based on a galaxy sample that is fully representative of
all conditions under which star formation is known to take place in the Local
Universe, and 2) it assembles as many different star formation tracers as possible
for the entire sample (\citealt{kouroumpatzakis,mahajan}).
The SFRS sample is drawn from the IRAS \PSCz{} \citep[][]{PSCz}, a catalogue of
18\,351 objects virtually complete down to its detection limit
(0.6~Jy in the IRAS 60~$\mu$m band; Figure \ref{figure:LogN-LogS}).
Because most \PSCz{} galaxies lie within 180~Mpc, the \PSCz{} can be considered an
essentially complete catalogue of nearby star-forming galaxies.

\begin{figure}
 \begin{center}
  \includegraphics[width=0.5\textwidth,angle=0]{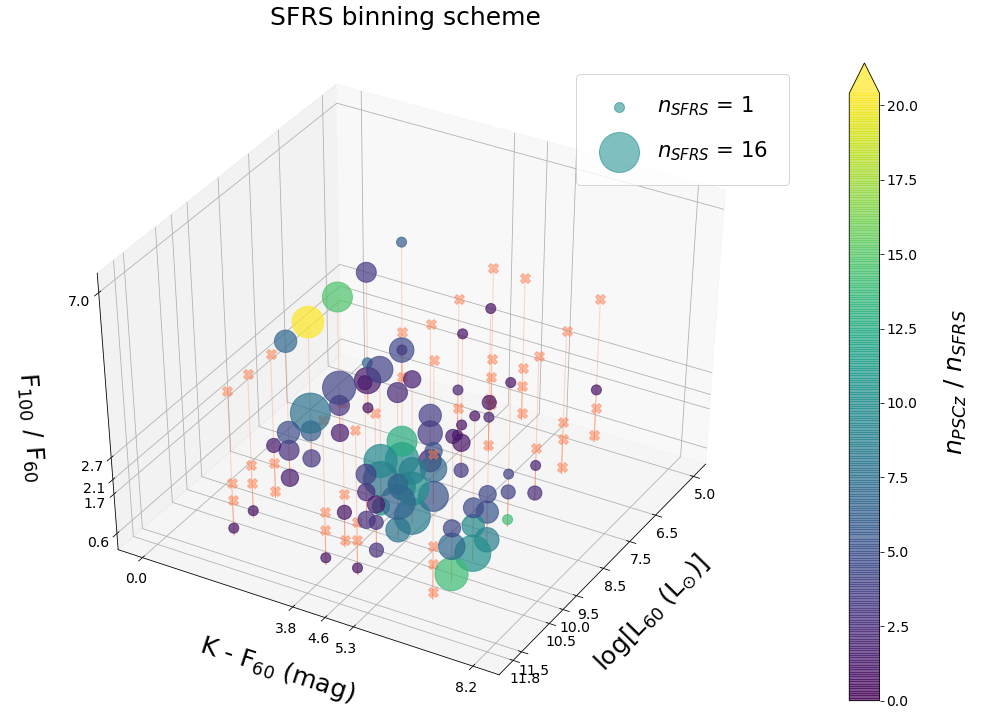}
  \caption[SFRS binning scheme]{
   Grid of the three-dimensional binning scheme used for the selection of the
   SFRS sample out of the \PSCz{} catalogue \citep{ashby}.  
   The size of each data point is proportional to the number of SFRS galaxies in
   that 3D bin.
   The data color shows the ratio of the \PSCz{} galaxies over the
   number of SFRS galaxies in the same bin.
   Empty bins --- i.e., potentially populated by \PSCz{} objects but not by 
   SFRS galaxies --- are indicated by a cross.
   \label{figure:SFRS_binning_scheme}
  }
 \end{center}
\end{figure}

In order to include representative numbers of all types of star forming galaxies
known to exist in the Local Universe,
the SFRS sample was assembled with reference to proxies for dust temperature, SFR,
and specific SFR (sSFR; i.e., the SFR per unit stellar mass).
This was achieved by sampling the three-dimensional parameter space defined
by our adopted proxies --- these are:
$F_{100}$/$F_{60}$, IRAS 60~$\mu$m luminosity, and $F_{60}$/$K_{s}$.
Here $F_{60}$ and $F_{100}$ are the flux densities in the two longest-wavelength
IRAS bands, and $K_{s}$ is the 2MASS $K_{s}$-band flux density, which effectively
traces stellar mass.
Figure \ref{figure:SFRS_binning_scheme} shows how the SFRS galaxies distribute
in the 3D parameter space.
Further details on the sample creation can be found in \cite{ashby}.

In practice, the SFRS selection scheme created a study sample of bright galaxies
which is of manageable size without sacrificing the dynamic range of the parent sample. 
Ultimately, SFRS consists of 369 galaxies spanning 5 orders of
magnitude in FIR luminosity and more than 2 orders of magnitude in sSFR.
The sampling of galaxies within the aforementioned parameter space allows
characterizing star formation activity with respect to the three properties
simultaneously without imposing a threshold on any of them.
Additionally, there exists a complete classification of the nuclear
activity of the SFRS galaxies
\citep[][see $\S$\ref{Best-fit model selection}]{maragkoudakis:2018}, so that
the presence of any AGN can be accounted for.
In summary, the relative small size of the SFRS sample, coupled with
its extensive data sets, allows to perform \emph{detailed} analysis while
still fully representing the larger population of star-forming galaxies.

The basic data for the SFRS galaxies are presented in Table 2 of \cite{ashby}.
When available, the distances were retrieved from the catalogue by
\cite{tully:SFRS}, which reports redshift-independent measurements for 127
SFRS objects.  For the remaining 242 galaxies, the distances were calculated
using the \PSCz{} heliocentric velocities, corrected for the gravitational
influence of Virgo, the Great Attractor and the Shapley supercluster, according
to the prescriptions by \cite{mould}.

\section{$K$-band data}
\label{K-band data}

In order to investigate the relation between star formation and galaxy stellar
mass and morphology, we first separated the contributions of the bulge and
disk components to the total stellar mass of each SFRS galaxy. 
The stellar mass of a galaxy is best described by its emission in the
$K$-band, which is largely dominated by the contribution of main sequence and giant
long-lived stars \citep[e.g.,][]{binney_tremaine} and is minimally affected by
the absorption by the interstellar medium \citep{devereux:K_mass}.
Young, massive stars are relatively minor contributors to galaxies' $K$-band
emission.
Although those objects significantly affect the optical emission of a
late-type galaxy (for stars along the main sequence $L$ $\propto$ $M^{3.5}$),
they do not play a significant role in the mass balance of
a galaxy, due to their relatively short lives and low number density 
\citep[e.g.,][]{binney_tremaine}.

To measure the stellar components, we obtained 
\href{http://www.ipac.caltech.edu/2mass/}{2MASS All Sky Survey} 
$K_{s}$-band images for the SFRS sources, on which we performed photometry
based on the total light derived by modelling their surface brightness.
The results were then used to calculate the total stellar mass of each galaxy
as well as the stellar mass of its disk and bulge sub-components when this
was possible.

\subsection[The 2MASS Extended Source Catalogue photometry]{The 2MASS Extended Source Catalogue photometry}
\label{The 2MASS Extended Source Catalogue photometry}

The 2MASS Extended Source Catalog \citep[2MASS-XSC;][]{jarrett} is 90\% complete
down to $K$ = 13.5~mag and provides uniform coverage of our target galaxies.
This catalogue does not provide information for the disk/bulge sub-components,
but it will be used, in this paper, as a comparison for the
integrated magnitudes.
The 2MASS-XSC essentially offers 2 kinds of extended fluxes, which correspond to
aperture photometry within the ``Kron'' and the ``total'' radii, respectively.
The former is derived from the intensity-weighted first-moment radius
\citep[e.g.,][]{sextractor}, while the
latter is derived by means of a curve-of-growth technique.

While the Kron radius is known to exclude a small (few percent) but constant
fraction of the flux, independently of the object magnitude \citep[e.g.,][]{kron},
the so-called ``total'' (extrapolated) magnitudes recover most of
the flux from the source and were therefore selected as a comparison benchmark
for the total luminosities derived in our work.
The extrapolated $K_{s}$-band magnitudes for all the SFRS targets are reported in
Table \ref{table:SFRS_2MASS_SDSS}, along with the corresponding $J$-band and
$H$-band magnitudes.
For $\sim$10 objects, no 2MASS-XSC counterparts were identified
due to the small angular size of the targets.
For these galaxies we report instead the magnitudes of the corresponding 2MASS
Point Source Catalogue (2MASS-PSC) counterparts, which we identified visually.

Both the Kron and total apertures are affected by several major issues.
In fact, the definition of both apertures in 2MASS relies
on a \Sersic{} fit to the surface brightness profile of the galaxies
between 5$\arcsec$ (a radial distance considered out of the PSF influence)
and the radius where the $S/N$ falls below 2\footnote{
  For more details on the definition of the 2MASS apertures, refer to the
  \href{http://www.ipac.caltech.edu/2mass/releases/allsky/doc/sec4\_5e.html}
  {2MASS documentation} or \cite{jarrett}\\
}.
Not only may a single S\'{e}rsic model be an over-simplification of the light
profile, but also, due to the modest average exposure times of the combined 2MASS
mosaics (7.8~sec; $\S$\ref{The 2MASS data for the SFRS targets}), the
$S/N \sim 2$ threshold may constrain the \Sersic{} fit only within a very small
galactocentric radial range and in turn yield an erroneous Kron/total radius.
Secondly, there is no certainty that the flux within the apertures won't include
a contribution by the background.
Finally, contaminating objects (mainly stars) may fall within the aperture and
contribute to the source flux if the automated masking did not work properly
due to source blending.
There is growing evidence that the 2MASS photometry is affected by
systematic biases which lead to severely underestimated (10\% to 40\%) isophotal
and total luminosities \citep[e.g.,][]{andreon:2002,kirby:2008,schombert},
even for bright ellipticals.

Our modelling of the light profiles of the galaxies (presented in
$\S$\ref{The 2D fit of SFRS galaxies}), which is primarily intended to perform
disk/bulge decomposition, also largely overcomes the limitations of the 2MASS
apertures by accurately evaluating the background component, by accounting for
the extended wings  of the surface brightness profiles of the sources, and
by carefully masking foreground stars and contaminating objects.

\subsection{The 2MASS data for the SFRS targets}
\label{The 2MASS data for the SFRS targets}

The 2MASS observational strategy mapped each piece of the surveyed
area with overlapping tiles \citep{beichman}, where each tile covers a projected
area in the sky of \mbox{6\degr $\times$ 8\arcmin.5}.
The main 2MASS scientific image products are the ``Atlas'' (or ``co-add'') images,
which were produced by the combination of 6 frames and have a size
of~\mbox{512 $\times$ 1024}~pixels, for an exposure time of 7.8~s
at each sky location.
In the combination process, the original pixel scale (2\arcsec/pixel) is resampled
to 1\arcsec{}/pixel, so that the area covered by an Atlas image is
\mbox{8\arcmin.5 $\times$ 17\arcmin}.

As part of the star-galaxy discrimination process, the 2MASS pipeline also created
``postage-stamp'' images for each of the extended sources detected in the co-add
images.
The postage-stamps were conservatively cropped from the Atlas images around the
target of interest, for a size which is a multiple of the Kron radius of the source,
so as to not miss any galactic light.
Despite the obvious advantage of dealing with source-specific images, we
preferred the larger Atlas products, because --- for calibration purposes
(see $\S$\ref{The 2D fit of SFRS galaxies}) --- we wanted to include as many field
stars as possible and have an independent estimation of the local background.
We retrieved 2MASS $K_{s}$-band Atlas images for all the SFRS sources from the
\href{http://irsa.ipac.caltech.edu/applications/2MASS/IM/}{\mbox{NASA-IPAC} Infrared Science Archive}.

\section[2D fit of SFRS galaxies]{The 2D fit of SFRS galaxies}
\label{The 2D fit of SFRS galaxies}
 
The core of our analysis was our derivation of separate disk and bulge
luminosities and masses, based on innovative two-dimensional (2D) fitting to
the $K_s$-band surface brightness in 2MASS $K_{s}$ Atlas images of the SFRS
galaxies.
These luminosities were then used to estimate the bulge and disk masses.
Out of the 369 SFRS targets, we excluded from this analysis \sampleUnfittable{}
interacting galaxies with strongly disturbed morphologies, for which the definition
of ``bulge'' or ``disk'' is ambiguous and whose surface brightness distribution
cannot be represented with a parametric function.
Nonetheless, these galaxies are included in the determination of the integrated
galaxy mass function (see \S\ref{SFRS stellar mass functions}). 
Moreover, we excluded \sampleAGN{} targets which host a dominant AGN and which
appear as point-like sources in the 2MASS data.
Therefore, the actual sample we will consider hereafter is composed of \sample{}
galaxies.

There are a number of available tools for the 2D fit of astronomical images,
of which the most widely used are arguably \textsc{Gim2D} \citep{GIM2D} and
\GALFIT{} \citep{GALFIT}.
In their testing review, \cite{haussler} discussed the treatment of the uncertainties
of the two codes and their robustness in crowded-field environments.
They concluded that \GALFIT{} has an advantage in terms of dealing with contaminating
objects (which can be fit independently), although both codes underestimate
uncertainties in the fitted parameters.
For our analysis, we therefore opted to use \GALFIT{} v.3\footnote{
 \url{http://users.obs.carnegiescience.edu/peng/work/galfit/galfit.html}
}.
This version of \GALFIT{} implements the possibility of fitting
complex and asymmetric galaxy features such as spiral arms, rings, bars, etc. via 
Fourier modification of standard models (exponential disk, \Sersic{}, etc.) or
coordinate rotations along the radial direction.
Although extremely appealing, we decided not to make use of these new features,
because --- as they are currently implemented --- they are mostly based on the
observer's interpretation and because the quality of the 2MASS data generally did
not allow such level of detailed modelling.

Due to the relatively large number of targets (\sample{}), the variety of morphological
models we intended to apply, and the preparatory steps (e.g., PSF creation, object
masking, etc.), we created a fully-automated pipeline based on \GALFIT{} to
perform the 2D fits.
The resulting code is flexible and with minor changes can be adapted
to different types of data, given that it essentially requires as input just an
estimate of the point-spread-function (PSF) FWHM (FWHM$_{PSF}$) and a list of
target positions.

\subsection[Fit procedure]{Fit procedure}
\label{Fit procedure}

\noindent
Our fitting pipeline operates according to the following procedure:

\begin{cutenumerate}[label={\arabic*})~]
 \addtolength{\itemindent}{0.05\linewidth}
 \itemsep-0.0em  

 \item it sets up \GALFIT{} input files and first-guess parameters
 \item it creates masks for the contaminating objects
 \item it creates the PSF used for the model convolution
 \item it fits the targets with different parametric models:

 \vspace*{-0.25cm}
 \begin{enumerate}[label=4{\alph*}]
  \addtolength{\itemindent}{0.05\linewidth}
  \itemsep-0.0em  
  
  \item --- it first fits a simple \Sersic{} function
  \item --- it uses the results as input parameters for the next step
  \item --- it fits nested disk, bulge, and AGN components
 
 \end{enumerate} 
 \vspace*{-0.2cm}

 \item it calculates the magnitude zero-point
 \item it converts integrated luminosities into masses

\end{cutenumerate} 

\noindent
The details of the preparatory steps 1, 2, and 3 are reported in
Appendix \ref{Pre-fitting procedures}, while the calibration steps
5 and 6 are presented in
Appendix \ref{Zero-points, luminosities, and stellar mass calculation}; in the
remainder of this section we will focus instead on the fitting process.

\medskip

\noindent
The fit procedure itself consisted of multiple steps.
First, we fitted all the objects using the S\'{e}rsic function, with the
following variants:

\smallskip
\noindent\begin{tabular}{l@{ }l}
 $\blacktriangleright$~\Sersic{} model with free \Sersic{} index $n$      & (\Sersic{})\\
 $\blacktriangleright$~\Sersic{} model with fixed \Sersic{} index $n$ = 1 & (\Sersicone{})\\
 $\blacktriangleright$~\Sersic{} model with fixed \Sersic{} index $n$ = 4 & (\Sersicfour{})\\
\end{tabular}
\smallskip

\noindent
We introduced the \Sersic{} models with fixed index (i.e., \Sersic{}$^{n=1}$, and
\Sersic{}$^{n=4}$) in order to reduce the number of degrees of freedom and assist
the fit of problematic targets for which the free-index \Sersic{} fit failed.
Apart from the \Sersic{} indices, we allowed all other model parameters to
vary, except for the ``boxiness'' (or ``diskyness'').
This parameter gives a measure of the deviation from a perfect elliptical shape,
a feature routinely observed in the isophotes of galaxies throughout the whole
Hubble sequence (see \citealt{profiler} for a mathematical description of
isophote shapes).
However, a second order parameter such as the boxiness is not realistically
quantifiable for the vast majority of our sample galaxies, whose effective radii
are just a few pixels on the 2MASS images (or just over the PSF size).
On the contrary, we observed that galaxies with low bulge to disk ratios
were erroneously reproduced by extremely disky models with a single component
if the parameter was left free to vary.
Given the limitations of our data, we therefore decided to fix this parameter
to the value of 1 (i.e., no boxiness) for the current analysis.
The list of ``frozen''/``thawed'' parameters for the different \Sersic{} models
is summarized in Table \ref{Sersic_parameters}.

\begin{table}
 \centering
  \begin{tabular}{lccc}
   \hline
   \multicolumn{4}{c}{\textsc{GALFIT \Sersic{} Parameters}} \\
   \hline
   \hline
   \addlinespace 
   \multicolumn{1}{c}{Parameter}  &  \Sersic{} & \Sersicone{} & \Sersicfour{} \\
   \multicolumn{1}{c}{\tiny{(1)}} & \tiny{(2)}  & \tiny{(3)}  & \tiny{(4)}    \\
   \hline
   \addlinespace 
\textsc{Center (x,y)}   & \Circle & \Circle & \Circle\\
\textsc{Magnitude}      & \Circle & \Circle & \Circle\\
\textsc{$R_{e}$}        & \Circle & \Circle & \Circle\\
\textsc{Axis ratio}     & \Circle & \Circle & \Circle\\
\textsc{P.A.}           & \Circle & \Circle & \Circle\\
\textsc{\Sersic{}} $n$  & \Circle & \CIRCLE & \CIRCLE\\
\textsc{Boxiness}       & \CIRCLE & \CIRCLE & \CIRCLE\\
   \hline
  \end{tabular}
  \caption[Free/fixed parameters for the \Sersic{} models]{
   Free/fixed parameters for the \Sersic{} models.
   \\
   $^{(1)}$  Parameter type.
   $^{(2)}$$^{(3)}$$^{(4)}$ The $\Circle$ and $\CIRCLE$ symbols indicate free and
                            fixed parameters, respectively, for the \Sersic{},
                            \Sersicone{}, and \Sersicfour{} models.
   \label{Sersic_parameters}
  }
\end{table}

In each fit, we also included a constant to fit the sky background,
whose value was derived as discussed in Appendix \ref{Setting up GALFIT input}.
Although \GALFIT{} offers the possibility to add a ``gradient'' parameter to the sky
model, we preferred not to complicate the fit by adding this extra degree of freedom,
because: \emph{a}) we might introduce degeneracies with the parameters of the galaxy
models (see discussion in Appendix \ref{Setting up GALFIT input}), \emph{b})
we already limited any large-scale background variation by considering a fit region
specifically cropped around each source, and \emph{c}) the $K_{s}$-band background of
the 2MASS images (which is due to thermal continuum) in standard observational
conditions does not present high-frequency structures (as it often happens for the
$J$ and $H$-bands).

\GALFIT{} is based on a least-squares minimization; the fit statistic of $\chi^{2}$
is defined as:

\vspace{-0.3cm}
\begin{eqnarray}
 \chi^{2} = \displaystyle\sum_{x=1}^{nx} \displaystyle\sum_{y=1}^{ny} { (f_{data}(x,y) - f_{model}(x,y))^2 \over \sigma(x,y)^2}
 \label{equation:GALFIT_Chi}
\end{eqnarray}

\noindent
where $f_{data}(x,y)$ are the counts of the ($x$,$y$) pixel of the input image,
$f_{model}(x,y)$ are the counts of the corresponding pixel in the PSF-convolved 
model image generated at each iteration,  $x$ and $y$ refer to the pixels in the
images, and $\sigma(x,y)$ is the uncertainty or ``sigma'' image.
The sigma image is used to weight the discrepancy of the model to the data,
given the statistical uncertainty of the observed counts of each pixel,
and it is calculated internally by \GALFIT{}.
The 1-$\sigma$ errors are calculated for all free parameters based on their fit covariance
matrix \citep[][]{GALFIT}.
As an intuitive indicator of the goodness of fit we adopted the \emph{reduced} $\chi^{2}$
($\chi^{2}_{\nu}$), defined as the $\chi^{2}$ divided by the number of degrees of freedom
($N_{\rm DOF}$), i.e.:

\vspace{-0.3cm}
\begin{eqnarray}
 \chi^{2}_{\nu} = {\chi^{2} \over N_{\rm DOF}}
 \label{equation:GALFIT_Chi^2_nu}
\end{eqnarray}

We set some mild constraints on the fit parameters and in particular 
on the \Sersic{} index, the effective radius, and, most importantly, on
the model total magnitude, having found \GALFIT{} particularly sensitive to this
parameter.
These limits were based on the initial results from the \textsc{SExtractor}
run on each field (Appendix \ref{SExtractor measurements}), and they were
imposed in order to minimize the possibility of a failed or wrong fit rather
than to actually constrain the fit.
In fact, all the models which hit the parameter limits were attributed an
arbitrarily large magnitude error and were rejected by the magnitude criterion
imposed in the process of \emph{best-fit} model selection
($\S$\ref{Best-fit model selection}).
Additionally, we constrained the positions of contaminating objects
(within about $\pm$1 FWHM$_{PSF}$) to avoid large shifts of their centers
during the fit, which would result in their taking the place of a sub-component
of the main target.

Contaminating objects were fit with different models, based on their spatial
extent: small, quasi point-like sources were fit with a Gaussian model, while
for the extended ones we adopted a \Sersic{} model.
Small objects were distinguished from extended ones based on their effective
radius ($R_e$) as calculated by \textsc{SExtractor}, with the separating
threshold set at $R_e$ $ = $ 1.5 $\times$ FWHM$_{PSF}$.

We used the results of the single \Sersic{} fits as first-guess parameters
to implement more complex models, intended to separately account for different
galaxy components.
In particular, we wanted to model bulges and disks with \Sersic{} and
exponential (exDisk) profiles respectively, and AGNs with a PSF component.
The evaluation of the fit results, and the disk/bulge separation procedure
based on them will be presented in Section \ref{Best-fit model selection}.
These additional, more complex models were:

\noindent\begin{tabular}{l@{ }l}
 $\blacktriangleright$~\Sersic{} + PSF                    & (\Sersic{} + psfAgn)\\
 $\blacktriangleright$~\Sersic{} + exponential disk       & (\Sersic{} + exDisk)\\
 $\blacktriangleright$~\Sersic{} + exponential disk + PSF & (\Sersic{} + exDisk + psfAgn)\\
 \end{tabular}

\noindent
An accurate estimate of the first-guess input magnitude of each model component
turned out to be critical to guarantee the convergence of these fits.
We calculated the brightness of each component by re-distributing the integrated flux
of the \Sersic{} model obtained above: in the case of \Sersic{} + psfAgn, the PSF
component was initially attributed 1/10 of the total flux; in the other cases
the flux was re-distributed equally among the components.

At this stage, the constraints for the contaminating objects were simply inherited
from the previous \Sersic{} run.
For the targets of interest, we instead imposed new, stricter rules, given that
we could now rely on the first-guess parameters from the \emph{best-fit} \Sersic{}
models.
In particular, we enforced that: \emph{a}) the centres of all the components
(either \Sersic{}, exDisk, or psfAgn) associated to a target were locked within
a range of $\pm$1/2 FWHM$_{PSF}$, and \emph{b}) the effective
radius of a disk should be at least 20\% more than that of the associated bulge.
The latter measure ensured that the bulge component, as we physically
interpret it, was automatically contained within the disk component.
By imposing this, we of course ignore embedded disks: those features are not
expected to be observable in the few early-type galaxies in our sample, given the
poor 2MASS resolution ($\S$\ref{The 2MASS data for the SFRS targets}).

\begin{figure*}
 \centering

 \begin{minipage}[b]{0.24\textwidth}
  \centering
  \textsc{\large Data}
 \end{minipage}
 \begin{minipage}[b]{0.24\textwidth}
  \centering
  \textsc{\large Model}
 \end{minipage}
 \begin{minipage}[b]{0.24\textwidth}
  \centering
  \textsc{\large Residuals}
 \end{minipage}
 \begin{minipage}[b]{0.24\textwidth}
  \centering
  \textsc{\large Projection}
 \end{minipage}
  
 \vspace{0.2cm}

 \makebox[\linewidth]{
  \includegraphics[align=c,width=0.24\textwidth]{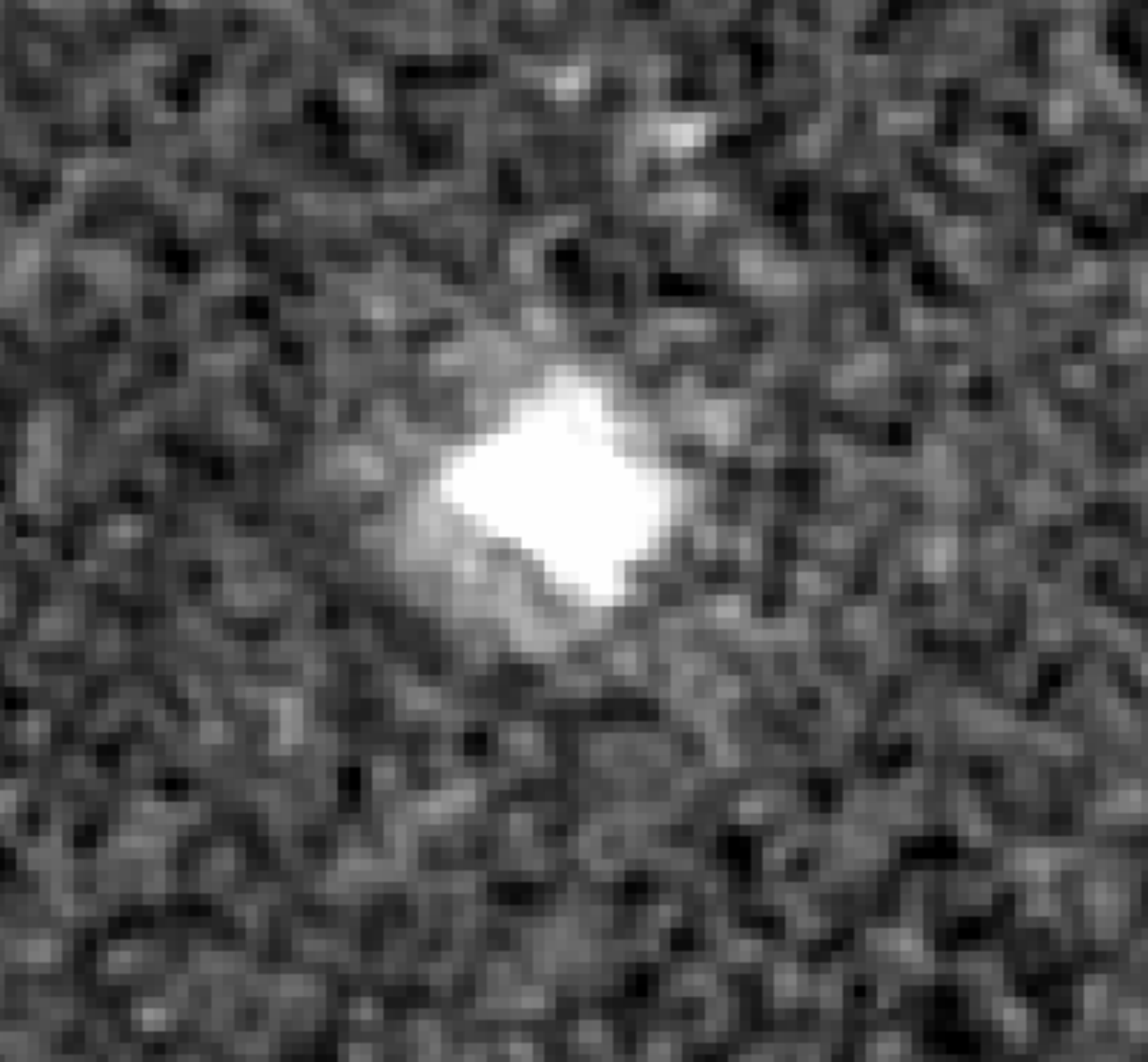}
  
  \begin{minipage}[c]{0.24\textwidth} 
   \begin{overpic}[align=c,width=1\textwidth]
    {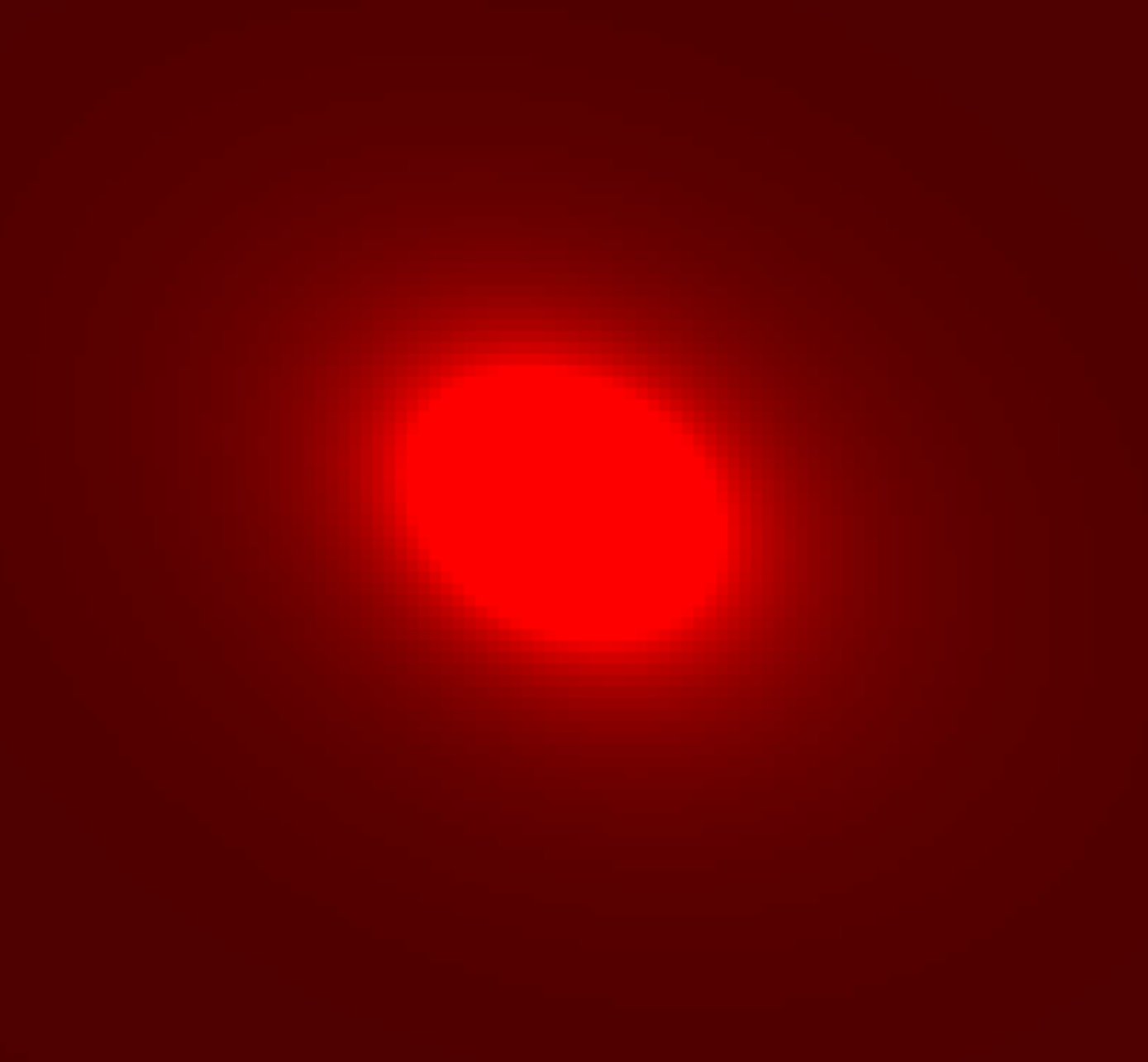}
    \put(50,10){\makebox(0,0){\textcolor{white}{\large \Sersic{}}}}
   \end{overpic}
  \end{minipage}
  
  \includegraphics[align=c,width=0.24\textwidth]{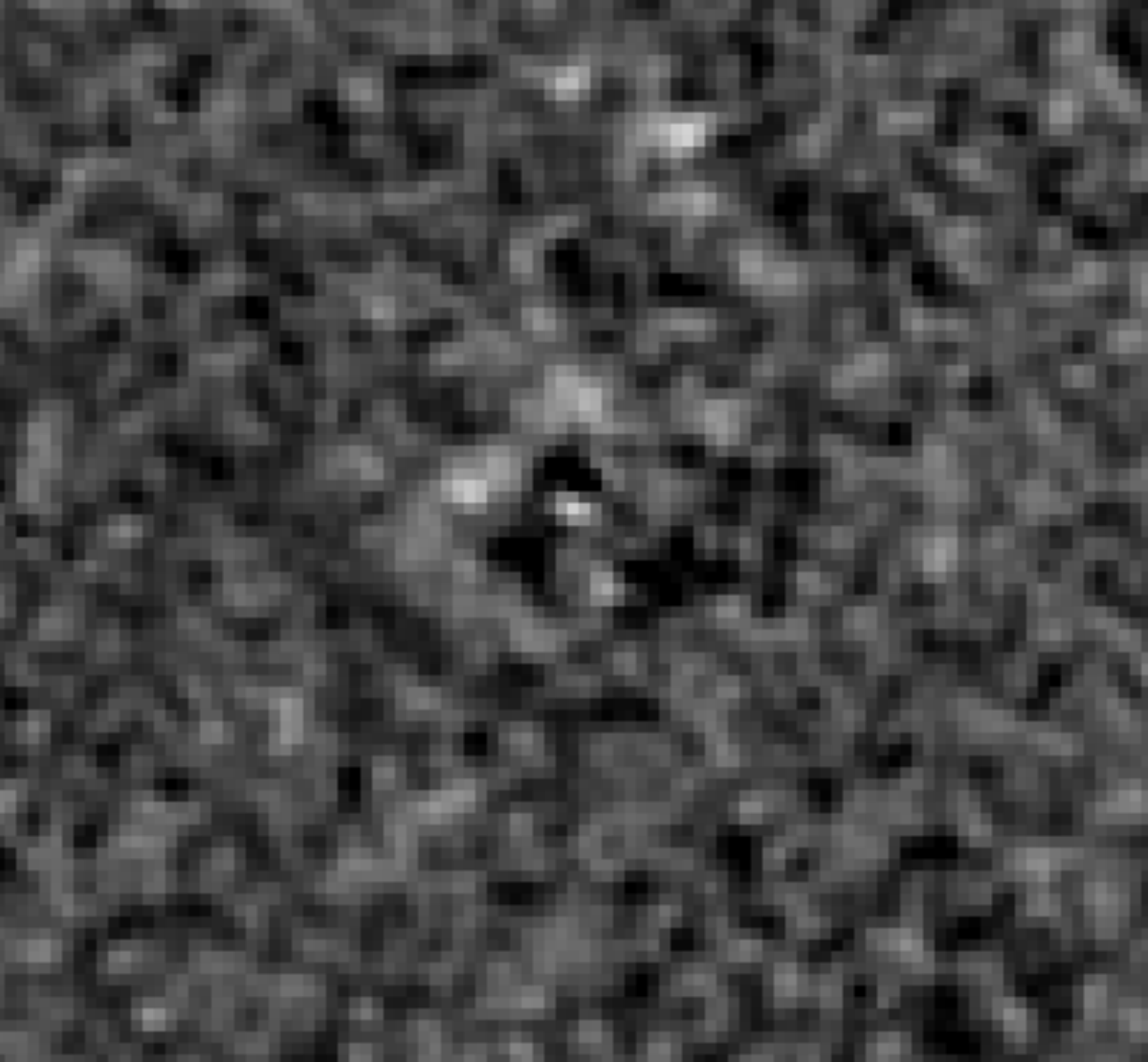}

  \fbox{\includegraphics[align=c,width=0.24\textwidth,height=0.155\textheight]{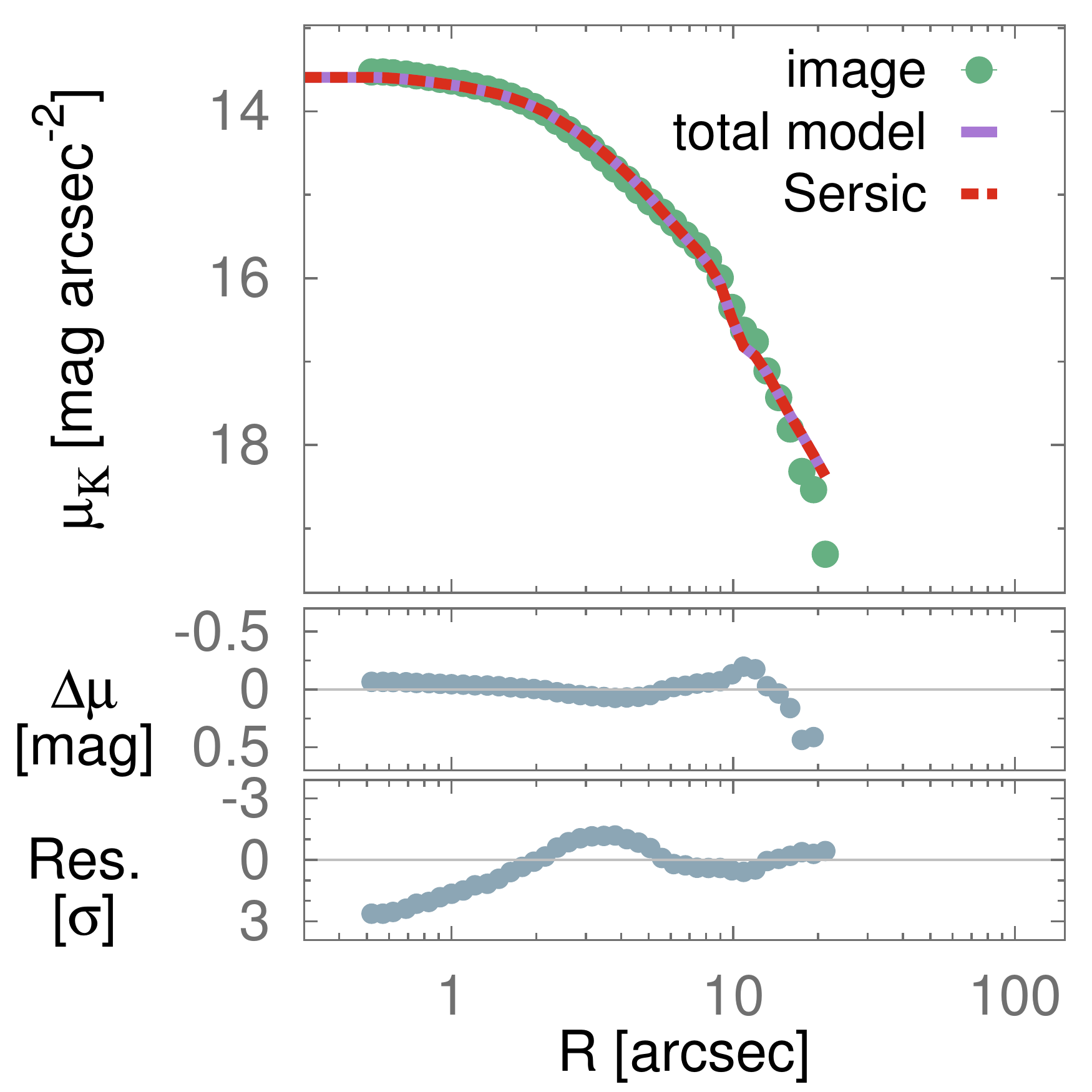}}
  
 }
 
 \vspace{0.2cm}

 \makebox[\linewidth]{
  \includegraphics[align=c,width=0.24\textwidth]{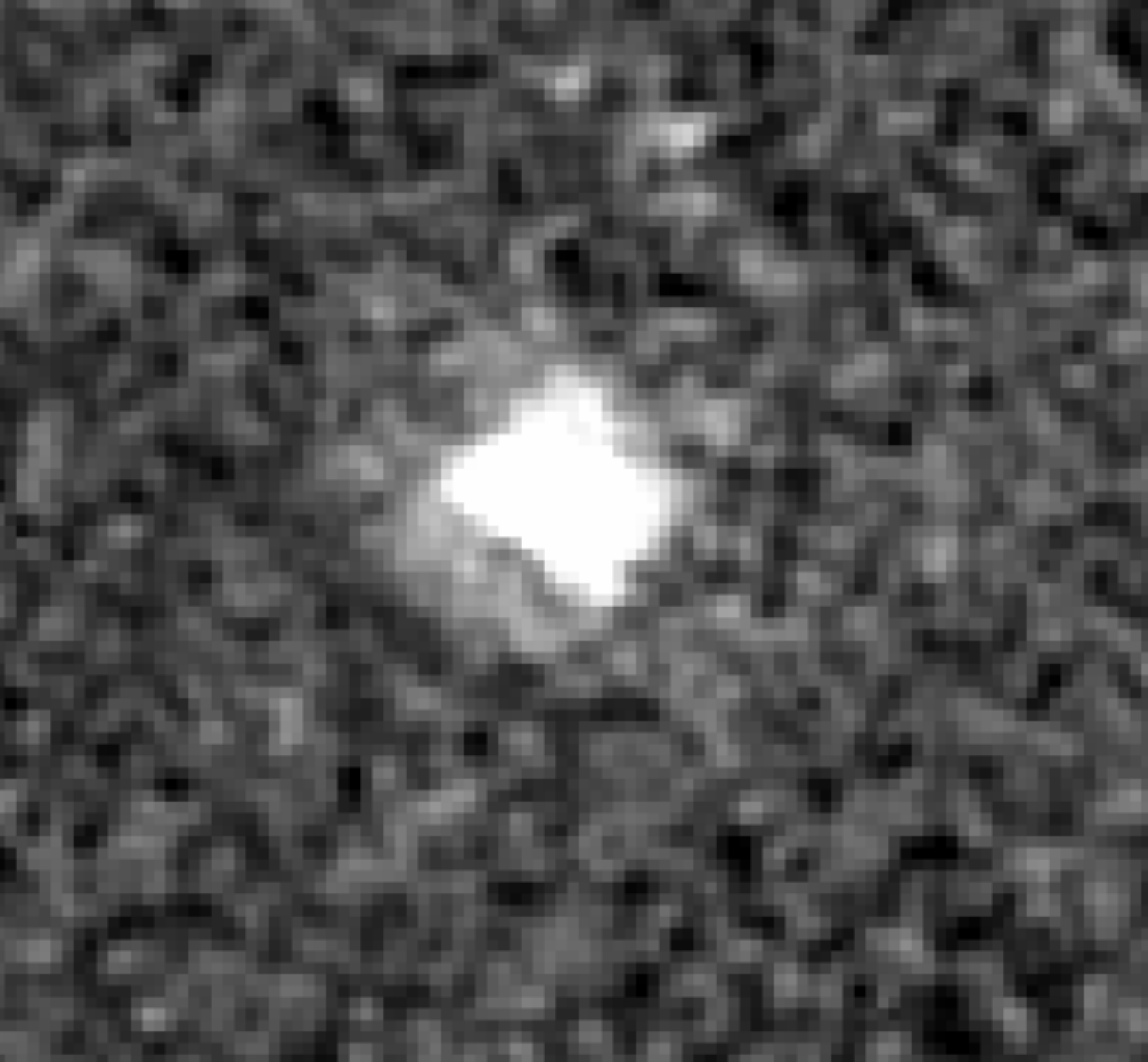}
  
  \begin{minipage}[c]{0.24\textwidth} 
   \begin{overpic}[align=c,width=1\textwidth]
    {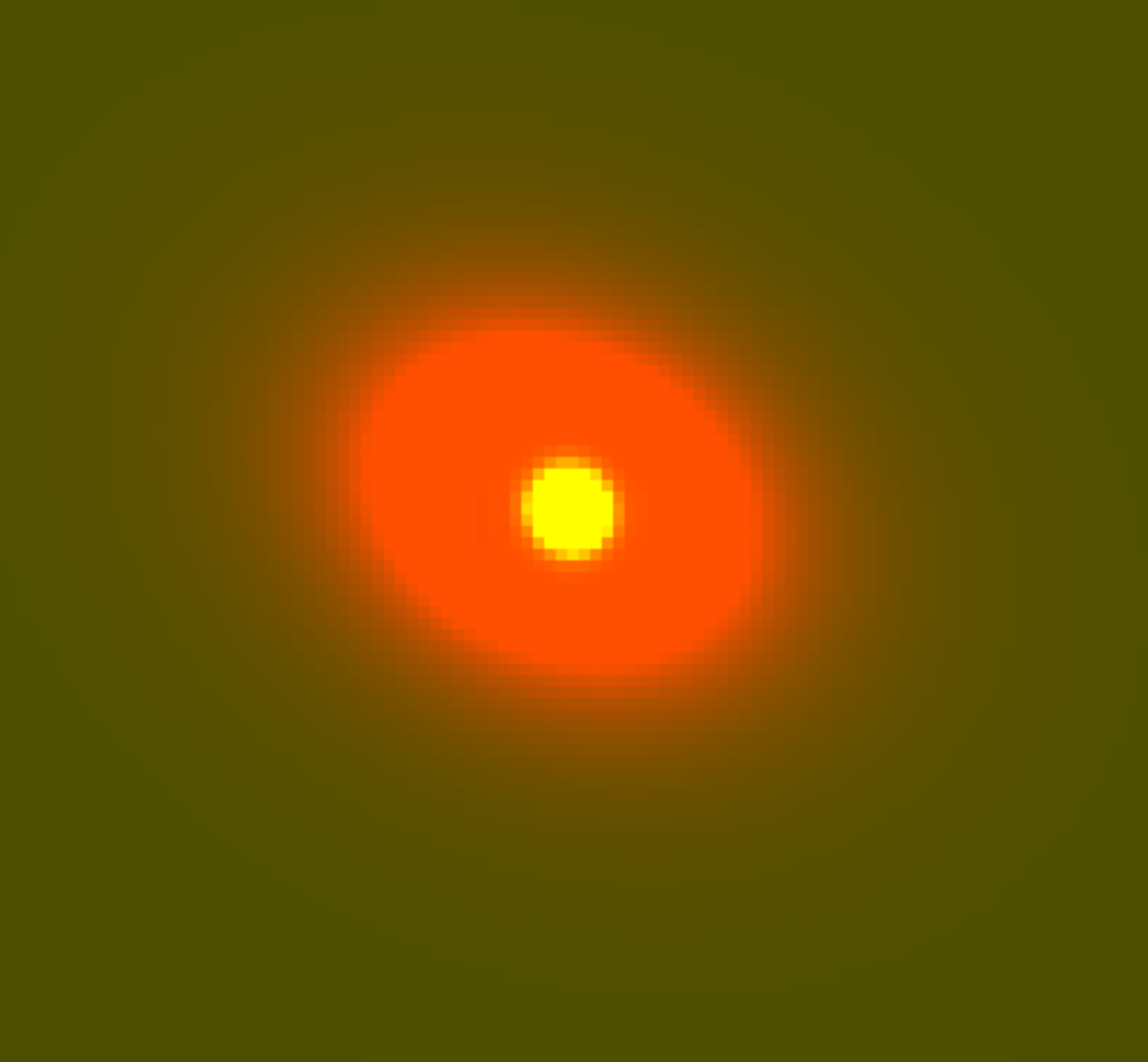}
    \put(50,10){\makebox(0,0){\textcolor{white}{\large \Sersic{} + psfAgn}}}
   \end{overpic}
  \end{minipage}
  
  \includegraphics[align=c,width=0.24\textwidth]{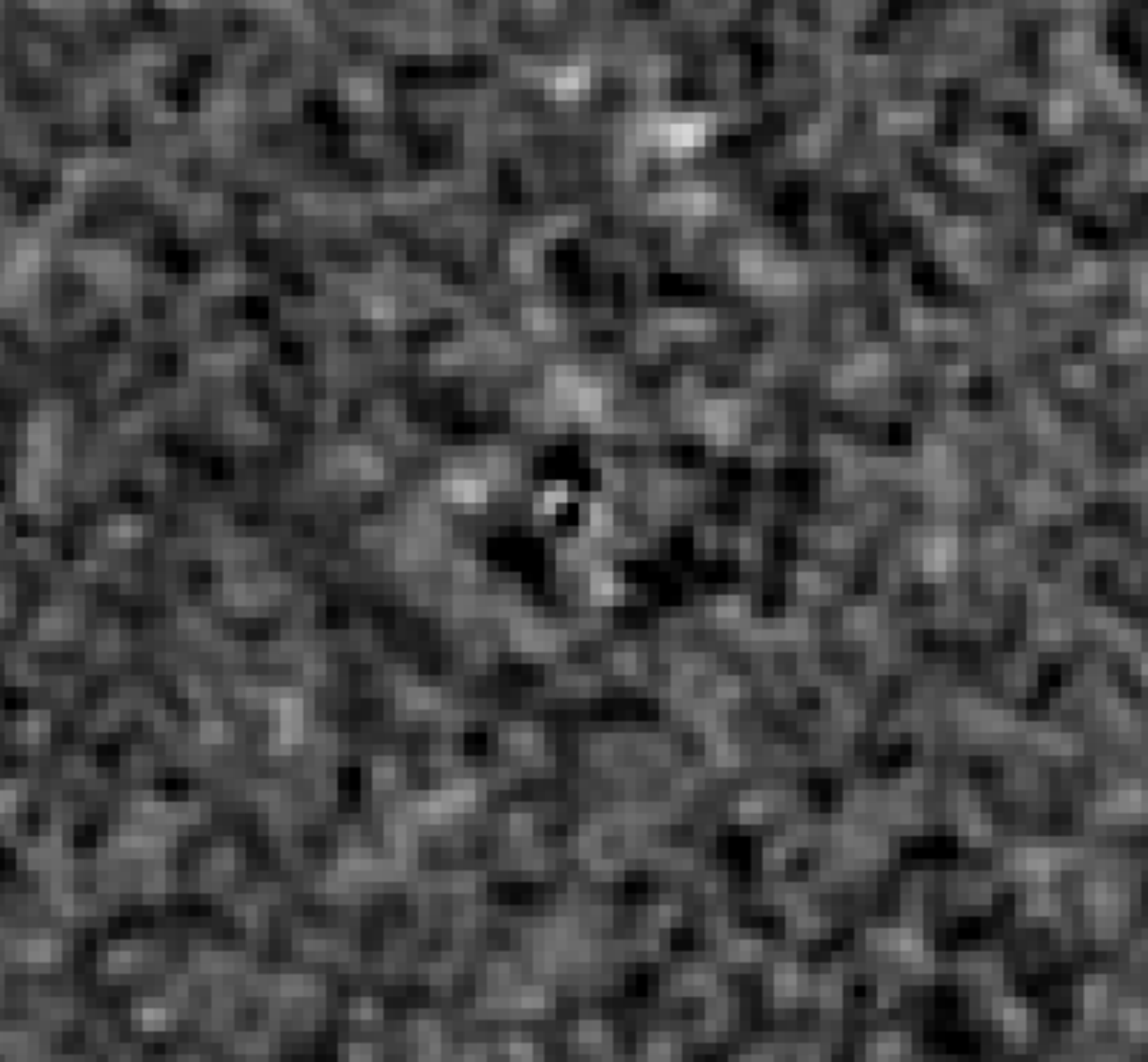}

  \fbox{\includegraphics[align=c,width=0.24\textwidth,height=0.155\textheight]{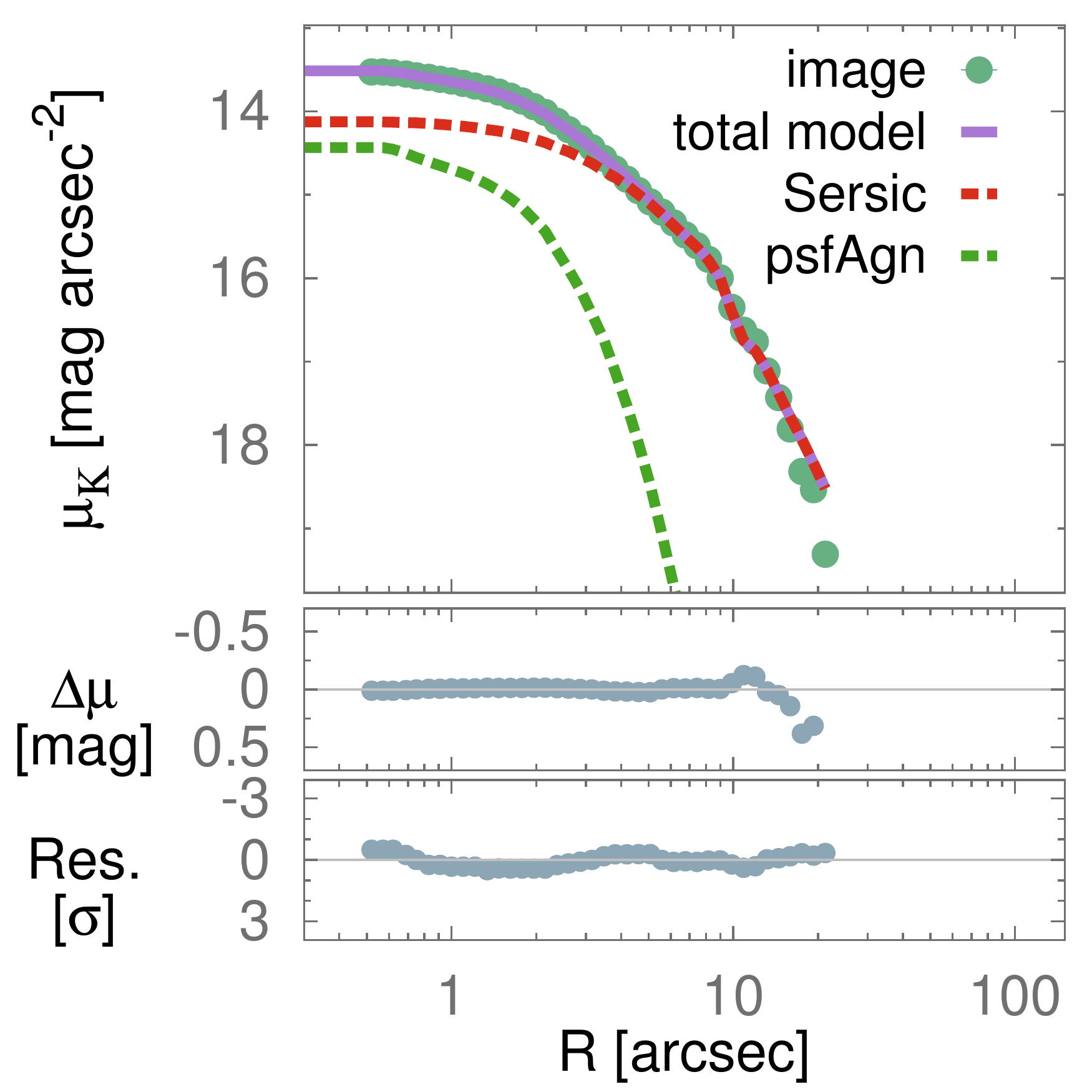}}
  
 }
 
 \vspace{0.2cm}
 
 \makebox[\linewidth]{
  \includegraphics[align=c,width=0.24\textwidth]{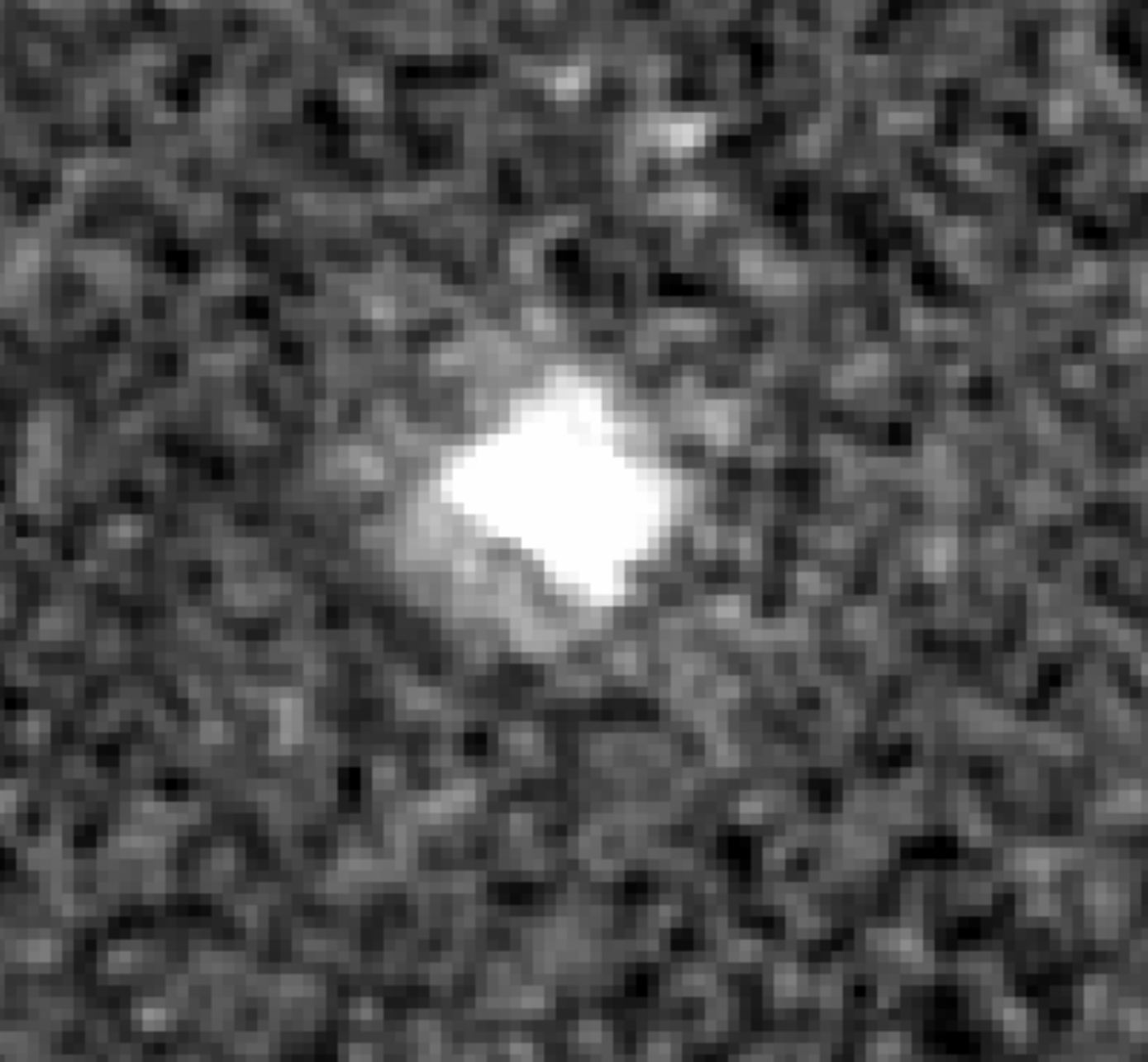}
  
  \begin{minipage}[c]{0.24\textwidth} 
   \begin{overpic}[align=c,width=1\textwidth]
    {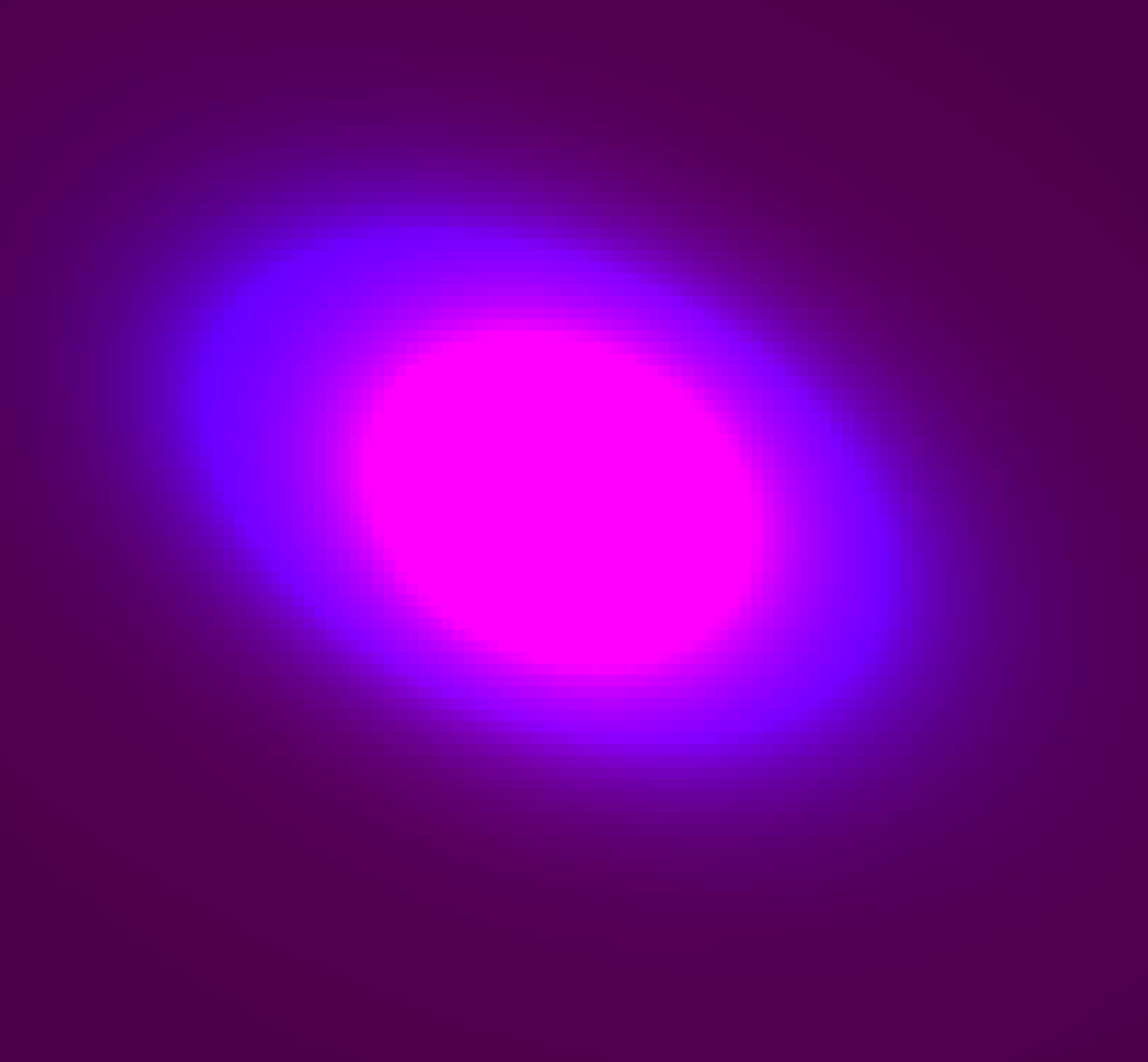}
    \put(50,10){\makebox(0,0){\textcolor{white}{\large \Sersic{} + exDisk}}}
   \end{overpic}
  \end{minipage}
  
  \includegraphics[align=c,width=0.24\textwidth]{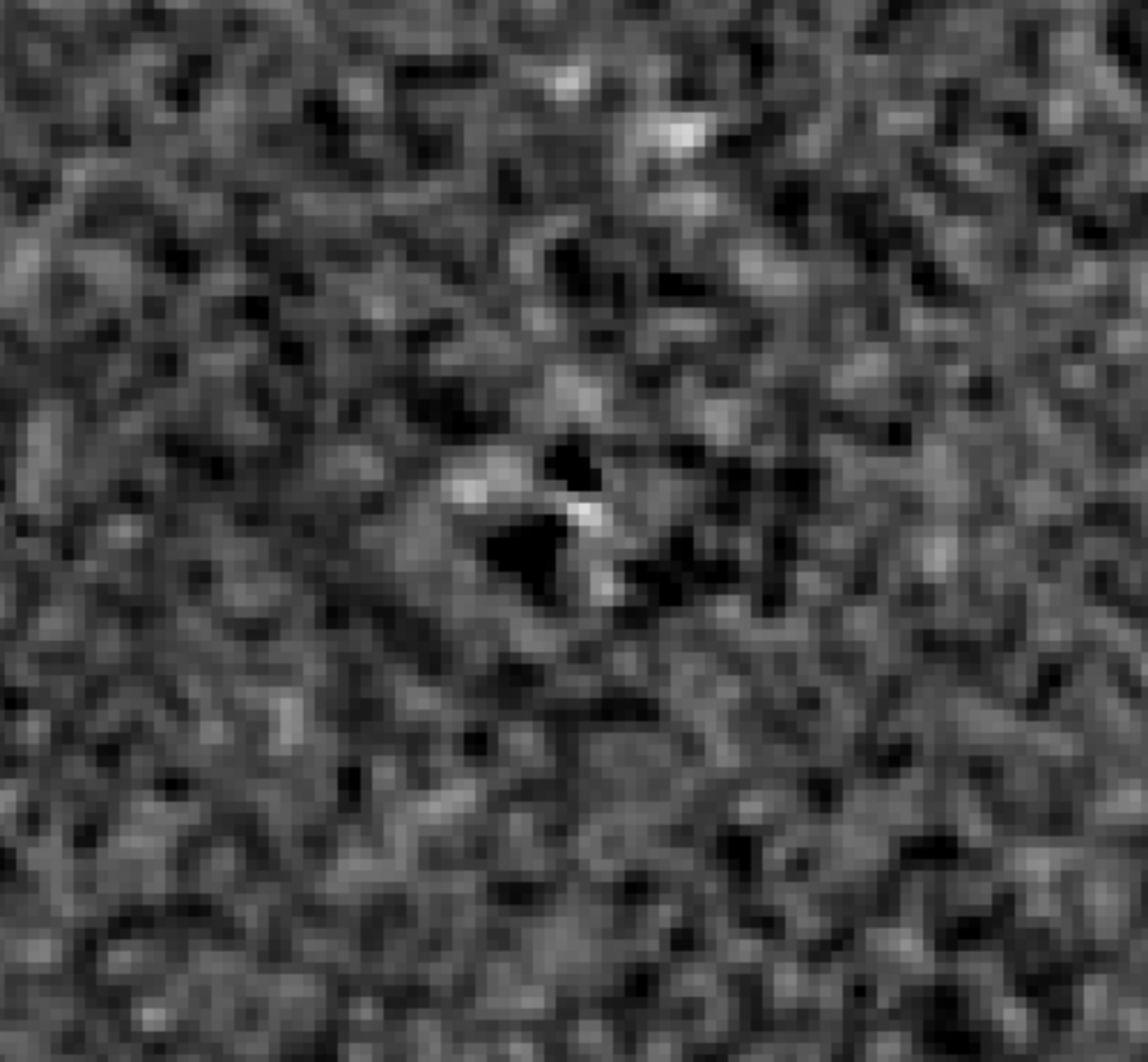}

  \fbox{\includegraphics[align=c,width=0.24\textwidth,height=0.155\textheight]{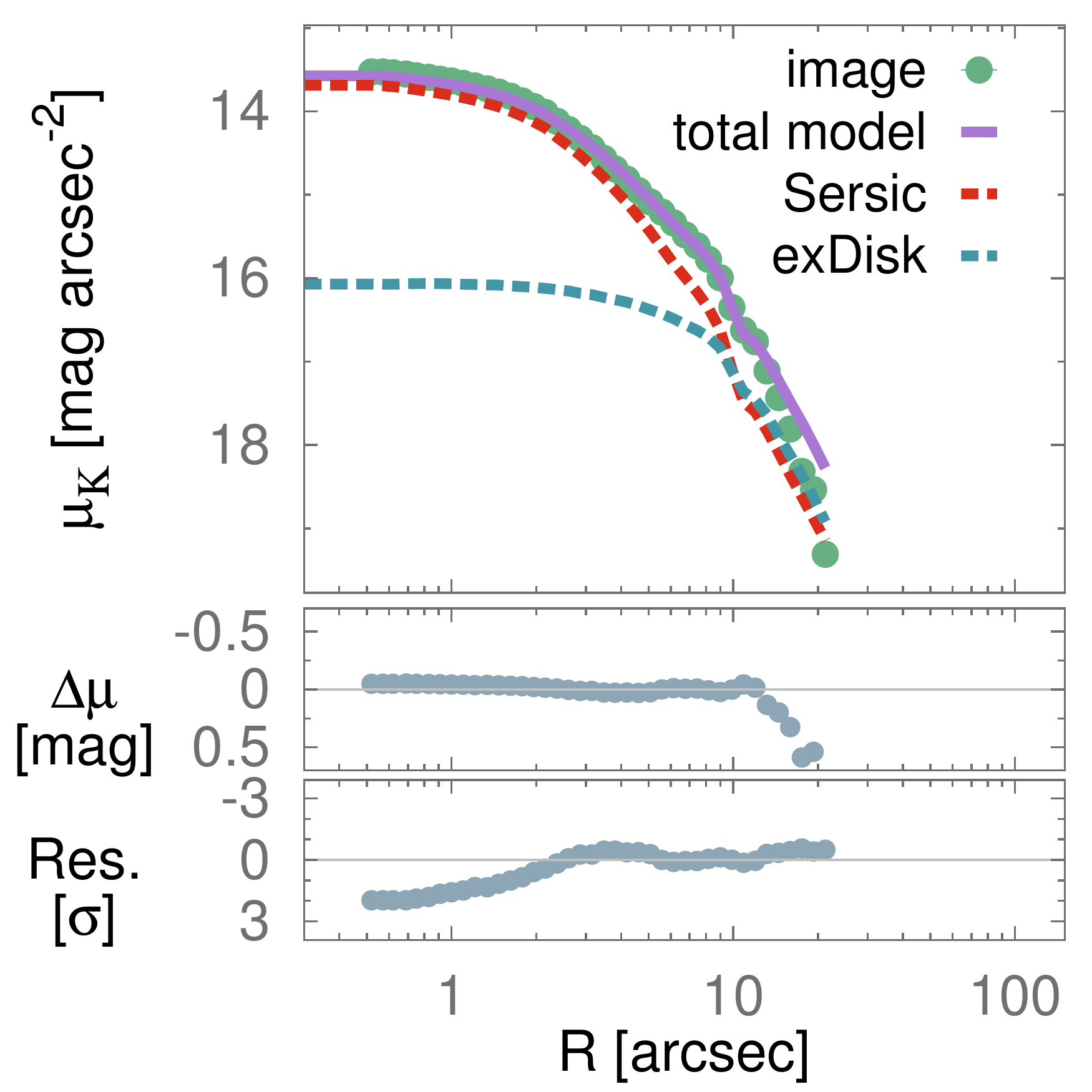}}
  
 }
 
 \vspace{0.2cm}
 
 \makebox[\linewidth]{
  \includegraphics[align=c,width=0.24\textwidth]{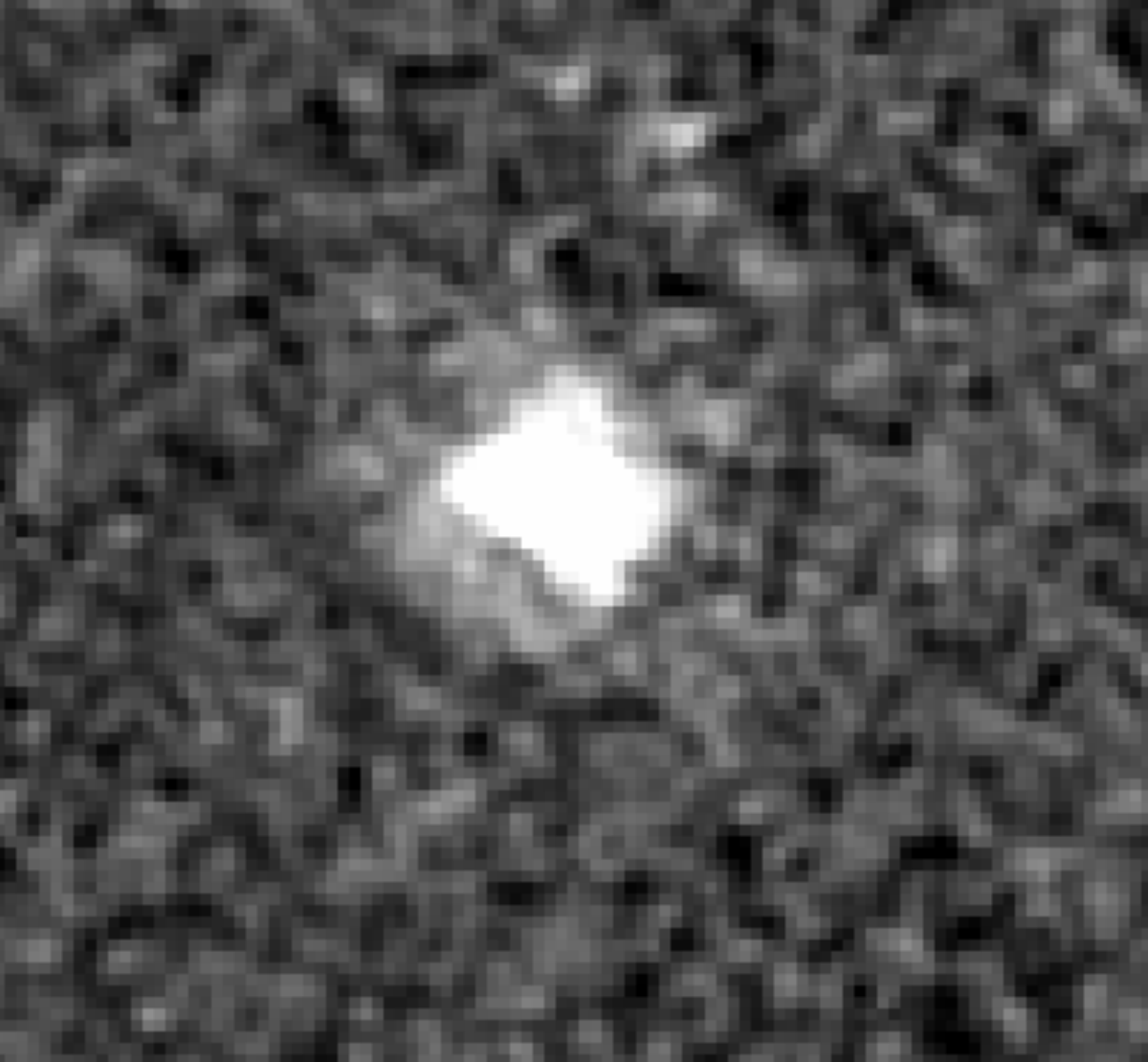}
  
  \begin{minipage}[c]{0.24\textwidth} 
   \begin{overpic}[align=c,width=1\textwidth]
    {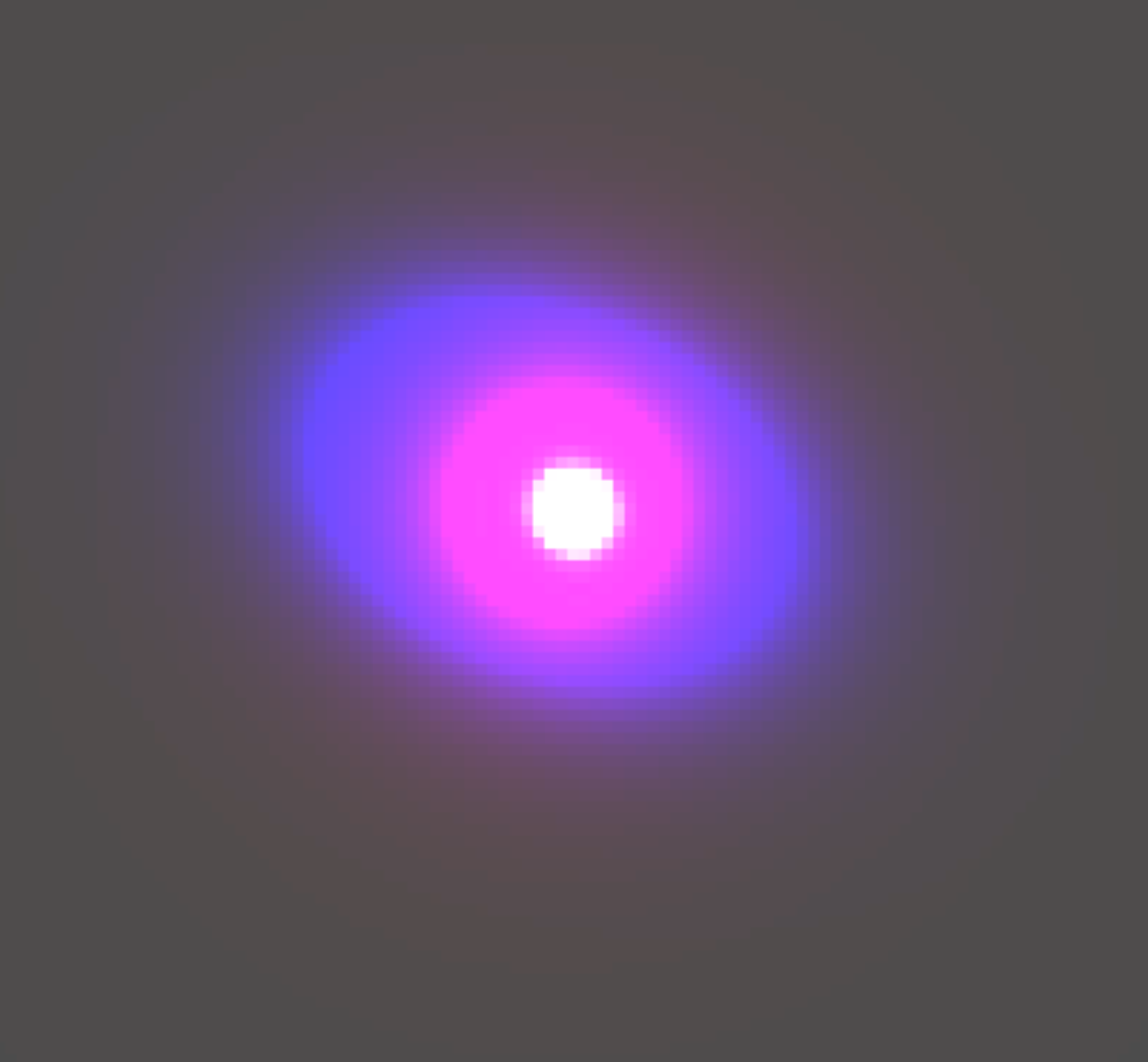}
    \put(50,10){\makebox(0,0){\textcolor{white}{\large \Sersic{} + psfAgn + Disk}}}
   \end{overpic}
  \end{minipage}
  
  \includegraphics[align=c,width=0.24\textwidth]{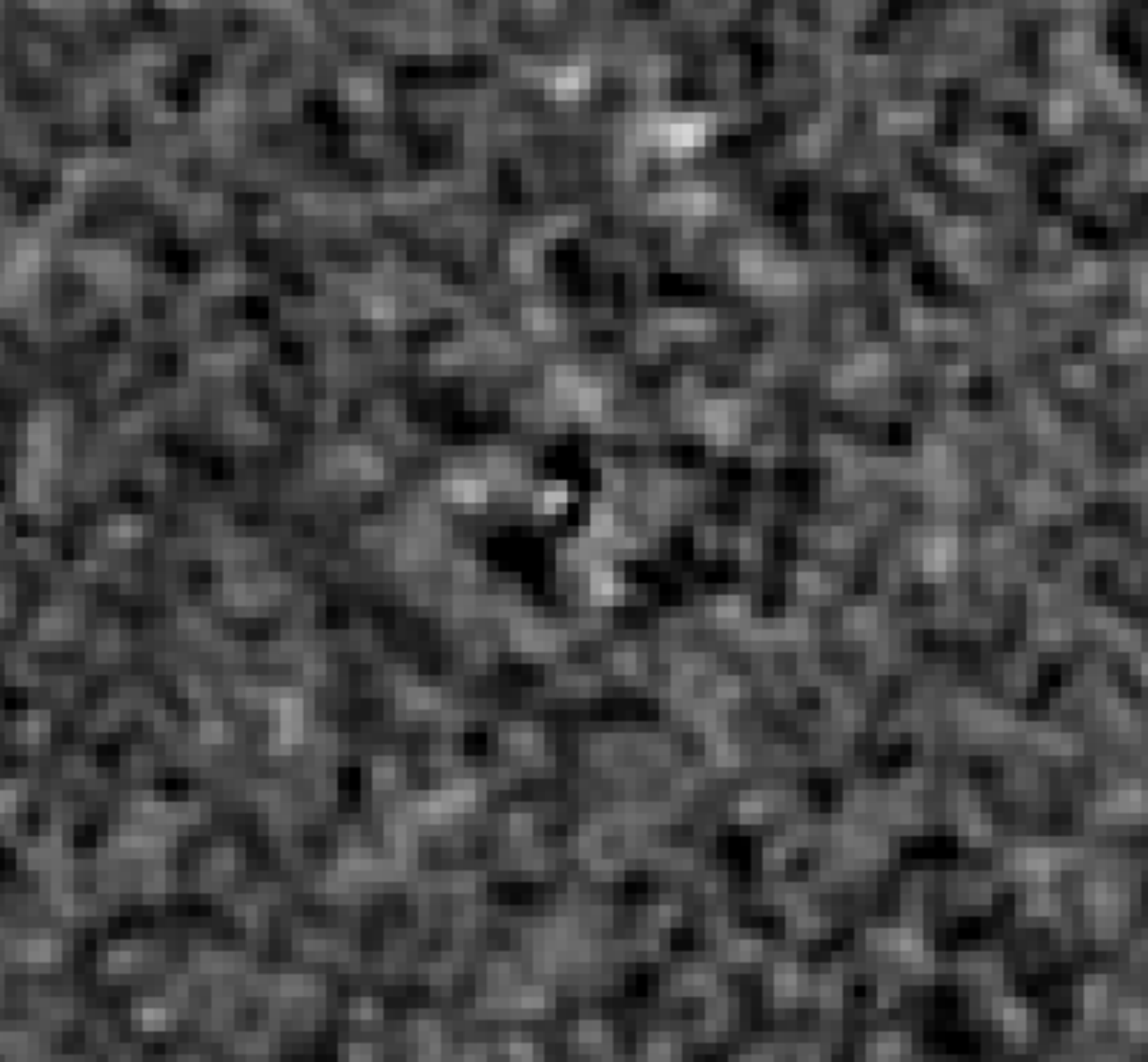}

  \fbox{\includegraphics[align=c,width=0.24\textwidth,height=0.155\textheight]{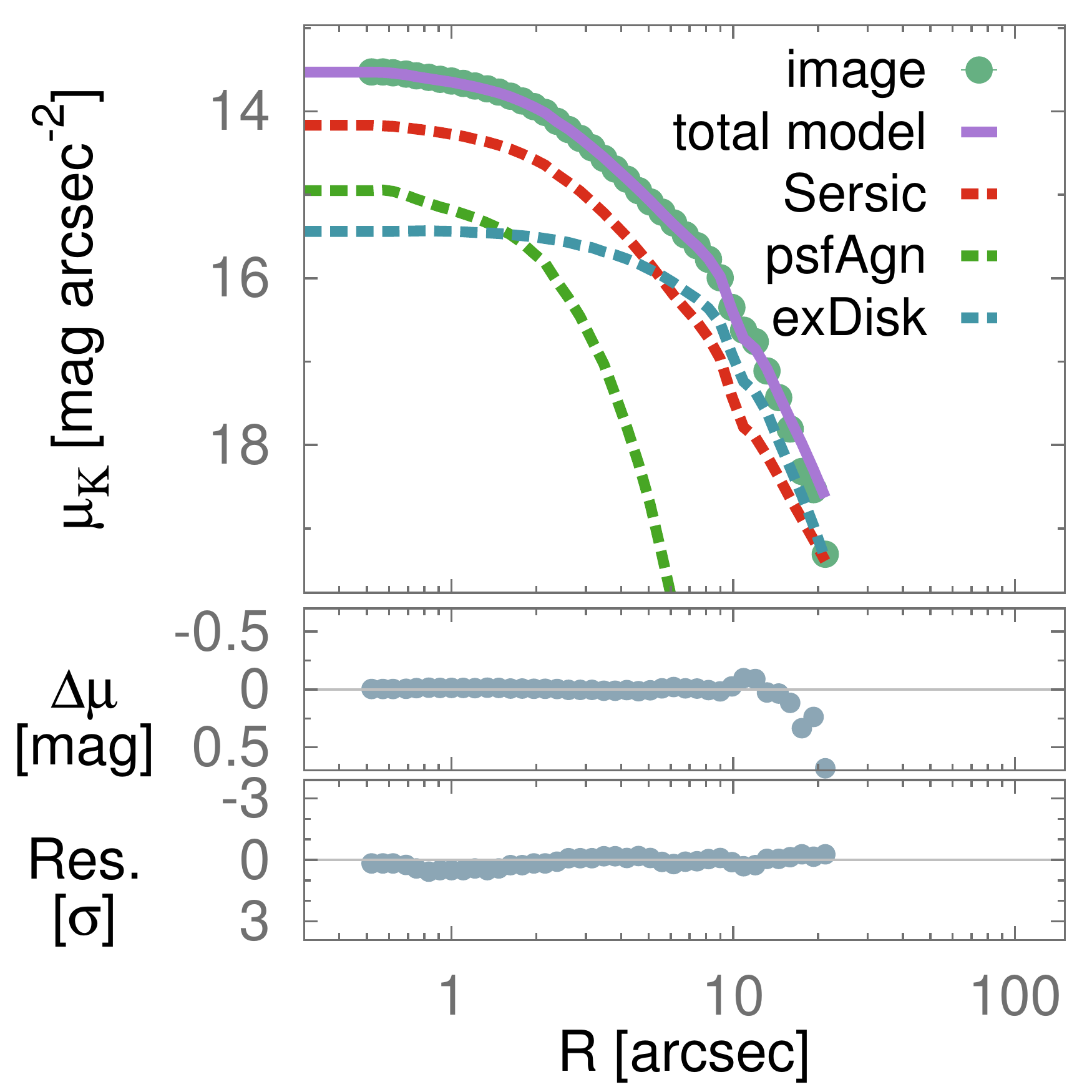}}
  
 }
 
 \vspace{0.2cm}
 
 \caption[Results of the fits for the object UGC~6625]{
  Results of the fits for the object UGC~6625, performed using the models presented
  in $\S$\ref{Fit procedure}.
  The first, second, and third columns from the left show the data, model, and
  residuals, respectively, for the (\emph{from top to bottom}) \Sersic{},
  \Sersic{} + psfAgn, \Sersic{} + exDisk, and \Sersic{} + psfAgn + exDisk models.
  The model images (center-left) are arbitrarily color-coded in order to highlight the
  separate \Sersic{} (red), psfAgn (green), and exDisk (blue) components and to show
  their relative intensity, extension, and centres.
  The right-hand column shows one-dimensional radial projections of the data, models,
  and residuals along concentric isophotes: the image data are shown in green, the total
  model is shown with a violet line, and its sub-components with dashed lines
  coloured according to the same  scheme as for the model image.
  We stress that the curves in the plots do \emph{not} represent the ``standard'' fits
  to the major-axis surface brightnesses but are rather 1D radial projections of
  the separate 2D images.
  These radial projections were obtained with the technique described in
  \cite{bonfini:Holm15A}.
  In brief: we first measured the galaxian surface brightness along the elliptical
  isophotes identified on the data image by the IRAF.\emph{ellipse} task \citep{ellipse};
  then, we obtained the radial profiles from the images of the 2D model (for the individual
  sub-components as well as their sum, i.e., the total emission) by measuring their surface
  brightnesses over exactly the same isophotes.
  Finally, the two lower panels of each plot present the projections of the residuals
  (gray points), first expressed in terms of mag/arcsec$^{2}$, and then in terms of
  data noise ($\sigma$).
  For this specific target, the \emph{best-fit} model (selected as described in
  $\S$\ref{Best-fit model selection}) is the \Sersic{} + psfAgn + exDisk.
  For this object we can ignore the \Sersicone{} and \Sersicfour{} models,
  because the [free index] \Sersic{} yielded a successful fit.
 }
 \label{figure:fit_models}

\end{figure*}

Figure \ref{figure:fit_models} illustrates our fitting procedure for
one typical galaxy, UGC~6625.
The fit statistics for each model of UGC~6625 are reported in Table
\ref{table:UGC6625}.
This specific target contains an AGN (see Table \ref{table:GALFIT_selected_models}),
and the \emph{best-fit} model turned out to be \Sersic{} + psfAgn + exDisk.
Figure \ref{figure:fit_models} highlights how difficult it is to distinguish by eye
the \emph{best-fit} model, especially because models with more
components will automatically yield better residuals without necessarily
providing a statistically significant improvement of the fit (i.e., without being
more representative).
Our model selection procedure, designed to overcome this issue, is described in
$\S$\ref{Best-fit model selection}.

\begin{table}
 \centering

 \begin{tabular*}{0.4\textwidth}{l @{\extracolsep{\fill}} cc}
  \hline
  \multicolumn{3}{c}{\textsc{Fit Results for UGC~6625}} \\
  \hline
  \hline
  \addlinespace 
  \multicolumn{1}{c}{Model}      & $\chi^2_{\nu}$                  & $\sigma_{XS}^2$               \\
  \multicolumn{1}{c}{\tiny{(1)}} & \multicolumn{1}{c}{\tiny{(2)}} & \multicolumn{1}{c}{\tiny{(3)}} \\
  \hline
  \addlinespace 
  \Sersic{}                   & 1.141 & 1.451 (0.111) \\
  \Sersic{} + psfAgn          & 1.136 & 1.185 (0.103) \\
  \Sersic{} + exDisk          & 1.142 & 1.608 (0.116) \\
  \Sersic{} + psfAgn + exDisk & 1.136 & 1.325 (0.107) \\
  \hline

 \end{tabular*}
 \caption[Statistical results of the \GALFIT{} fits to the object UGC~6625]{
  Statistical results of the \GALFIT{} fits to the object UGC~6625, for the
  models shown in Figure \ref{figure:fit_models}.
  \\
  $^{(1)}$ Fit model.
  $^{(2)}$ Reduced $\chi^{2}$.
  $^{(3)}$ Excess variance (described in Section \ref{Best-fit model selection}).
  \label{table:UGC6625} 
 }
\end{table}

\section[Best-fit model selection]{Best-fit model selection}
\label{Best-fit model selection}

\noindent
For each of the \sample{} SFRS galaxies composing our sample, our pipeline attempted
a fit using all of the following models (see $\S$\ref{Fit procedure}):

\smallskip

\noindent\begin{tabular}{l@{ }l}
 $\blacktriangleright$~\Sersic{} model with free  index $n$     & (\Sersic{})\\
 $\blacktriangleright$~\Sersic{} model with fixed index $n$ = 1 & (\Sersicone{})\\
 $\blacktriangleright$~\Sersic{} model with fixed index $n$ = 4 & (\Sersicfour{})\\
 $\blacktriangleright$~\Sersic{} + PSF                          & (\Sersic{} + psfAgn)\\
 $\blacktriangleright$~\Sersic{} + exponential disk             & (\Sersic{} + exDisk)\\
 $\blacktriangleright$~\Sersic{} + exponential disk + PSF       & (\Sersic{} + exDisk + psfAgn)\\
 \end{tabular}

\smallskip

\noindent
Not all of the fits were necessarily successful for each of the targets,
meaning that in some cases \GALFIT{} did not converge (hence providing unreliable
parameter values) or it crashed due to any of the parameters exceeding the hard-coded
thresholds.
Actually, for \sampleNotFitted{} objects we could not obtain any fit with the
aforementioned models.
These galaxies are OJ~287, IRAS~13218+0552, and IRAS~11069+2711: a visual inspection
of their 2MASS images revealed that these objects are only marginally resolved,
hence not allowing any meaningful spatial fit.
For an additional \sampleBadMag{} galaxies, the recovered magnitudes from all the
models differed by more than 1~mag from the corresponding 2MASS counterpart (see below).
These are galaxies affected by severe contamination by neighbouring objects,
an issue which is properly accounted for in our routines but not in the 2MASS
pipelines.
For a robust consistency check with the 2MASS catalogue we discarded the above
\sampleBadMag{} galaxies (accounting for only $\sim$3\% of the sample), hence
applying our analysis to the remaining \sampleBestFits{} objects.

Among the valid, successful fits (which often had very similar \Chisq{} and residual
patterns; see e.g.,\ Table \ref{table:UGC6625} and Figure \ref{figure:fit_models}),
we had to determine the ``best-model'' (i.e., the most appropriate among the
\emph{best-fits} of the 6 models presented above).
The selected model automatically defines the separation of any bulge or disk
component of the galaxy (see $\S$\ref{Separation of disk/bulge components}).
Therefore, the selection procedure had to take into account both the fit results
and morphological considerations.

\medskip
\noindent
The selection sequence we applied to determine the best-model from the original
pool of models was.

\begin{cutenumerate}[label=\arabic*)~]
 \itemsep0.5em  

 \newcounter{point}

 \stepcounter{point}
 \item[\arabic{point})~] \emph{Rejection of unphysical results ---}
   Figure \ref{figure:K_mag_GALFIT_2MASS}
   compares the 2MASS $K_{s}$-band magnitudes against those computed by \GALFIT{}.
   There is, in general, good agreement between the 2MASS and the \GALFIT{} data,
   although a few models present (often at the same time) two kinds of
   suspect behaviours: (1) unrealistically large uncertainties in their integrated
   magnitudes\footnote{
    Calculated as the sum in quadrature of the uncertainty on the fit magnitude
    and the zero-point uncertainty.
   }
   compared to the magnitude uncertainties of other model fits to the same
   target; (2) significant discrepancy with respect to the integrated 2MASS
   magnitudes.
   The former issue is related to non-converging fits or numerical issues during
   the \GALFIT{} iterations, while the latter is due to the presence of
   contaminating objects (not properly treated by the 2MASS pipeline) or
   to models not accounting for the central AGN component (see point 2).
   
\begin{figure*}

 \centering
 
  \begin{minipage}[t]{0.48\textwidth}
   \includegraphics[width=\textwidth,angle=0]{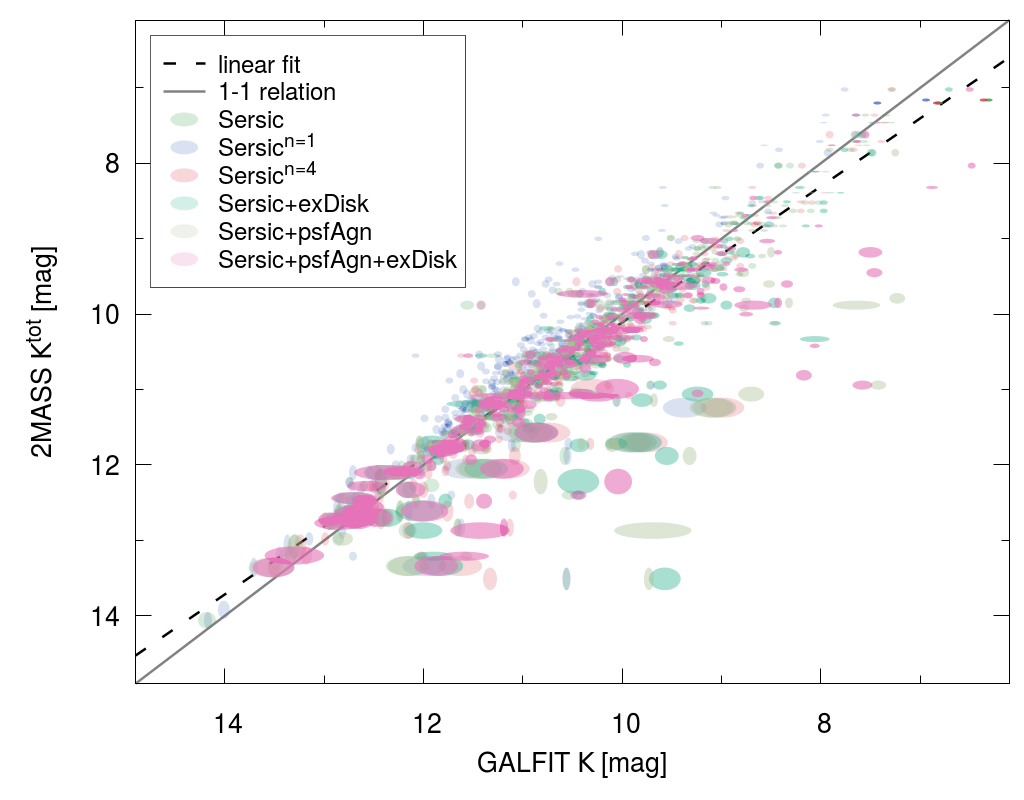}
   \caption[Comparison between 2MASS and \GALFIT{} magnitudes - All models]{
    Comparison of catalogued 2MASS total magnitudes to those derived by our
    \GALFIT{}-based fitting procedure described in $\S$\ref{The 2D fit of SFRS galaxies}
    for all the targets and all the models.
    The results plotted here refer to all successful fits for all SFRS galaxies
    (the 6 different models applied to each object are coded by different colors).
    The sizes of the ellipses represent the statistical uncertainties.  
    The solid line shows the 1-to-1 relation, while the dashed black line is
    a linear fit to the data.
    \label{figure:K_mag_GALFIT_2MASS} 
   }
 \end{minipage}
 \hspace*{0.2cm}
 \begin{minipage}[t]{0.48\textwidth}

   \vspace*{-7.2cm} 
   \includegraphics[width=1.1\textwidth,angle=0]{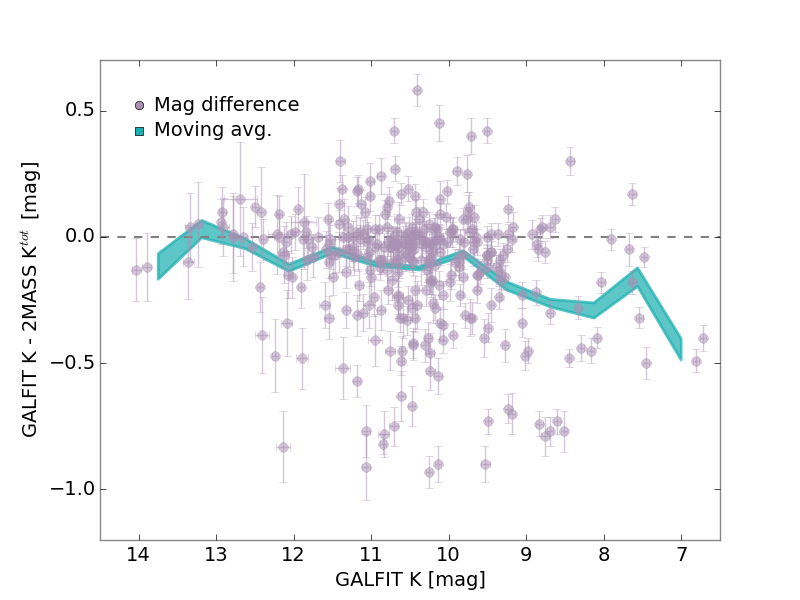}
   \vspace*{0.2cm} 
   \caption[Comparison between 2MASS and GALFIT magnitudes - Selected models]{
    Comparison between the 2MASS total magnitudes and the ones computed by \GALFIT{},
    for the selected models.
    The data points represent the magnitude difference as a function of the \GALFIT{}
    magnitude, while the shaded area shows the moving average (and its uncertainty),
    which has been evaluated adopting a smoothing box of 0.5~mag in size.
    The upper and lower bounds of the curve represent the 1-$\sigma$ width of the
    distribution of moving averages calculated over 500 mock samples generated
    by Monte-Carlo sampling the data points within their respective errors.     
    \label{figure:masses_selected_Dmag_plot}
   }
 \end{minipage}

\end{figure*}
   
   \smallskip
   Although many of these erroneous results would be rejected in the following
   steps of our analysis (because they are not trustworthy due to their low
   statistics; e.g.,\ low $\chi_{\nu}^2$), we imposed sharp limits on the magnitude
   uncertainties ($\delta{M_K}$ $<$ 1~mag) and on their difference from the 2MASS
   magnitudes (\mbox{$|m_{GALFIT}$ - $m_{2MASS}|$ $<$ 1~mag}) in order to exclude
   the most deviant models.
   The latter margin was based on the scatter of the points on the \GALFIT{}--2MASS
   comparison (Figure \ref{figure:K_mag_GALFIT_2MASS}) around their best-fit line, and it
   corresponds to $\sim$6 times the average uncertainty of the faintest 2MASS
   objects.
   We chose this conservative value in order not to excessively penalize our
   photometry, which has been performed with higher accuracy than that of 2MASS.

 \item[2.a)~] \emph{Accounting for AGNs ---}
   As mentioned at the beginning of $\S$\ref{The 2D fit of SFRS galaxies}, we excluded
   from our sample any SFRS galaxy unresolved in the 2MASS images
   and hosting an AGN due to the impossibility of performing a bulge/disk
   decomposition (namely, we excluded: IC~486, 3C~273, UGC~8058, IRAS~13144+4508,
   and UGC~8850).
   However this is not sufficient, because the uncertainty about the presence of
   an AGN can severely influence the modelled near-IR morphology of a galaxy
   and in turn cause an incorrect estimation of the galaxy/bulge mass.

   \smallskip
   When fitting a galaxy hosting a central AGN with a \Sersic{} + exDisk model,
   the central \Sersic{} might account for  the AGN by acquiring extreme
   \Sersic{} $n$ and effective radius values (so that the profile effectively
   resembles a point-like source).
   When fitting it with a single \Sersic{} component, the AGN can shift the
   surface brightness normalization upwards (see point 1) or --- as for the
   previous case --- force an unrealistically large \Sersic{} index.
   Even a low-luminosity AGN, if overlooked, can still affect the total luminosity
   budget of a model by e.g.,\ mimicking a  bulge component.
   Conversely, our models including a psfAgn component need to be
   accurately screened in order to avoid misinterpreting an unresolved bulge
   (i.e., a PSF) as an AGN.

   \smallskip
   It is therefore of vital importance to ensure that the point-like AGN component
   is isolated.
   To this purpose, we took advantage of the activity classification of the
   SFRS nuclear sources provided by \cite{maragkoudakis:2018}.
   The activity classification was based on a combination of optical
   emission line diagnostics (the 3 variations of the BPT diagrams) and
   IR-colour diagnostics \citep{stern}.
   In addition, to account for host-galaxy light contamination in the nuclear
   colours of the galaxies, the authors performed matched aperture photometry
   to the nuclear regions of all the IRAC images and designed a
   ``nuclear-color Stern diagram'', which then revealed obscured AGN not
   identified with the standard integrated IRAC colours Stern plot.
   The SFRS activity classification consists of: 269 (73\%)
   star-forming galaxies, 50 (13\%) Seyferts (Sy; including 3C~273 and OJ~287),
   33 (9\%) LINERs, and 17 (5\%) transition objects (TOs).
   Because this classification is based on star-light subtracted spectra it minimizes
   the effect of dilution of the AGN by the stellar component of the host galaxy
   \citep[see][]{maragkoudakis:2014}.
   For the current study, we considered as ``secure'' AGNs only the 50
   SFRS galaxies identified as Seyferts.

   \smallskip
   For these targets, the models \emph{not} including a psfAgn component were
   automatically rejected (unless they were the only successful fits).
   For the remaining objects, we excluded at this stage the \Sersic{} + psfAgn + exDisk
   model, but we kept all the others, including the \Sersic{} + psfAgn (but in this case we
   assumed that the central PSF represented an unresolved bulge;
   see $\S$\ref{Separation of disk/bulge components}).

 \item[2.b)~] \emph{Accounting for TOs ---}
   Transition objects are defined as sources of composite contribution
   from both star-forming and AGN activity, falling in between the H\,{\sevensize II}
   and AGN regions in the [N\,{\sevensize II}]/H$\alpha$ BPT diagram.
   \cite{maragkoudakis:2018}, following the combined classification
   scheme from all three BPT diagnostics, applied stricter conditions in
   the definition of TOs.
   Specifically, a galaxy was assigned a TO classification when all of
   the following three conditions applied: (i) it was defined as a TO by the
   [N\,{\sevensize II}]/H$\alpha$ diagnostic; (ii) it was attributed a Sy or LINER classification
   by one of the other BPTs; and (iii) it was attributed a H\,{\sevensize II}
   classification by the remaining diagrams.
   For these objects, if a \Sersic{} + psfAgn + exDisk model was successful,
   we ignored the \Sersic{} + psfAgn one. 
   The reason is that, in TOs, the nuclear component could host contributions
   from both AGN and star formation emission \citep{ho}.
   Given the rules set by point 2, adopting the \Sersic{} + psfAgn model would
   have implied that \emph{all} the nuclear emission was associated with an AGN.
   The \Sersic{} + psfAgn + exDisk model is instead more suitable to redistribute
   the nuclear emission between star-forming regions and AGN.

 \setcounter{point}{3}
 \item[\arabic{point})~]  \emph{Evaluation of \Sersic{}$^{n=1}$ and \Sersic{}$^{n=4}$ models ---}
   We introduced the \Sersic{} models with fixed index (i.e., \Sersic{}$^{n=1}$
   and \Sersic{}$^{n=4}$) only to assist the fit of those $\sim$10 problematic
   targets for which the free-index \Sersic{} fit failed.
   Therefore, in case the \Sersic{} fit with free index was successful, these
   models were not considered.

 \setcounter{point}{4}
 \item[\arabic{point})~] \emph{Selection of model with best statistics ---}
   The statistic minimized by \GALFIT{} in order to converge to the best-fit
   parameters is the \Chisq{}.
   However, this statistic cannot be used as a parameter to decide
   the best model, given that the models contain a different number of components.
   In these cases, it is common to adopt the $F$-test \citep[e.g.,][]{bevington}
   in order to assess the importance of an additional component (e.g., a bulge or
   a PSF component), but this test has been shown to be problematic in the case
   of fits within a bounded parameter space
   \citep[as in our case where we constrain the range of the model parameters;
   see e.g.,][]{protassov:f_test}.
   We therefore preferred to define a metric based on the residual image, which
   gives a direct picture of how well a model represents the spatial distribution of
   the source intensity and accounts for its total flux (which is our goal in this
   effort).

   \smallskip
   The absolute value of the sum of the residual counts over the fitted area is
   not indicative of the quality of the fit.
   The reason is that a model may be a poor fit but generate an equal amount of
   positive and negative residual counts, so that the sum is close to 0.
   A similar argument holds against the average value of the residuals (i.e., the sum
   of the residuals divided by the area of the fit).
   A better representation of the deviation of the residuals is given in their RMS
   variation within the area of the fit. 
   However, one has to consider the RMS of the residuals generated only by
   the subtraction of the model from the galaxy light profile, excluding the noise
   due to the fit of the background.
   Following \citet[their equation 8]{vaughan}, we defined an \emph{excess variance}
   statistic:

   \vspace{-0.3cm}
   \begin{eqnarray}
    \sigma_{XS}^2  = {\sigma^2}_{objects} -{\sigma_{sky}}^2 
    \label{equation:excess_variance} 
   \end{eqnarray}

   \noindent
   where $\sigma_{objects}$ and $\sigma_{sky}$ represent the RMS of
   the residual image evaluated over the fitted target(s) and the sky areas,
   respectively.
   This quantity evaluates the amplitude of the variations in the residuals
   at the area of an object, after removing the statistical variations purely
   due to the background component (which gives a reference baseline noise
   in a source-free region).
   The uncertainty on $\sigma_{XS}^2$ is given by the same authors
   (their equation 11), as:
   
   \vspace{-0.3cm}
   \begin{eqnarray}
    \delta{\sigma_{XS}^2} = \sqrt{ {2 \over N_{objects}} \cdotp ({{{\sigma_{sky}}^2}})^2 +  { {\sigma_{sky}} \over N_{objects} } \cdotp 4{\sigma^2_{XS}} }~~,
    \label{equation:excess_variance_err} 
   \end{eqnarray}
   
   \noindent
   where $N_{objects}$ is the number of pixels corresponding to the object(s)
   area.
   For the cases in which the objects residuals $\sigma_{objects}$ are close to the
   background  variation ($\sigma_{objects}$ $\approx$ ${{\sigma_{sky}}^2}$),
   Equation \ref{equation:excess_variance_err} simplifies into:

   \vspace{-0.3cm}
   \begin{eqnarray}
    \delta{\sigma_{XS}^2} = \sqrt{ {2 \over N_{objects}} \cdotp ({{{\sigma_{sky}}^2}})^2 }~~.
    \label{excess_variance_err_approx} 
   \end{eqnarray}

   \noindent
   To correctly calculate $\sigma_{objects}$ from the residual image, we used
   a ``negative mask'' in which the only valid pixels are those corresponding to the
   objects of interest.
   The \emph{center-right} panel of Figure \ref{figure:masks} shows the negative mask
   for NGC~4438: the companion galaxy NGC~4435 (N-W of the target), despite
   being part of our sample, is masked-out because its fit is performed within a separate
   region and its residuals must not affect the model selection results fir our target
   of interest.
   Similarly, to calculate $\sigma_{sky}$ we adopted a ``negative background mask''
   which hid all the sources (see e.g.,\ Figure \ref{figure:masks}, \emph{right} panel)\footnote{
    Notice that this ``negative background map'' is \emph{not} a reverse of the
    conservative background mask image described in \ref{Setting up GALFIT input},
    and used to evaluate the sky statistics.
    Instead, it is a [roughly] complementary image to the ``negative mask'' for the object.
   }.
   For each target, we selected the \emph{best-statistic} model as the one having
   the smallest $\sigma_{XS}^2$.   
   
 \setcounter{point}{5}
 \item[\arabic{point})~] \emph{Selection of candidate best-models ---}
   In order to account for statistical uncertainties, a pool of additional ``good-fit''
   models with $\sigma_{XS}^2$  around the \emph{best-statistic} model excess
   variance ($\sigma_{XS,best}^2$) with a tolerance set by $\delta\sigma_{XS,best}^2$
   was selected.

 \setcounter{point}{6}
 \item[\arabic{point})~] \emph{Screening for the simplest models ---}
   The models which ``survived'' up to this stage were considered equivalent.
   Given this assumption, there is no reason to prefer a more complex model
   when a simpler one can give an equally valid interpretation of the data.
   Therefore, among the surviving models, the simplest models were selected
   (i.e., the one with fewest number of components).
   If the surviving models had the same number of components they were
   all kept.
   In any case though, given the selection scheme and the composition of
    the available models
   (i.e., \Sersic{}, \Sersic{}$^{n=1}$, \Sersic{}$^{n=4}$, \Sersic{} + psfAgn, etc.),
   there were no more than 2 models left in the \emph{final} pool.

 \setcounter{point}{7}
 \item[\arabic{point})~] \emph{Final choice of best-fit model ---}
   In these few cases with more than one final model, the selected \emph{best-fit} model
   was the one with the smallest $\sigma_{XS}^2$ within the \emph{final} pool.
   Notice that, due to point 6, it did not necessarily correspond to the
   $\sigma_{XS,best}^2$ model.

\end{cutenumerate} 

\noindent
Following the case of the example galaxy UGC~6625 (Figure \ref{figure:fit_models}),
our routine operated in this way: at point 1, no model was rejected;
at point 2, because the nuclear source was classified as TO, the \Sersic{} + psfAgn model
was excluded; at point 3, the \Sersic{} model being successful, the \Sersicone{}
and \Sersicfour{} models were excluded; at point 4, the \Sersic{} + psfAgn + exDisk
model was selected due to the best $\sigma_{XS}^2$; at point 5, no model was added to
the pool (being every other model either excluded or having a $\sigma_{XS}^2$
larger than $\sigma_{XS}^2 + \sigma_{XS,best}^2$), hence prematurely selecting the
\Sersic{} + psfAgn + exDisk combination as the \emph{best-model}.
Notice that this model does \emph{not} correspond to the best \Chisq{} nor
$\sigma_{XS}^2$ (see Table \ref{table:UGC6625}) because our model selection
procedure considers several more criteria.

\medskip

\noindent
The summary of our \GALFIT{} fit results, along with the corresponding masses
for the \emph{best-models} are reported in Table \ref{table:GALFIT_selected_models}.
Figure \ref{figure:masses_selected_Dmag_plot} shows the difference between the 2MASS
total magnitudes and the \emph{best-model} ones computed by \GALFIT{}.
The 2MASS magnitudes are on average $0.12 \pm 0.24$~mag fainter than
our \GALFIT{} magnitudes.
This is consistent with the systematic $\sim$0.3~mag underestimation of the
catalogued 2MASS magnitudes found by \cite{schombert}.
\cite{lauer} argue that one possible reason for the underestimation lies in the
fact that the 2MASS-XSC calculation of total magnitudes erroneously
``\emph{assumes that the galaxies essentially have exponential profiles}''.
This potentially explains why we retrieve a smaller magnitude discrepancy than
\cite{schombert}.
In fact, while their sample is composed of elliptical galaxies, the SFRS objects
are predominantly late-type galaxies and hence closer to satisfying the 2MASS-XSC
assumption of having pure exponential profiles.

\section[Separation of disk/bulge components]{Separation of disk/bulge components}
\label{Separation of disk/bulge components}

\begin{table*}

 \definecolor{LightGray}{rgb}{0.92,0.92,0.92}
 \definecolor{White}{rgb}{1.0,1.0,1.0}
  
 \centering
  \renewcommand{\arraystretch}{1.5}
  \begin{tabular}{|l|cccc|}
   \hline
   \multicolumn{5}{c}{\textsc{Decisional Algorithm for Bulge/Disk Decomposition}} \\
   \hline
   \hline
   \textsc{Model} & \textsc{Bulge component} & \textsc{Disk Component} & \textsc{Mixed Component} & \textsc{AGN Component} \\
   \hline

\rowcolor{LightGray}
\Sersic{}$^{n=1}$                   &  --                           &\Sersic{}$^{n=1}$              & --                            & -- \\
\hline
\rowcolor{White}
\Sersic{}$^{n=4}$                   & \Sersic{}$^{n=4}$             & --                            & --                            & --\\
\hline
\rowcolor{LightGray}
\Sersic{}                           & --                            & -                             & \Sersic{} $n > \nthreshold{}$ & -- \\
\rowcolor{White}
                                    & --                            & \Sersic{} $n < \nthreshold{}$ & --                            & -- \\
\hline
\rowcolor{LightGray}
\Sersic{} + psfAgn                  & --                            & \Sersic{} $n < \nthreshold{}$ & --                            & psfAgn\\
\rowcolor{White}
                                    & --                            & --                            & \Sersic{} $n > \nthreshold{}$ & psfAgn\\
\rowcolor{LightGray}
                                    & psfAgn                        & \Sersic{}                     & --                            & -- \\
\hline
\rowcolor{White}
\Sersic{}$^{n=4}$ + exDisk          & \Sersic{}$^{n=4}$             & exDisk                        & --                            & -- \\
\hline
\rowcolor{LightGray}
\Sersic{}$^{n=4}$ + exDisk + psfAgn & \Sersic{}$^{n=4}$             & exDisk                        & --                            & psfAgn\\
\hline

  \end{tabular}

  \begin{minipage}{0.8\linewidth}
   \caption[Decision Tree for disk/bulge decomposition]{
    Components interpretation as bulge, disk, ``mixed'', or AGN, according to each of
    the possible fit models.
    \label{table:segregation}
   }
  \end{minipage}
\end{table*}

After the \emph{best-fit} model selection was completed, we proceeded with
separating the bulge from the disk components.
This was a trivial task in case the \emph{best-fit} model turned out to be
the \Sersic{} + exDisk + psfAgn\footnote{
 As mentioned in $\S$\ref{Best-fit model selection}, this model could be selected
 only if the galaxy hosted a Seyfert or TO central source.
}.
In this case the \Sersic{} component naturally represents the bulge while the
exponential component represents the disk, and, if the galaxy hosts a central
Seyfert source, the psfAgn component obviously represents the AGN.
If instead the central object was spectroscopically classified as TO
(i.e., of uncertain nature; see $\S$\ref{Best-fit model selection}), it would
be here re-classified as ``AGN''.
For all the other \emph{best-fit} models, deciding what each component represents
was more complicated, as detailed in the following scheme:

\begin{cutenumerate}[label=$\blacktriangleright$~]
 \itemsep0.5em  

 \item psfAgn component

 \begin{enumerate}[label=$\hookrightarrow$~]
   \itemsep0.2em  
   
   \item considered an ``AGN'' if the object is known to host a Seyfert or a TO
         central source
   \item considered a ``bulge'' otherwise
        
 \end{enumerate}

 \item \Sersic{} component

 \begin{enumerate}[label=$\hookrightarrow$~]
   \itemsep0.2em  
   
   \item always considered a ``disk'' if $n \le$ \nthreshold{}\footnote{
          This threshold was set at approximately the mode of the \Sersic{} $n$
          distribution for disk-dominated galaxies in Figure \ref{figure:Sersic_n}:
          below this value any \Sersic{} component can be safely considered
          representative of a disk.
         }

 \end{enumerate}
   
 \vspace{-0.1cm} 
 \noindent
 or else:   
 \vspace{-0.1cm} 

 \begin{enumerate}[label=$\hookrightarrow$~]
   \itemsep0.2em  
   
   \item considered a ``mixed'' component in \Sersic{} \emph{best-fit} model and,
         if the object is known to host a Seyfert, also in \Sersic{} + psfAgn
   \item considered a ``disk'' in \Sersic{} + psfAgn \emph{best-fit} model if
         psfAgn is considered a ``bulge''
   \item considered a ``bulge'' otherwise
         
 \end{enumerate}
 
 \item exDisk component

 \begin{enumerate}[label=$\hookrightarrow$~]
   \itemsep0.2em  
   
   \item always considered a ``disk''

 \end{enumerate}
 
\end{cutenumerate}

\noindent
Table \ref{table:segregation} summarizes the scheme described above.
Here the label ``mixed'' reflects the impossibility --- due to the limitation
imposed by the 2MASS data --- to unambiguously decompose that \Sersic{} component
into a bulge and a disk: this problem is addressed in
$\S$\ref{Further decomposition of mixed components}.
Out of the \sampleBestFits{} \emph{best-fit} models, we were able to directly
decompose \sampleDecomposed{} ($\sim$80\%), while \sampleNotDecomposed{}
($\sim$20\%) contain a ``mixed'' component.
For the decomposed galaxies, the bulge-to-total ratio ($B/T$) was defined
as the ratio of the luminosities of the respective components, while for the
``mixed'' \Sersic{} objects it was defined using the concentration index,
as explained in $\S$\ref{Further decomposition of mixed components}.
Table \ref{table:summary} summarizes the statistics for the outcome of the 2D fitting
procedure and for the results of the decomposition algorithms.

\begin{table}

 \centering

 \begin{tabular*}{0.48\textwidth}{l @{\extracolsep{\fill}} r}
  \hline
  \multicolumn{2}{c}{\textsc{Fit and Decomposition Summary}} \\
  \hline
  \hline

  \addlinespace 
  \textsc{Sample Galaxies} & \sample{}\\
  \addlinespace 

  \hline

  \addlinespace 
  \textsc{Excluded Galaxies} & 10 \\
  \addlinespace 
  $\hookrightarrow$ All models failed                 & \sampleNotFitted{} \\
  $\hookrightarrow$ All fits exceeded magnitude limit & \sampleBadMag{}    \\
  \addlinespace 

  \hline

  \addlinespace 
  \textsc{Galaxies with a Best-Fit Model} & \sampleBestFits{} \\
  \addlinespace 
  $\hookrightarrow$ Decomposed     & \sampleDecomposed{}    \\
  $\hookrightarrow$ Not decomposed & \sampleNotDecomposed{} \\
  \addlinespace 

  \hline

 \end{tabular*}
 
 \begin{minipage}{0.48\textwidth}
  \caption[Outcome and bulge/disk decomposition]{
    Outcome of the fitting procedure and of the related
    bulge/disk decomposition algorithm.
    For the \sampleNotDecomposed{} galaxies we could not decompose via 2D fitting,
    we used the concentration of the surface brightness to infer the bulge-to-total
    ratio $B/T$ (see $\S$\ref{Further decomposition of mixed components}).
   \label{table:summary}
 } 
 \end{minipage}
 
\end{table}


\subsection[Estimation of uncertainty on bulge-to-total ratio (B/T)]{Estimation of uncertainty on bulge-to-total ratio (B/T)}
\label{Estimation of uncertainty on bulge-to-total ratio (B/T)}

\noindent
To evaluate the uncertainty on a galaxy bulge-to-total ratio ($B/T$) we performed
--- for each object --- a set of simulated fits in which the $B/T$ was varied
around the actual \emph{best-fit} value.
In practise, at each trial we arbitrarily re-distributed the galaxy total
luminosity into a bulge and disk luminosities (L$_{bulge}^{sim}$ and
L$_{disk}^{sim}$ respectively) in order to sample a simulated $B/T^{sim}$
such that:

\vspace{-0.3cm}
\begin{eqnarray}
 B/T^{sim} = L_{bulge}^{sim} / (L_{bulge}^{sim} + L_{disk}^{sim}).
\end{eqnarray}

\begin{figure}
  \begin{center}
   \includegraphics[width=0.48\textwidth,angle=0]{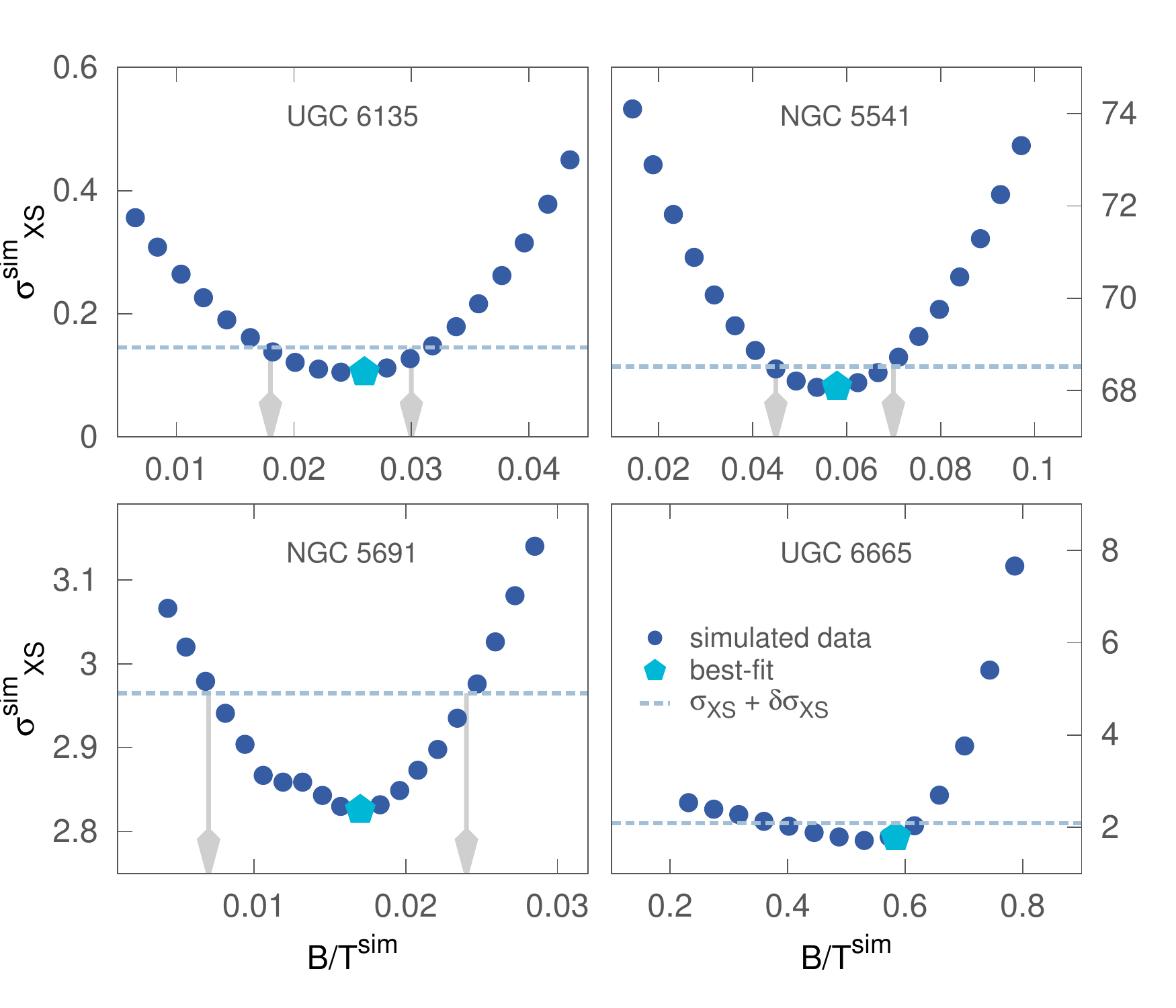}
   \caption[]{
    Examples of the technique used to estimate the uncertainty on $B/T$.
    Simulated \GALFIT{} fits were performed by sampling values around the actual
    \emph{best-fit} $B/T$ and locking the luminosity of the bulge and disk
    components at the selected $B/T^{sim}$.
    The resulting $\sigma_{XS}^{sim}$ measured at each iteration is recorded
    and then plotted against $B/T^{sim}$ (blue points): the uncertainty on the
    \emph{best-fit} $B/T$ is estimated from the values (grey arrows)
    at which the curve hits the $\sigma_{XS}$ + $\delta\sigma_{XS}$ limit
    (dashed horizontal line).
    Notice that the \emph{best-fit} $\sigma_{XS}$ (light blue pentagon) does not
    necessarily fall exactly at the bottom of the curve (e.g., UGC~6665)
    because the fits are minimized on \Chisq{}, rather than on $\sigma_{XS}$.
    \label{figure:sigma_XS_vs_BT} 
   }
  \end{center}
 \end{figure}

\noindent
The magnitudes of the \emph{best-fit} bulge and disk components  were fixed to
the new values, and a new \GALFIT{} fit was performed for each $B/T^{sim}$,
for which we recorded the resulting $\sigma_{XS}^{sim}$, calculated as usual
according to Equation \ref{equation:excess_variance}.
When $\sigma_{XS}^{sim}$ is plotted against $B/T^{sim}$, it roughly traces a
parabolic curve: representative examples of such plots are shown in
Figure \ref{figure:sigma_XS_vs_BT}.
The uncertainty on $B/T$ was defined as the $B/T$ at which the curve crosses the
($\sigma_{XS}$ + $\delta\sigma_{XS}$) limit, indicative of the upper statistical
uncertainty on the fit.
Although the uncertainties are not necessarily symmetrical, we assumed the
average between the lower and the upper $B/T$ values.
We stress that the uncertainties obtained in this way do not strictly represent
a Gaussian 1-$\sigma$ (or 68\%) confidence level because the $\Delta\sigma_{XS}$
has not been calibrated as an absolute statistic.
   
The disk and bulge luminosities were kept fixed throughout the fit; however,
the other bulge and disk parameters (e.g., \Sersic{} $n$, R$_{e}$) were let free to
vary, in order not to excessively penalize the ``artificial'' $B/T$ fits.
Similarly, all the parameters of the other components of the SFRS target
(e.g., the central AGN), as well as those of the other ``contaminating'' sources
fit along, were frozen to prevent them from compensating for the less-than-optimal
$B/T^{sim}$.

\subsection[Further decomposition of mixed components]{Further decomposition of ``mixed'' components}
\label{Further decomposition of mixed components}

\noindent
\noindent
The ``mixed'' components are those \Sersic{} components to which our
pipeline was unable to attribute a bulge or disk classification (see Table
\ref{table:segregation}) because they represent an unknown mixture
of the two.
In order to infer the relative mass of the bulge (or equivalently, the
bulge-to-disk ratio $B/D$) for the ``mixed'' components, one would be
tempted to use the \Sersic{} index $n$ as a summarizing parameter for the overall
radial profile shape.
However, several studies have highlighted that an improper \Sersic{} fit with
a high index $n$ can mimic e.g., a compact bulge superimposed on an extended disk.
As a consequence, the \Sersic{} index is not a reliable
indicator of the galaxy Hubble type \citep[e.g.,][their Figure 3]{graham:BD}.
This issue is also present in our data.

\begin{figure}
  \begin{center}
   \includegraphics[width=0.45\textwidth,angle=0]{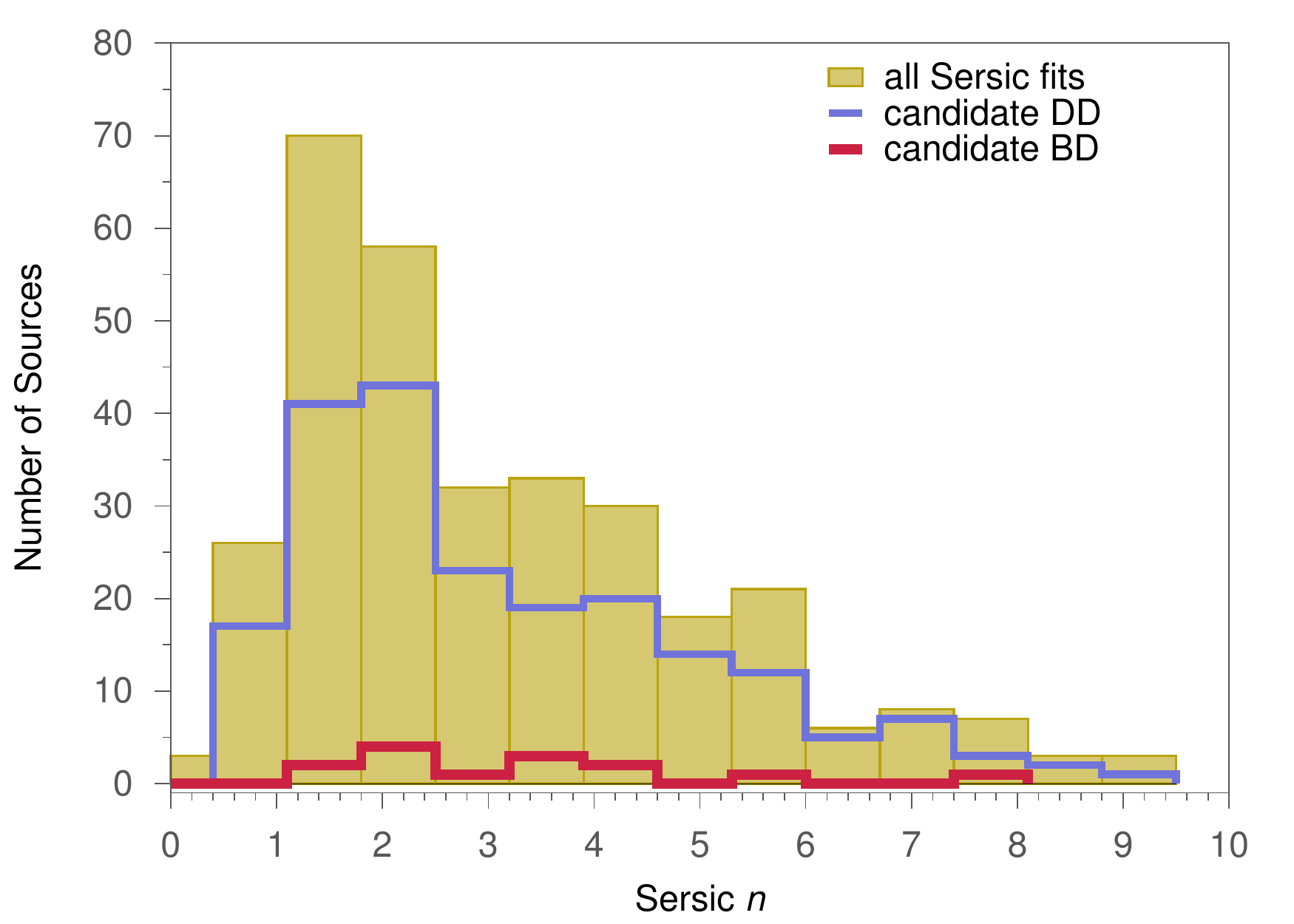}
   \caption[\Sersic{} index distribution for single \Sersic{} fit of the candidate disk/bulge-dominated galaxies]{
    \Sersic{} index ($n$) distribution for single \Sersic{} fit to the candidate
    disk-dominated (DD) galaxies (blue), candidate bulge-dominated (BD) galaxies (red),
    and for the \Sersic{} fit to all the objects (gold); bin size is $\Delta~n$ = 0.7.
    The candidate DD/BD galaxies are the objects in which the disk
    ($L_{disk}$ $>$ $L_{bulge}$) or the bulge ($L_{bulge}$ $>$ $L_{disk}$) 
    luminosity is dominant, respectively
    (see $\S$\ref{Further decomposition of mixed components}).
    The distributions for the candidate DD and candidate BD galaxies do
    \emph{not} sum up to the total sample because the galaxies we were not able
    to decompose do not appear in either of the sub-samples.
    \label{figure:Sersic_n} 
   }
  \end{center}
 \end{figure}

To check whether the \Sersic{} index traces the bulge/disk decomposition
obtained as described above, we considered two sub-samples: the
``candidate disk-dominated (DD) galaxies'' and the
``candidate bulge-dominated (BD) galaxies''.
These are the galaxies in which the disk ($L_{disk}$ $>$ $L_{bulge}$) or the bulge
($L_{bulge}$ $>$ $L_{disk}$) component is dominant, respectively.
Figure \ref{figure:Sersic_n} presents the \Sersic{} index distribution from
the single \Sersic{} fit to these two samples along with the corresponding
distribution for the \Sersic{} fit to the whole sample.
As expected, the vast majority of our targets are disk-dominated because the sample
has been selected on the basis of star formation activity and hence biased towards
late-type galaxies.
There is a significant number of candidate DD galaxies which have been fit
with an extremely large \Sersic{} index (up to $n \sim 9$).
This is indicative of the issue mentioned above, i.e., that in several cases an
unphysically large $n$ has erroneously compensated for an extended disk
accompanied by a marginally resolved/bright bulge or a central AGN
($\sim$80\% and $\sim$20\% of the cases, respectively, for the DD galaxies
with \Sersic{} index larger than 4).
This precludes any chance to use our single-\Sersic{}-fit $n$ to derive
the missing $B/D$.

For galaxies with a ``mixed'' \Sersic{} component we resorted instead to the
correlation between the concentration of the surface brightness and the $B/T$.
In particular, we followed the method of e.g., \cite{gadotti} and \cite{lackner},
who used SDSS data to calibrate their $B/T$ with respect to the concentration index
$C$ = $R_{90}/R_{50}$ (see their Figure 5), where $R_{90}$
and $R_{50}$ are the radii enclosing 90\% and 50\% of the Petrosian flux,
respectively \citep{petrosian}.
We calculated the concentration indices for the SFRS galaxies using the
SDSS-DR12 R$_{90}$ and R$_{50}$ derived in the $z$-band (the closest SDSS
band to the 2MASS $K_{s}$-band) and calibrated them against the $B/T$
for the targets for which we had a valid decomposition.
A linear fit yielded:

\vspace{-0.3cm}
\begin{eqnarray}
 B/T = 0.08 \times C(R_{90}/R_{50}) - 0.07
 \label{equation:BT}
\end{eqnarray}

\noindent
with an RMS scatter of $\sim$0.15 around the best-fit line.
In order to consider only unbiased concentration indices,
we did not include in this analysis the sources hosting an AGN or those
whose effective radius is comparable to the point-source FWHM
(i.e., with R$_e$ < 3.5\arcsec).
The calibration provided by Equation \ref{equation:BT} was used to derive
$B/T$ for the \sampleNotDecomposed{} galaxies lacking a direct \GALFIT{}
decomposition, and the relevant bulge and disk masses were calculated as the
$B/T$ fraction of the ``mixed'' component mass.
The final decision on the model decomposition of each target is reported in
Table \ref{table:GALFIT_selected_models}.

\subsection[Disk/bulge optical colors]{Disk/bulge optical colors}
\label{Disk/bulge optical colors}

\begin{figure*}

 \includegraphics[width=0.48\textwidth,angle=0]{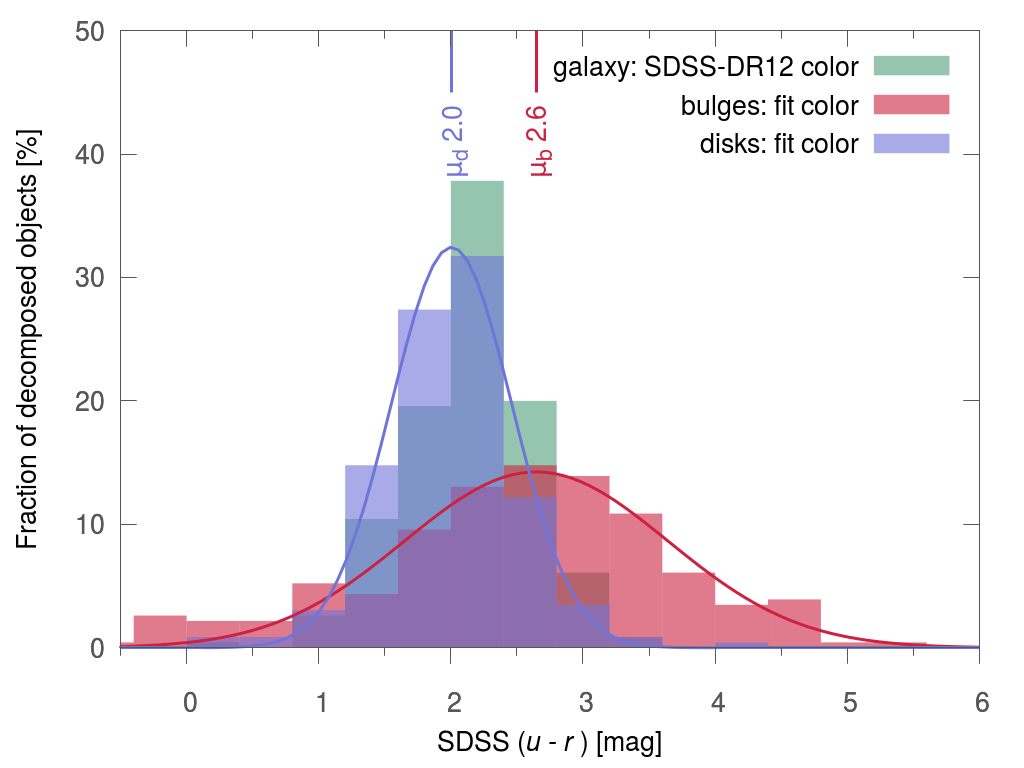}
 \includegraphics[width=0.495\textwidth,angle=0]{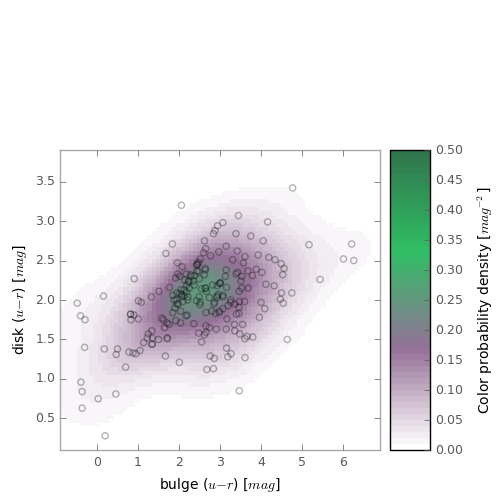}
 \caption[Disk/bulge colors]{
  \emph{Left ---}
  ($u - r$) colour distribution of disks (blue histogram) and bulges (red histogram)
  yielded by our 2D fit of the SDSS-DR12 images for the \sampleDecomposed{} SFRS
  galaxies hosting both a disk and a bulge. 
  The blue and red curves represent Gaussian fits to the distributions of disk
  and bulges, respectively; their means ($\mu_{d}$ and $\mu_{b}$) are indicated
  at the top of the figure.
  The green histogram shows the distribution of the global (``average'') galaxy
  color, directly obtained from the SDSS-DR12 catalogue.
  Note that all these magnitudes are expressed in the SDSS ``maggy''  native system,
  very close to the AB system.
  \emph{Right ---}
  The open circles present the measured bulge and disk colors for the \sampleDecomposed{}
  decomposed SFRS galaxies.
  The diffuse color gradient shows the bivariate ($u - r$) color density distribution,
  providing the probability of a given galaxy to host a disk \emph{and} a bulge with
  a given color combination.
  \label{figure:colors}
 }
\end{figure*}

\noindent
For the \sampleDecomposed{} SFRS galaxies with a successful \GALFIT{} decomposition,
we retrieved the separate ($u - r$) colors for the disks and the bulges by
performing 2D adaptive photometry on SDSS-DR12 images.
Namely, we fit the $u$ and $r$-band SDSS images by allowing the sub-components
of each \emph{best-fit} \GALFIT{} model (derived in the 2MASS $K_{s}$-band;
$\S$\ref{Best-fit model selection}) to independently vary their magnitudes,
while keeping all the other parameters (e.g., \Sersic{} $n$, axis ratio, P.A.,
etc.) constant, and resizing the radial scale lengths to match the SDSS
pixel scale.
Retaining consistent structural parameters comes at the cost of ignoring their
wavelength dependence \citep[e.g.,][]{haussler:2013}\footnote{
 Notice though that the work of \cite{haussler:2013} did not perform a full
 bulge/disk decomposition, but only adopted single \Sersic{} models,
 and therefore it is more sensitive to wavelength variations.
},
but this is justified because here we were only interested in retrieving
integrated magnitudes, in which \GALFIT{} has proven to excel \citep[][]{haussler}.
For the SDSS fitting we produced object masks similar to those created for the
2MASS analysis ($\S$\ref{Fit procedure} and Appendix \ref{Mask creation});
in brief, we identified the sources in the SDSS images with \textsc{SExtractor},
and then masked all the contaminating objects except for the blended ones,
which were fit simultaneously with the galaxies.

The left panel of Figure~\ref{figure:colors} shows the distributions of the bulge
and disk $u-r$ colours produced by our analysis.
A Gaussian fit to the distributions yielded average colors $(u-r)_b = 2.6$~mag
and $(u-r)_d = 2.0$~mag with standard deviations $\sigma(u-r)_b = 1.0$~mag and
$\sigma(u-r)_d = 0.46$~mag for the bulges and disks respectively.
Figure~\ref{figure:colors} also shows colours from SDSS-DR12\footnote{
 For the definition of the SDSS magnitudes refer to:
 \url{http://www.sdss.org/dr12/algorithms/magnitudes/\#mag_model}.}:
these magnitudes encompass the flux from whole galaxies and therefore represent the
average colours of each galaxy's bulge and disk.
As SFRS is by construction a disk-dominated sample, it is not surprising that the
global colour distribution is very close to that of the disks both in its peak
($(u-r)_{\rm SFRS} =2.2$~mag) and width($\sigma(u-r)_{\rm SFRS} = 0.43$~mag).
The bulges present a wider colour distribution than disks, a reflection of the
variety of secular and merger processes that lead to mass assembly in bulges.
Nevertheless, the bulges' average colours match those of red-sequence galaxies
(\citealt{strateva}, their Figure~2; \citealt{baldry:2006}, their Figure~7).
The SFRS disks are on average $\sim$0.5~mag redder than typical blue-cloud galaxies.
This difference is at least in part attributable to the sample selection: the SFRS includes
star-forming galaxies sampled to cover almost uniformly 3 orders of magnitude in SFR,
and it therefore includes objects living at the red edge of the blue cloud and in
the green valley.
The sample weighting will appropriately reduce the effects of these galaxies on
luminosity and mass functions, but in unweighted numbers, they can bias the colors.
Another effect might be that our bulge--disk separation was done at 2.2~$\mu$m and
might not be appropriate for determining $u-b$ colors.
Nonetheless, the maximum effect of an 0.5~mag colour error is a modest 11\% in
stellar mass difference (Equation~\ref{ML}).

The right panel of Figure \ref{figure:colors} shows the bivariate ($u - r$)
probability density of disks and bulges, calculated by means of a Gaussian
kernel-density estimate.
This plot provides the joint probability of a galaxy to host a disk \emph{and}
a bulge with a given color combination.
The distribution exhibits a weak correlation, with redder disks being associated to
redder bulges, indicating that star-forming galaxies evolve in parallel throughout their
spatial extent.
This is in line with our previous results showing that the local activity of a galaxy
scales similarly to its global activity \citep[][]{maragkoudakis:main_sequence}.

\medskip
\noindent
The disk/bulge colors derived in this section were used as stellar population
proxies for the mass-to-light conversion of the disk/bulge $K_{s}$-band
luminosities (Appendix \ref{Zero-points, luminosities, and stellar mass calculation}).
For the galaxies which were not directly decomposed by our pipeline, but rather
attributed a $B/T$ through the procedure outlined in
$\S$\ref{Further decomposition of mixed components},
we adopted the average color of the disk or bulge sub-population
($\mu_{(u - r),d}$ or $\mu_{(u - r),b}$) as the representative color of their
disk or bulge fraction.

\section[Luminosity and Mass Functions]{Luminosity and Mass Functions}
\label{Luminosity and Mass Functions}

\noindent
We built the luminosity function (LF) and stellar mass functions (MFs) adopting
a variant of the $1/V_{MAX}$ method \citep[e.g.,][]{rowan-robinson,kafka,schmidt},
performed according to the following procedure.

 \begin{cutenumerate}[label=\alph*)~]
  \itemsep-0.0em  
  \item Count the number of sources per magnitude/mass bin.
  \item Correct for the selection effects (which we will refer to as ``incompleteness'').
  \item Account for the Malmquist bias (i.e., the preferential detection of
        intrinsically bright objects). This is performed by dividing the counts in
        each bin by the surveyed volume $V_{MAX}$.
\end{cutenumerate} 
 \vspace{-0.3cm}

\medskip
\noindent
Steps \emph{a} and \emph{b} take into account that the SFRS sample has been defined
based on the \PSCz{} catalogue \citep{PSCz}, which is a 60$\mu$m flux density limited
catalogue.
Therefore, we need to calculate the completeness for both
the SFRS and the \PSCz{} samples and the volume correction for the latter.
The \emph{actual} selection procedure that brought us from the \PSCz{} down to the
SFRS sample \citep{ashby} used a binning scheme that was not based solely on
luminosity but involved also colors.
Because we were interested in evaluating the completeness \emph{only}
as a function of luminosity, we had to derive a different selection function.

One potential bias related to the $1/V_{MAX}$ method regards an overestimation
of the LF at its faint end in case of non-homogeneous spatial distribution of
sources (\citealt{willmer}; \citealt{takeuchi}; see also \citealt{LF_review} for
an extended review), when compared to other methods such as e.g.,
the step-wise maximum likelihood \citep[SWML;][]{SWML} or the $C^{-}$ \citep[][]{C-}.
However our analysis did not suffer from such bias, because the redshift
distribution of our sample galaxies does not show any significant clustering,
but it is rather a monotonically decreasing function of $z$
\citep[see][]{ashby}.
As described later in this section, the $V_{MAX}$ was calculated on the \sample{}
sample galaxies, and only \emph{subsequently} corrected
for selection effects, thus maintaining the spatial uniformity.

In $\S$\ref{V/V_MAX test} we prove the reliability of our method by
applying the $V/V_{MAX}$ test on the full \PSCz{} sample.
Similarly, the azimuthal distribution of the SFRS galaxies is not biased,
since the sources are spread uniformly over the $>$20$^\circ$ northern
Galactic cap \citep[see][]{ashby}.

\subsection[The PSCz sample]{The PSCz sample}
\label{The PSCz sample}
 
 \noindent
 As mentioned in $\S$\ref{The Star Formation Reference Survey (SFRS)}, the
 SFRS sample was selected from the \PSCz{} catalogue \citep{PSCz} of IRAS sources.
 Owing to the fact that the \PSCz{} is itself complete down to its detection limits
 (0.6~Jy; see Figure \ref{figure:LogN-LogS}), it was used to assess
 the completeness of the SFRS sample.
 In order to calculate the selection function of SFRS with respect to its parent
 sample we first needed to obtain the distances of the \PSCz{} objects
 in the same way as we did for the SFRS objects\footnote{
  In this context, we applied to the \PSCz{} catalogue the same coordinate and distance
  cuts that were originally applied to define the SFSR sample \citep[see][]{ashby}.
 }, and apply the relevant K-correction \citep{oke:K_corr}.

 The data for the full \PSCz{} sample have been obtained through the \emph{Vizier}
 service.
 The heliocentric velocities reported in the catalogue were corrected for the
 effects of local velocity fields (namely, the flows towards: Virgo, the Great
 Attractor, and the Shapley supercluster) using the recipe of \cite{mould}.
 These ``cosmic velocities'' were then used to recalculate the redshifts of the
 galaxy sample ($z$ = v$_{cosmic}$/c).
 For consistency, we applied the same correction to the SFRS sample: the
 velocities considered in this paper are therefore an updated version of the ones 
 used in \citep{ashby}.

 We applied a K-correction to the \PSCz{} 60 $\mu$m flux assuming a power-law
 flux density spectrum \mbox{$f_{\nu}(\nu)$ $\propto$ $(\nu)^{-\alpha}$} and
 applying the prescription of \cite{peterson}:

 \vspace{-0.3cm}
 \begin{eqnarray}
  F_{int} = F_{obs}~(1 + z)^{\alpha - 1}
  \label{equation:K_correction}
 \end{eqnarray}

 \noindent
 where $F_{int}$ is the intrinsic (true) flux emitted by the source
 within the rest-frame bandwidth, while $F_{obs}$  is the
 received flux.
 We assumed a spectral slope $\alpha$ $\sim$ 2, thus simplifying Equation
 \ref{equation:K_correction} to:

  \vspace{-0.3cm}
  \begin{eqnarray}
   F_{int} = F_{obs}~(1 + z).
   \label{equation:K}
  \end{eqnarray}

 \noindent
 The chosen value for $\alpha$ is appropriate for IR galaxies \citep[e.g.,][]{lawrence},
 and it has been previously adopted by \cite{saunders:LF} to construct the 60$\mu$m
 luminosity function of \PSCz{} galaxies (see Figure \ref{figure:LF_60mu}).
 At the median \PSCz{} redshift ($z \sim 0.02$), the K-correction
 yielded by Equation \ref{equation:K} is negligible (of the order of 2\%).

 The \PSCz{} catalogue contains sources of different nature (including e.g., stars,
 H\,{\sevensize II} regions, or planetary nebulae), out of which galaxies were
 selected by using the \PSCz{} CLASS flag, which characterizes
 the object type\footnote{
  We considered bona-fide galaxies only the \PSCz{} objects flagged as "\emph{go}",
  "\emph{gf}", or "\emph{gr}"
 }.
 We excluded 124 \PSCz{} galaxies having \mbox{$v_{cosmic}$ < 500~km s$^{-1}$}
 (corresponding to $d$ < 6.8~Mpc, according to the Hubble law).
 Such low velocities are heavily affected by peculiar motions,
 and therefore the distances (and luminosities) of these galaxies cannot be calculated
 reliably from the Hubble law \citep[e.g.,][]{binney_tremaine}.
 We accounted for this lost survey volume in our analysis
 (Section \ref{Calculation of V_MAX}).
 After this screening we obtained the relevant SFRS parent sample, comprising
 14\,635 galaxies (as opposed to the 18\,351 objects within the \PSCz{} catalogue):
 in the remainder we will refer to this sample when mentioning the ``\PSCz{}''.

\subsection[]{Correction for sky coverage }
\label{Correction for sky coverage }

 \noindent
 The completeness of the SFRS sample is fully determined by the
 ``weights'' calculated in \citet[][see Table \ref{table:GALFIT_selected_models}]{ashby}.
 These weights provide the ``representativeness'' of each SFRS source, i.e.,
 how many galaxies it represents in the sky area of the \PSCz{} survey, from
 which the SFRS sample is drawn.
 Specifically, the weights are calculated during the selection procedure,
 when the \PSCz{} catalogue is split into 3D bins of SFR, sSFR and dust temperature,
 and a number of SFRS sources are selected out of each bin
 (see $\S$\ref{The Star Formation Reference Survey (SFRS)}).
 The sum of all the SFRS weights therefore gives the total number of sources in
 the parent sample.
 
 When constructing e.g., the $K_{s}$-band magnitude function, the completeness
 correction for a magnitude bin is simply the sum of all the weights of
 the SFRS galaxies falling within that bin.
 In practice, by doing this one marginalizes the SFRS galaxies' weights
 along the variable  of interest --- in this example, the $K_{s}$-band magnitude.
 The same procedure was used in this work (in $\S$\ref{SFRS stellar mass functions})
 to construct the completeness correction for the mass functions, this time
 by marginalizing over the mass variable.

 To be precise, the actual completeness correction requires an additional
 factor due to the fact that the \PSCz{} survey covers only 84\% of the sky.
 Given that this spatial coverage was restricted by conditions independent
 from the $K_{s}$-band luminosity or the mass of the galaxies (nominally,
 it was limited by the IRAS coverage and spectroscopic follow-up; \citealt{PSCz}),
 the additional correction factor can be considered constant.
 Then, assuming a uniform distribution of the sources over the whole
 sky, the factor simply amounts to 1 / 84\%.

 \medskip
 \noindent
 The reliability of the completeness calculated following the procedure outlined
 above is discussed in Appendix \ref{Completeness test}.

\subsection[Calculation of V$_{MAX}$]{Calculation of V$_{MAX}$}
\label{Calculation of V_MAX}

 \noindent
 The $V_{MAX}$ represents the maximum volume up to which a source of luminosity $L$
 can be observed given the sample limiting flux $F_{lim}$.
 That is:

 \vspace{-0.3cm}
 \begin{eqnarray}
  V_{MAX}(L) ={ {4\pi \over 3} ({L \over 4\pi F_{lim}} })^{3/2}~~.
  \label{V}
 \end{eqnarray}

 \noindent
 Because the SFRS sample is selected from the \PSCz{} catalogue, we adopted ---
 as $F_{lim}$ --- the IRAS 60$\mu$m flux limit for \PSCz{}:

 \vspace{-0.3cm}
 \begin{eqnarray}
  F_{lim} = \nu_{0} \cdot f_{60\mu m}^{lim} \cdot 10^{-23} = 3\times 10^{-11} [\mbox{erg s$^{-1}$ cm$^{-2}$}]~~,
 \end{eqnarray}

 \noindent
 where $\nu_{0}$ is the central wavelength of the 60$\mu$m filter
 (\mbox{5 $\times$ 10$^{12}$~Hz}), $f_{60\mu m}^{lim}$ is the limiting flux density
 for the survey \citep[$\sim$0.6~Jy;][]{PSCz}, and the term 10$^{-23}$ is the
 conversion factor from Jy to erg s$^{-1}$ cm$^{-2}$ Hz$^{-1}$.

 To construct the LF with the V$_{MAX}$ method it is common to bin the data in
 luminosity, and for each bin calculate a $V_{MAX}$ and
 adopt the central value of the bin as the luminosity defining
 the $V_{MAX}$.
 However, because the central value of a luminosity bin is not an accurate
 representation of a \emph{non}-uniform distribution of the targets luminosities
 \emph{within} the bin, nor does it take into account the different redshifts
 of the sources, we calculated the $V_{MAX}$ on a source-by-source basis.

 First, consider that the luminosity $L$ appearing in Equation \ref{V}
 is that of the fiducial source at luminosity distance $d_{MAX}$ (corresponding
 to $V_{MAX}$).
 As mentioned in $\S$\ref{The PSCz sample}, we calculated the true luminosities
 $L_{corr}$ as they would be measured in the local frame of reference (by K-correcting
 for the \emph{actual} distances of the sources).
 Imagine now to start ``moving'' the source away from the local frame of reference,
 up to $d_{MAX}$.
 The source's true luminosity $L_{corr}$ will decrease when observed back at the
 local frame of reference.
 To account for this, we introduced an ``additional'' K-correction.
 At any intermediate redshift $z$ (between 0 and $z_{MAX})$, the observed luminosity
 $L_{obs}$ will be:

 \vspace{-0.3cm}
 \begin{eqnarray}
  L_{obs}(z) = L_{corr} / K(z)
  \label{L_obs}
 \end{eqnarray}

 \noindent
 and the observed flux $F_{obs}$ of a source at luminosity distance $d$ will
 be:
 
 \vspace{-0.3cm} 
 \begin{eqnarray}
  F_{obs}(z) = \frac{L_{obs}(z)}{4\pi d^2} = \frac{L_{corr} / (1+z)}{4\pi d^2}~~,
  \label{F_obs}
 \end{eqnarray}
 
 \noindent
 where we assumed $K$($z$) $\sim$ (1+$z$) as in $\S$\ref{The PSCz sample}.
 Using the simple cosmology $d = cz/H_{0}$, Equation \ref{F_obs} can be expressed
 as:

 \vspace{-0.3cm}
 \begin{eqnarray}
  F_{obs}(d) = { {L_{corr} / (1 + d \cdot H_{0}/c)} \over 4\pi d^2 }~~.
  \label{F_obs_d}
 \end{eqnarray}

\begin{figure}
  \begin{center}
   \includegraphics[width=0.48\textwidth,angle=0]{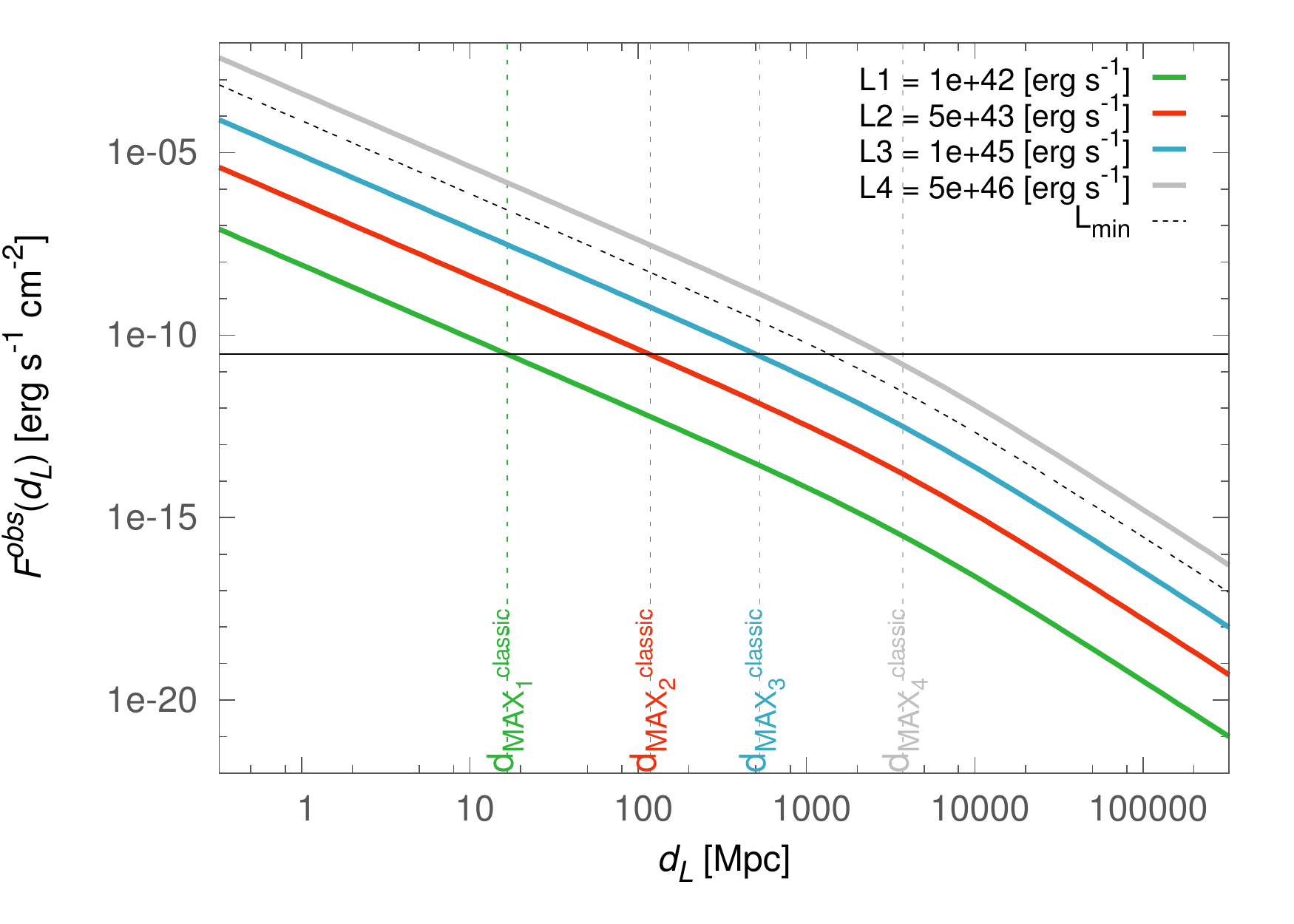}
   \caption[Graphical demonstration of the derivation of $d_{MAX}$]{
    Graphical demonstration of the derivation of $d_{MAX}$.
    The continous lines correspond to evaluations
    of Equation \ref{F_obs_d} as a function of the
    luminosity distance $d_{L}$ for different values of source luminosity $L_{corr}$
    as shown in the legend.
    Each curve represents the observed flux $F_{obs}$ of a source at different
    luminosity distances $d_{L}$ after applying the relevant K-correction.
    The range of the plotted luminosities ($L_{corr}$) is representative of the
    luminosity range of the SFRS sample.
    The dashed black line corresponds to the minimum luminosity $L_{min}$ for
    which Equation \ref{F_obs_d} has real roots.
    The maximum distance $d_{MAX}$ at which a source of luminosity $L_{corr}$ can
    be detected is defined by when the corresponding $F_{obs}$ curve hits the
    detection limit $F_{lim}$ (black horizontal line).
    Dashed vertical lines show the $d_{MAX}$ defined according to the ``classical''
    definition (i.e., without accounting for the additional K-correction;
    dashed vertical lines).
    \label{d_max}
   }
  \end{center}
 \end{figure}

 \noindent
 In order to find $d_{MAX}$, one should solve Equation \ref{F_obs_d} for $d$
 after setting \mbox{$F_{obs}(d)$ = $F_{lim}$}.
 We solved this equation numerically.
 Figure \ref{d_max} shows an example of the numerical evaluation of
 Equation \ref{F_obs_d} for a few sources of different $L_{corr}$, representative
 of the luminosity range of the SFRS sample (continuous lines).
 The maximum distance $d_{MAX}$ at which a source of
 luminosity $L_{corr}$ can be detected is that for which the observed flux $F_{obs}$
 curve (parametrized by Equation \ref{F_obs_d}) hits the detection limit $F_{lim}$
 (black horizontal line). 
 In this plot we also show the maximum distance that we would have calculated if we did
 not account for the additional K-correction ($d_{MAX,classic}$).
 Notice that the discrepancy arising from the additional factor ${1 \over {(1 + d \cdot H_{0}/c)}}$
 in Equation \ref{F_obs_d} increases for more distant objects.
 The slight bending of the $F_{obs}$ curves kicks in when the additional K-correction
 becomes significant, and it is the cause of the aforementioned discrepancy between
 $d_{MAX}$ and $d_{MAX,classic}$ at high luminosities.

 \bigskip
 \noindent
 Similarly to \cite{schmidt}, we evaluated the contribution of each SFRS target
 $i$ to the source density within each bin in terms of the ``density element''
 ${1 \over V_{MAX}}$.
 The total source density ${dN \over dV}^{bin}$ within a magnitude bin is then:
 \vspace{-0.3cm}
 \begin{eqnarray}
  {dN \over dV}^{bin} = \displaystyle\sum_{i=1}^{N^{bin}_{SFRS}}{w^{i} \over V_{MAX}^{i}(L^{i}_{60})}~~,
  \label{equation:dN_dV}
 \end{eqnarray}

 \noindent
 where the summation is performed over all the sources contained in the bin, and
 $w^{i}$ is the weight associated to the galaxy $i$ (see $\S$\ref{Correction for sky coverage }).
 Note that we explicitly wrote $V_{MAX}^{i}$ as $V_{MAX}^{i}(L^{i}_{60})$
 because the volume element for source $i$ is determined by the 60$\mu$m luminosity of
 that source, $L^{i}_{60}$, as discussed earlier in this section.
 In practice, for each galaxy $i$ in a given $K$-band bin we considered the
 corresponding 60$\mu$m luminosity $L^{i}_{60}$ and used it to calculate the relevant
 $V_{MAX}^{i}(L^{i}_{60})$.

\begin{figure}
  \begin{center}
   \includegraphics[width=0.48\textwidth,angle=0]{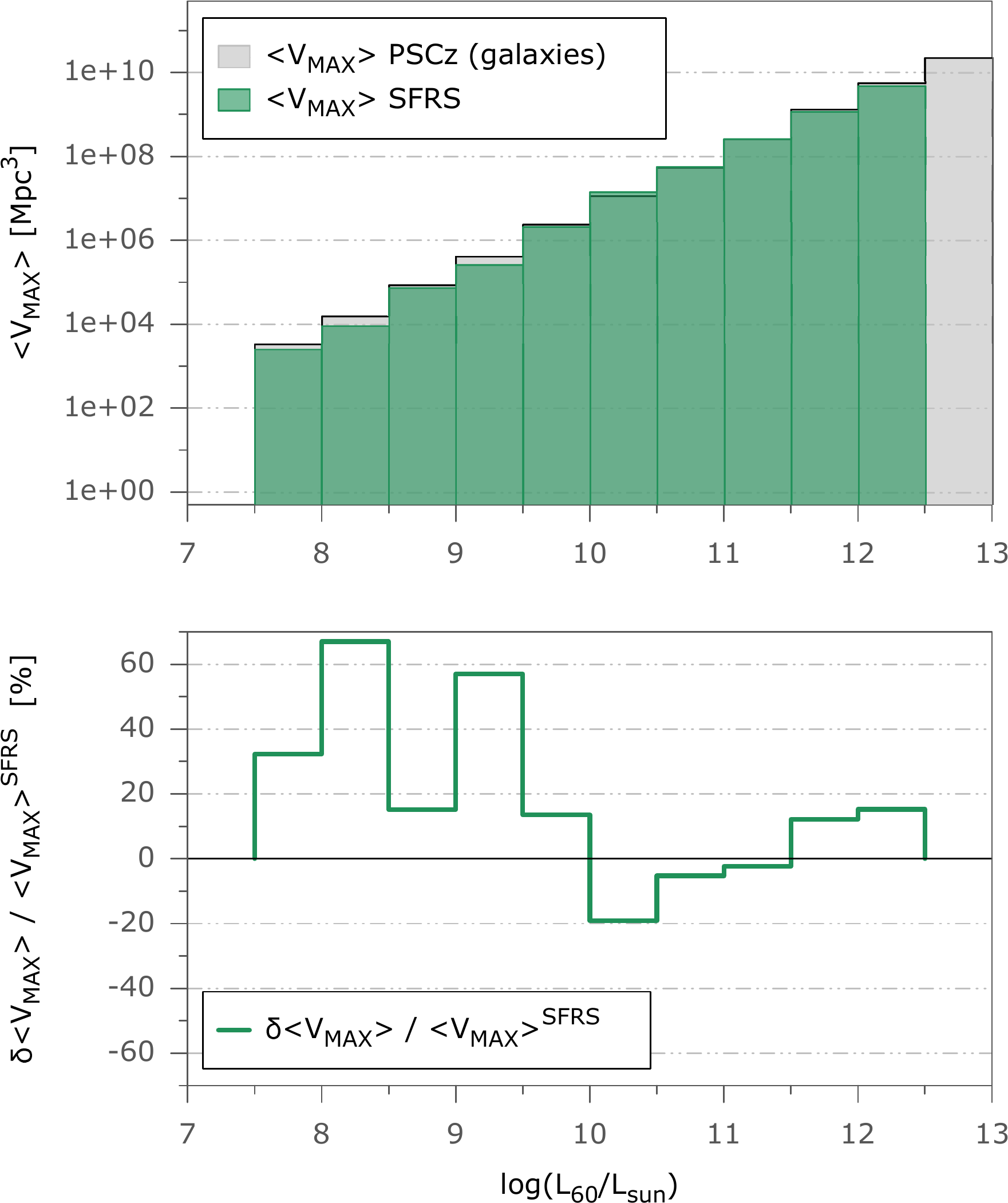}
   \caption[Distribution of $V_{MAX}$ for the SFRS and for its parent \PSCz{} sample]{
    Distribution of $V_{MAX}$ for the SFRS and for its parent \PSCz{} sample.
    \emph{Top ---}
    Distribution of the median $V_{MAX}$ calculated within each luminosity bin,
    for the SFRS (green) and the \PSCz{} (grey) samples.
    \emph{Bottom ---}
    Relative difference ($\delta\langle V_{MAX}^{bin} \rangle$; Equation \ref{equation:V_max_eff})
    between the median $V_{MAX}$
    distributions shown above.
    \label{figure:volume_correction}
   }
  \end{center}
 \end{figure}

 However, since the SFRS was drawn from the \PSCz{} by selecting the 60$\mu$m \emph{brightest}
 objects in each bin of the three-dimensional sample selection scheme
 (see $\S$\ref{The Star Formation Reference Survey (SFRS)} and \citealt{ashby}), the
 estimated volume is slightly biased with respect to the $V_{MAX}$ we would have calculated
 if the SFRS had been randomly drawn from the \PSCz{} sources in each bin.
 This discrepancy is illustrated in Figure \ref{figure:volume_correction}, which
 shows (top panel) the distribution of the median $V_{MAX}$ ($\langle V_{MAX} \rangle$).
 The bottom panel shows the difference $\delta \langle V_{MAX}^{bin} \rangle$:
 
 \vspace{-0.3cm}
 \begin{eqnarray}
  \delta\langle V_{MAX}^{bin}\rangle ~=~\langle V_{MAX}^{bin}\rangle ^{PSCz} - ~\langle V_{MAX}^{bin}\rangle ^{SFRS}~~.
  \label{delta} 
 \end{eqnarray}

 \medskip
 \noindent
 We correct for this bias by adding
 the appropriate $\delta{}V_{MAX}^{i}$ to the $V_{MAX}^{i}$ of each SFRS object.
 For this purpose we simply used the results shown in Figure \ref{figure:volume_correction},
 and for each galaxy $i$ we picked the $\delta{} \langle V_{MAX}^{bin} \rangle$ corresponding
 to the $L_{60}$ bin in which the galaxy resides. 
 Ultimately, the \emph{effective} $V_{MAX}$ for galaxy $i$ ($V_{MAX,eff}^{i}$)
 actually adopted in our analysis is:

 \vspace{-0.3cm}
 \begin{eqnarray}
  V_{MAX,eff}^{i} = V_{MAX}^{i}(L_{60}^{i}) + \delta \langle V_{MAX}^{bin} \rangle|_{L_{60}^{i} \in bin}
  \label{equation:V_max_eff} 
 \end{eqnarray}

 \noindent
 Finally, from this volume we additionally subtracted a constant value 
 corresponding to the minimum distance imposed during the selection of our
 sample ($\sim$6.8~Mpc; $\S$\ref{The PSCz sample}).

\subsection[$V/V_{MAX}$ test]{$V/V_{MAX}$ test}
\label{V/V_MAX test}

\begin{figure}
  \begin{center}
   \includegraphics[width=0.48\textwidth,angle=0]{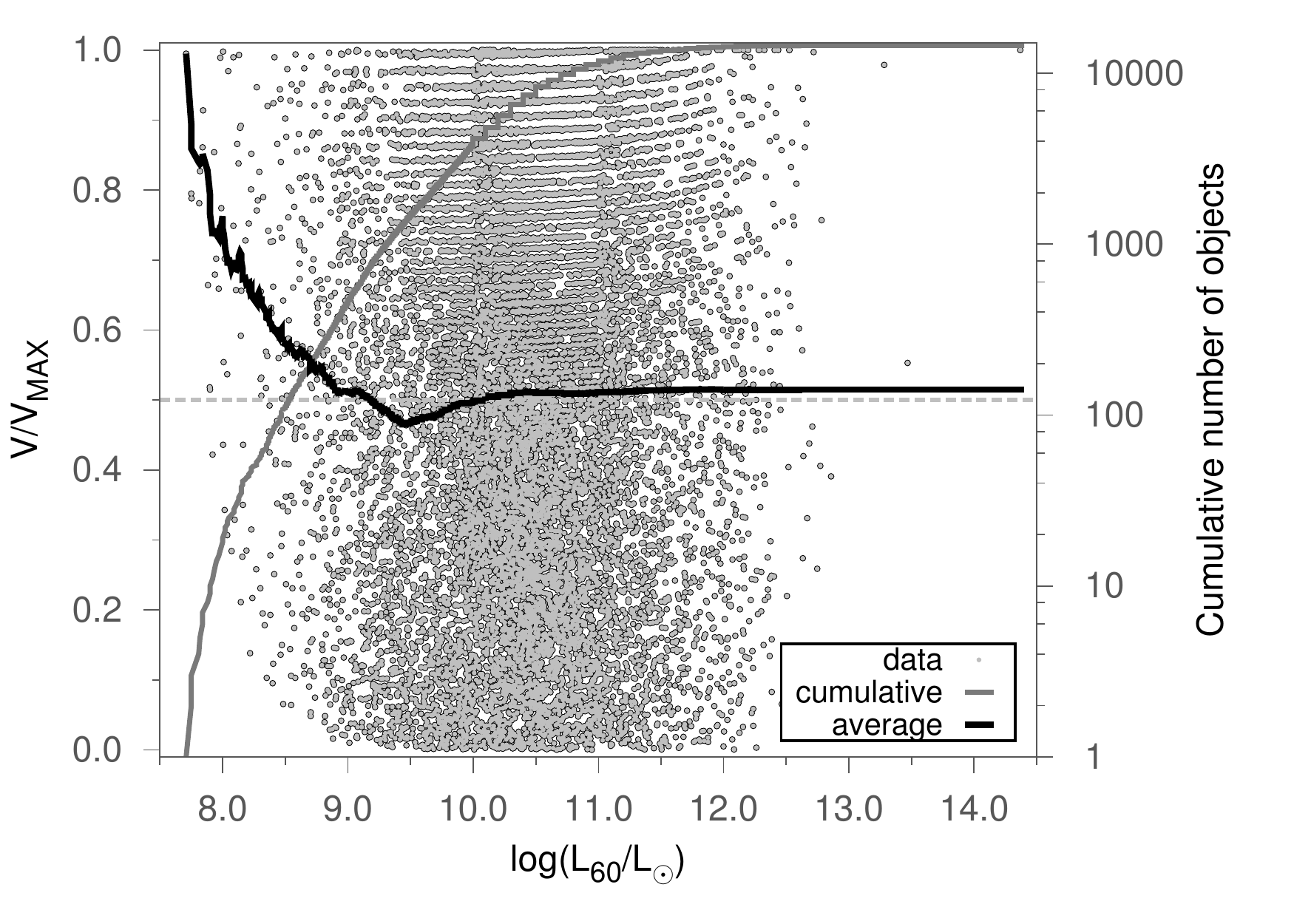}
   \caption[$V/V_{MAX}$ test for the \PSCz{} sample]{
    $V/V_{MAX}$ test for the \PSCz{} sample (grey dots).
    The black line shows the running average of the $V/V_{MAX}$ ratio as a function
    of 60$\mu$m luminosity.
    To show how many sources the running average is considering at a given
    luminosity, we also overplot a curve showing the cumulative number of objects
    (grey line, right ordinate).
    The running average reaches the $V/V_{MAX}$ = 0.5 value once $\sim$1\%
    of the sample is included.
    \label{V_V_max}
   }
  \end{center}
 \end{figure}

 \noindent
 The $V/V_{MAX}$ test \citep{schmidt} helps detect whether there are significant
 deviations from the uniform distribution of sources along the line of sight
 (as in the case of cluster of galaxies), which is a basic assumption in the
 definition of LF we are using.
 If we define $V$ as the volume contained within the distance of a given source,
 then it is expected that, in a uniformly distributed sample, the average
 $V/V_{MAX}$ is close to 0.5 (for a large number of sources).
 Figure \ref{V_V_max} reports the results of the $V/V_{MAX}$ test
 (where $V_{MAX}$ was calculated according to the ``classical'' definition
 --- Equation \ref{V}) for the whole \PSCz{} sample.
 The uniform distribution of the data points shows that the \PSCz{} sample is
 nearly unaffected by distribution biases.
 $V/V_{MAX}$ slightly increases towards bright luminosities,
 shifting the average $V/V_{MAX}$ slightly above 0.5.
 This effect is primarily due to the assumptions on the K-correction, which mostly
 affects the more distant (and hence brighter) objects.
 Nonetheless, this difference is very small  ($\sim$3\%), and it does not
 significantly affect our analysis.

\subsection[SFRS $K_s$-band luminosity function]{SFRS $K_{s}$-band luminosity function}
\label{SFRS $K_s$-band luminosity function}
 
\begin{figure}

 \includegraphics[width=0.48\textwidth,angle=0]{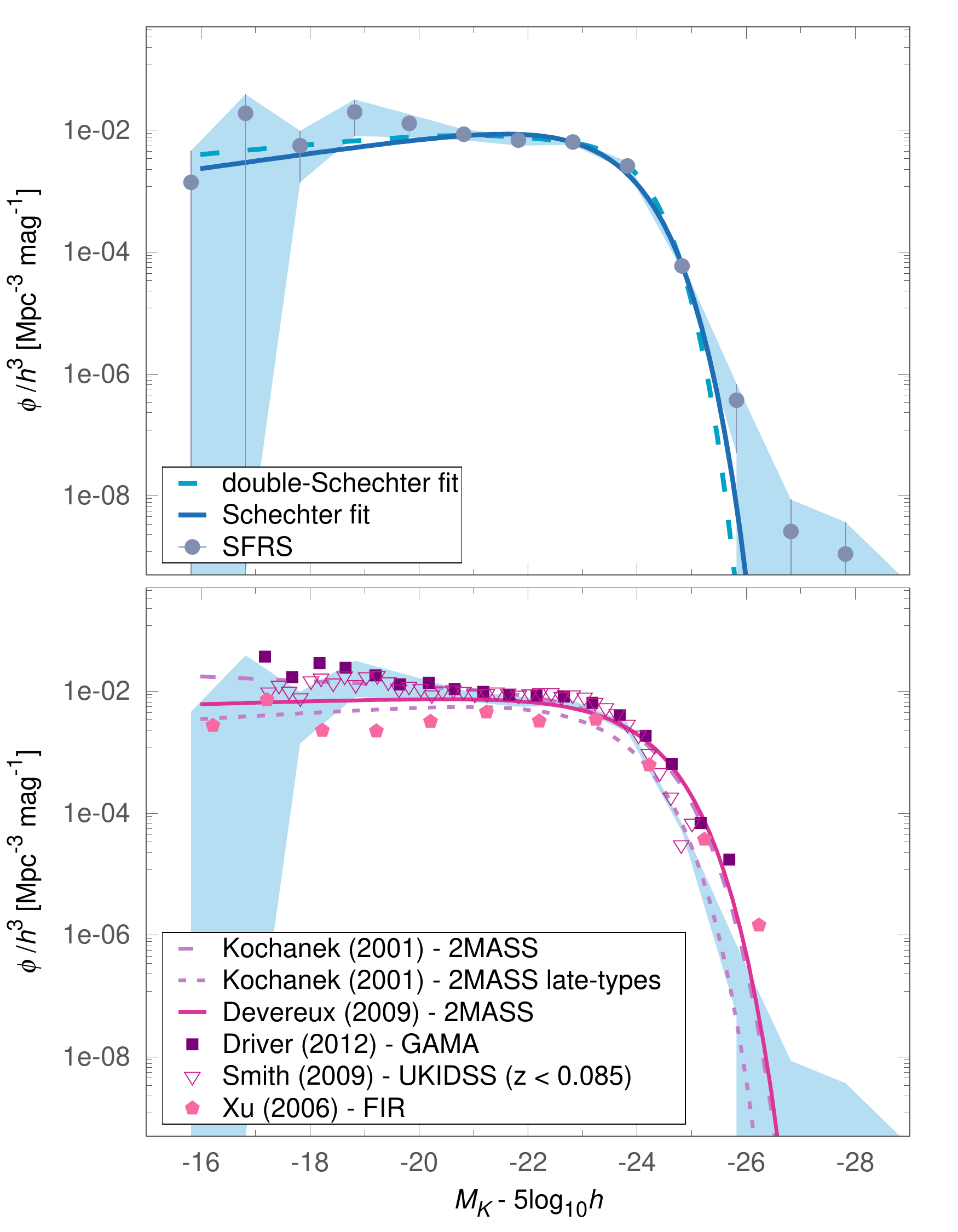}
 \vspace{-0.5cm}
 \caption[$K$-band luminosity function for the SFRS sample]{
  $K_{s}$-band luminosity function for the SFRS sample (blue dots), binned in
  ${\Delta}M_{K}$ = 1.0~mag bins.
  \emph{Top ---}
  Schechter (blue solid line) and double-Schechter (light blue dashed line)
  fit to the SFRS LF.
  \emph{Bottom ---} 
  Same data as for the above panel, now compared with other studies.
  The curves refer to the 2MASS-based samples of \citet[][solid]{devereux}
  and \citet[][long-dashed: full sample, short-dashed: late-type sub-sample]{kochanek}.
  The pentagons show the LF of FIR-selected galaxies of \cite{xu}, while the GAMA
  \citep{driver} and the UKIDSS \citep{smith} LFs are represented by squares and
  triangles, respectively.
  Data points represent the actually measured LFs and the bands indicate their
  uncertainties, while the curves represent fits to Schechter functions
  extrapolated within our magnitude range.  
  \label{figure:LF_K}
 }
 
  \begin{tabular*}{0.48\textwidth}{c @{\extracolsep{\fill}} ccc}
  \hline
  \multicolumn{4}{c}{\textsc{Luminosity Function}} \\
  \hline
  \hline
  \addlinespace 
  $M_{K}$                & $dN$      & $\phi$                                    & $\delta\phi$                             \\
  ~[Mag $-$ 5log($h$)] &             & [$dN$~$h^{3}$ Mpc$^{-3}$ dex$^{-1}$] & [$dN$~$h^{3}$ Mpc$^{-3}$ dex$^{-1}$] \\
  \tiny{(1)}              & \tiny{(2)} & \tiny{(3)}                                  & \tiny{(4)}                                  \\

  \hline
  \addlinespace 
  $-$27.82 & 1 & 1.1e$-$09 & 2.6e$-$09 \\
  $-$26.82 & 1 & 2.6e$-$09 & 6.0e$-$09 \\
  $-$25.82 & 4 & 3.7e$-$07 & 3.2e$-$07 \\
  $-$24.82 & 42 & 5.9e$-$05 & 8.6e$-$06 \\
  $-$23.82 & 111 & 2.6e$-$03 & 4.6e$-$04 \\
  $-$22.82 & 87 & 6.4e$-$03 & 6.6e$-$04 \\
  $-$21.82 & 41 & 6.9e$-$03 & 1.3e$-$03 \\
  $-$20.82 & 35 & 8.6e$-$03 & 1.7e$-$03 \\
  $-$19.82 & 22 & 1.3e$-$02 & 5.3e$-$03 \\
  $-$18.82 & 12 & 2.0e$-$02 & 1.2e$-$02 \\
  $-$17.82 & 4 & 5.6e$-$03 & 4.2e$-$03 \\
  $-$16.82 & 4 & 1.9e$-$02 & 2.0e$-$02 \\
  $-$15.82 & 1 & 1.4e$-$03 & 3.2e$-$03 \\
  \hline
  \addlinespace 
  \multicolumn{4}{c}{$j$ = 1.72 $\pm$ 0.93 $\times$ 10$^{9}$ [L$_{\odot}$ $h^{-1}$ Mpc$^{-3}$]}\\
  \addlinespace 
  \hline
 \end{tabular*}

 \captionof{table}{
  Values for the SFRS LF (Figure \ref{figure:LF_K})
  and luminosity density $j$.
  \label{table:LF}
 }
 \
 $^{(1)}$ Bin central value ($K$-band absolute magnitude, Vega system);
 $^{(2)}$ Number of SFRS sources in the bin;
 $^{(3)}$ Source density;
 $^{(4)}$ Poisson uncertainty on source density.


\end{figure}

 \noindent
 \noindent
 Figure \ref{figure:LF_K} shows the source density expressed in terms of 
 \mbox{$h$ = $H_{0}$ / 100}.
 The uncertainties were calculated by propagating the Poisson noise over the source
 counts in each bin. For small numbers of counts ($N < 5$), we calculated the
 uncertainties using the \cite{gehrels} approximation $G_{N}$:
 
 \vspace*{-0.3cm}
 \begin{eqnarray}
   G_N = 1+ \sqrt{0.75 + N}~~.
  \label{equation:Gehrels}
 \end{eqnarray}
 
 \noindent
 The top panel shows two fits to the LF; the first (blue solid line) is a fit to
 the Schechter function \citep{schechter} in its magnitude form \citep{felten}:

 \vspace{-0.3cm}
 \begin{eqnarray}
  \begin{aligned}
   \phi(M)dM = ~& 0.4 ln(10) ~ \phi^{*} ~ 10^{0.4(M^{*}-M)(\alpha+1)} \\
                & \times exp\left(-10^{0.4(M^{*}-M)}\right)dM~~.
  \end{aligned}
  \label{equation:Schechter_LF}
 \end{eqnarray}
 
 \noindent
 This function behaves as a power law of index $\alpha$ at low luminosities,
 and as an exponential at the high end.
 The value of $M^{*}$ indicates the magnitude of transition between the two
 components, while $\phi^{*}$ is a normalization factor.
 The second curve in the top panel (light blue dashed line) represents instead
 a fit to a double-Schechter function \citep[e.g.,][]{baldry}: 

 \vspace{-0.3cm}
 \begin{eqnarray}
  \begin{aligned}
   \phi(M)dM & = \phi_{1}(M)dM + \phi_{2}(M)dM \\
             & = 0.4 ln(10) \\
             & ~~~~\times \left( \phi^{*}_{1} 10^{0.4(M^{*}-M)\alpha_{1}} + \phi^{*}_{2} 10^{0.4(M^{*}-M)\alpha_{2}} \right)\\
             & ~~~~\times exp\left(-10^{0.4(M^{*}-M)}\right) 10^{0.4(M^{*}-M)} dM~~.
  \end{aligned}
 \end{eqnarray}
 
 \noindent
 This function implements an outer exponential-like profile connected to a
 double inner power-law with logarithmic slopes $\alpha_{1}$ and $\alpha_{2}$,
 such that $\alpha_{1} < \alpha_{2}$ (i.e., $\alpha_{1}$ regulates the behaviour
 at the faintest magnitudes).
 Here, $\phi^{*}_{1}$ and $\phi^{*}_{2}$ are the normalization factors of the
 separate Schechter components $\phi_{1}(M)$ and $\phi_{2}(M)$, while $M^{*}$ 
 retains its meaning of transition point.
 The best-fit parameters for both the single- and double-Schechter fit are
 reported in Table \ref{table:Schecter_fit}.
 The two functions provide a similar fit to the data, and hence we will focus,
 for the rest of the paper, on the simpler (single-)Schechter function.
 Although consistent within the uncertainties, we observe an excess of brightest
 sources with respect to the Schechter fit in the last 2 magnitude bins.
 The most plausible explanation involves low-number statistics; in fact, before
 applying the completeness correction, these bins contain
 only 1 source each.
 This feature is discussed further in $\S$\ref{Comparisons against other samples}.

\begin{table*}
 \centering
  \begin{tabular}{lcccccc}
   \hline
   \multicolumn{7}{c}{\textsc{Schechter Fits to the LF and MF}} \\
   \hline
   \hline
   \addlinespace 
   \multicolumn{1}{c}{Model}      & $M^{*}_{K}$          & M$^{*}$                     & $\alpha^{}_{1}$ & $\phi^{*}_{1}$ & $\alpha^{}_{2}$ & $\phi^{*}_{2}$ \\
                                  & [Mag $-$ 5log($h$)]  & [log($h^{2}$ M/M$_{\odot}$] &                                  & [$dN$ $h^3$ Mpc$^{-3}$ dex$^{-1}$] &             & [dN $h^3$ Mpc$^{-3}$ dex$^{-1}$]  \\
   \multicolumn{1}{c}{\tiny{(1)}} & \tiny{(2)}           & \tiny{(3)}                  & \tiny{(4)}                       & \tiny{(5)}        & \tiny{(6)} & \tiny{(7)} \\
   \hline
   \addlinespace 
   \multicolumn{7}{c}{\textsc{Luminosity Function}} \\
   \hline
   \addlinespace 
   Schechter          & $-$22.84 & -- & $-$0.7 & 0.015 & --  & --    \\
   double-Schechter   & $-$22.38 & -- & $-$0.8 & 0.017 & 1.2 & 0.006 \\
   \hline
   \addlinespace 
   \multicolumn{7}{c}{\textsc{Mass Function}} \\
   \hline
   \addlinespace 
   Schechter          & -- & 10.41 & $-$0.7 & 0.017 & --  & --    \\
   \hline
  \end{tabular}
  \begin{minipage}{0.8\textwidth}
   \caption[]{
    Results of the single- and double-Schechter fits to the LF and MF.
    \\
    $^{(1)}$   Model function.
    $^{(2)}$   Transition magnitude.
    $^{(3)}$   Transition mass.
    $^{(4,6)}$ Inner power law.
    $^{(5,7)}$ Normalization.
    \label{table:Schecter_fit}
   }
 \end{minipage}
\end{table*}

 Due to this discrepancy with the Schechter fit, we calculated the integrated
 luminosity density $j$ from the direct summation of the 1/$V_{MAX}$
 weighted luminosities of the SFRS sources rather than from the integration of
 the Schechter curve.
 Namely:
 \vspace{-0.3cm}
 \begin{eqnarray}
  j = \displaystyle\sum_{i=1}^{N}{{w^{i} \over V_{MAX,eff}^{i}} \times 10^{-0.4(M^{i}_{K} - M_{\odot})}}~~,
  \label{equation:j} 
 \end{eqnarray}
 \noindent
 where the first term in the summation represents the contribution to the LF
 from source $i$, which has $K_{s}$-band magnitude $M^{i}_{K}$.
 This yields $j$ = 1.72 $\pm$ 0.93 $\times$ 10$^{9}$ L$_{\odot}$ $h^{-1}$ Mpc$^{-3}$.
 
\begin{figure}

 \includegraphics[align=c,width=0.48\textwidth,angle=0]{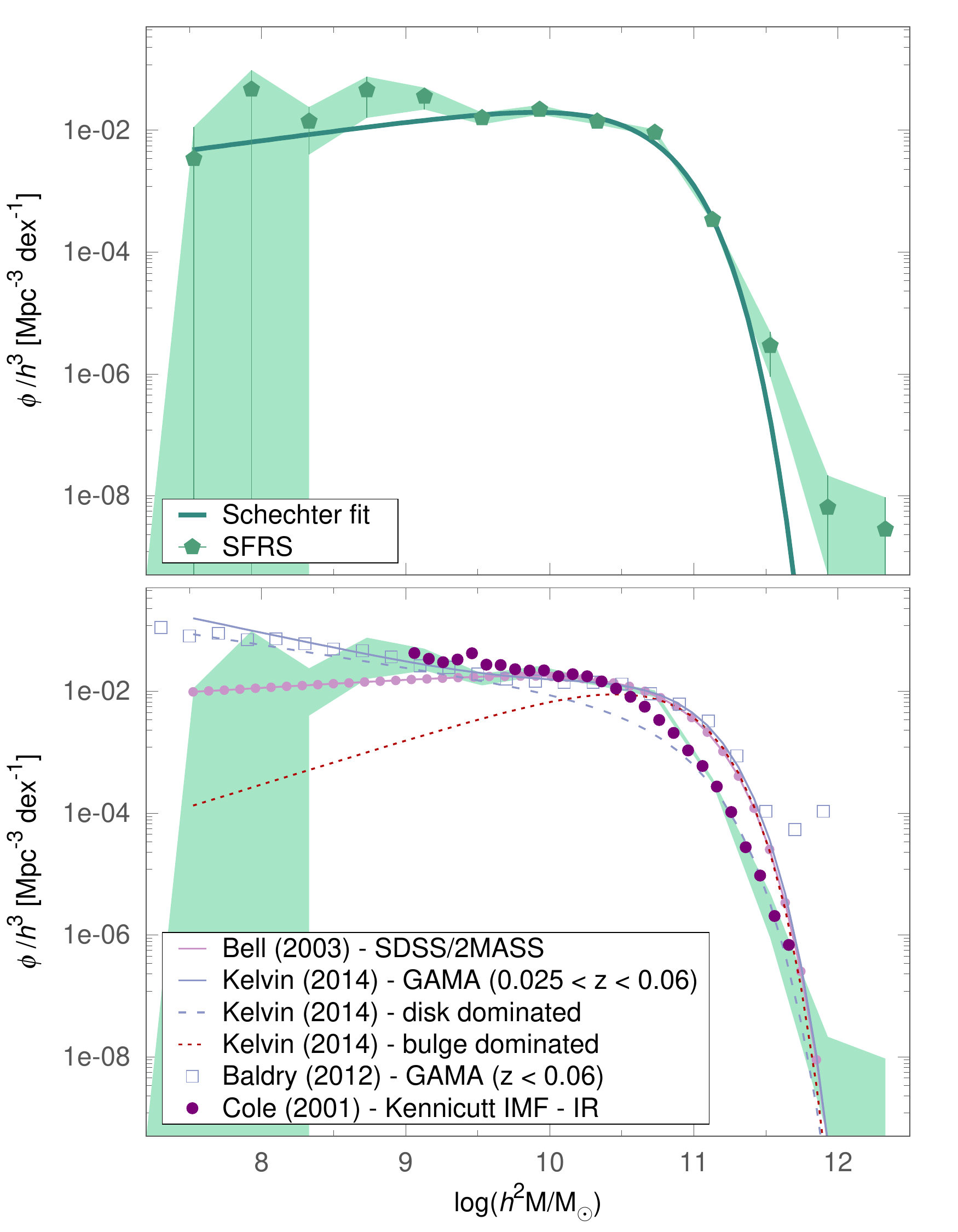}
 \caption[Stellar mass function for the SFRS sample]{
  Stellar mass function for the SFRS sample (green data points), binned in
  0.4~dex bins.
  \emph{Top ---}
  Schechter fit to the SFRS MF (green line).
  \emph{Bottom ---}
  As a comparison, we overplot the results from the studies of the mass functions
  of galaxies in  the Local Universe by \citet[][dotted pink curve]{bell:2003},
  the GAMA MFs by \citet[][dashed line]{kelvin} and \citet[][empty squares]{baldry},
  and the MF of IR-selected galaxies by \citet[][violet data points]{cole}.
  The MFs of \cite{kelvin} is further subdivided into the MFs for their disk-
  and bulge-dominated systems.
  Data points represent the actually measured MFs and the bands indicate their
  uncertainties, while the curves represent fits to Schechter functions extrapolated
  within our mass range.  
  \label{figure:MF}
 }
 
  \begin{tabular*}{0.48\textwidth}{c @{\extracolsep{\fill}} ccc}
  \hline
  \multicolumn{4}{c}{\textsc{Mass Function: Total}} \\
  \hline
  \hline
  \addlinespace 
  M                                  & $dN$      & $\phi$                                    & $\delta\phi$                             \\
  ~[log($h^{2}$ M/M$_{\odot}$)] &             & [$dN$~$h^{3}$ Mpc$^{-3}$ dex$^{-1}$] & [$dN$~$h^{3}$ Mpc$^{-3}$ dex$^{-1}$] \\
  \tiny{(1)}                        & \tiny{(2)} & \tiny{(3)}                                  & \tiny{(4)}                                  \\

  \hline
  \addlinespace 
  7.53 & 1 & 3.4e$-$03 & 8.0e$-$03 \\
  7.93 & 4 & 4.7e$-$02 & 5.1e$-$02 \\
  8.33 & 4 & 1.4e$-$02 & 1.0e$-$02 \\
  8.73 & 11 & 4.6e$-$02 & 3.0e$-$02 \\
  9.13 & 23 & 3.6e$-$02 & 1.4e$-$02 \\
  9.53 & 28 & 1.6e$-$02 & 3.5e$-$03 \\
  9.93 & 41 & 2.2e$-$02 & 4.0e$-$03 \\
  10.33 & 74 & 1.4e$-$02 & 1.6e$-$03 \\
  10.73 & 111 & 9.2e$-$03 & 1.2e$-$03 \\
  11.13 & 58 & 3.4e$-$04 & 3.6e$-$05 \\
  11.53 & 7 & 2.9e$-$06 & 2.0e$-$06 \\
  11.93 & 1 & 6.4e$-$09 & 1.5e$-$08 \\
  12.33 & 1 & 2.8e$-$09 & 6.6e$-$09 \\
  \hline
  \addlinespace 
  \multicolumn{4}{c}{$\rho_{M}$ = 4.61 $\pm$ 2.40 $\times$ 10$^{8}$ [M$_{\odot}$ $h^{-1}$ Mpc$^{-3}$]}\\
  \addlinespace 
  \hline
 \end{tabular*}

 \captionof{table}{
  Values for the SFRS MF (\mbox{Figure \ref{figure:MF}})
  and mass density $\rho_{M}$.
  \label{table:MF}
 }
 \
 $^{(1)}$ Mass bin central value;
 $^{(2)}$ Number of SFRS sources in the bin;
 $^{(3)}$ Source density;
 $^{(4)}$ Poisson uncertainty on source density.


\end{figure}

\subsection[SFRS stellar mass functions]{SFRS stellar mass functions}
\label{SFRS stellar mass functions}

 \noindent
 The total stellar masses ($M$) of the SFRS galaxies, as well as their disk and bulge
 sub-components (defined as summarized in Table \ref{table:segregation}),
 were calculated from their $K_{s}$-band luminosities by assuming a M/L ratio, as
 described in Appendix \ref{Zero-points, luminosities, and stellar mass calculation}.
 The stellar masses for the \emph{best-fit} models are reported in
 \mbox{Table \ref{table:GALFIT_selected_models}}.
 For the faint and/or noisy galaxies for which we could not obtain a fit to their 2D
 surface brightness, we used the total $K_{s}$-band luminosities from 2MASS
 (listed in Table \ref{table:SFRS_2MASS_SDSS}), and we only report their total stellar
 mass (see $\S$\ref{Separation of disk/bulge components}).

\begin{figure*}

 \includegraphics[width=0.48\textwidth,angle=0]{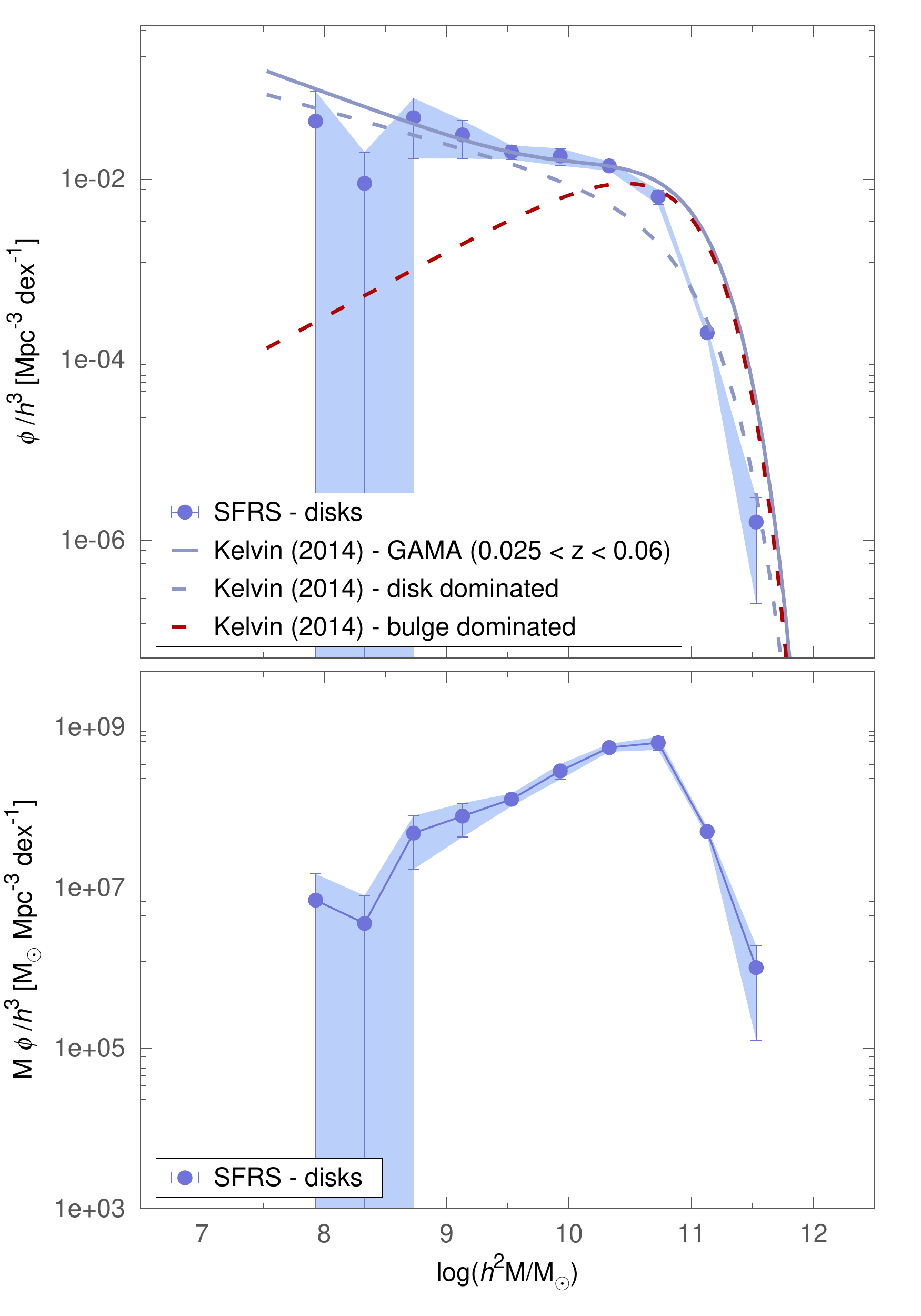}
 \includegraphics[width=0.48\textwidth,angle=0]{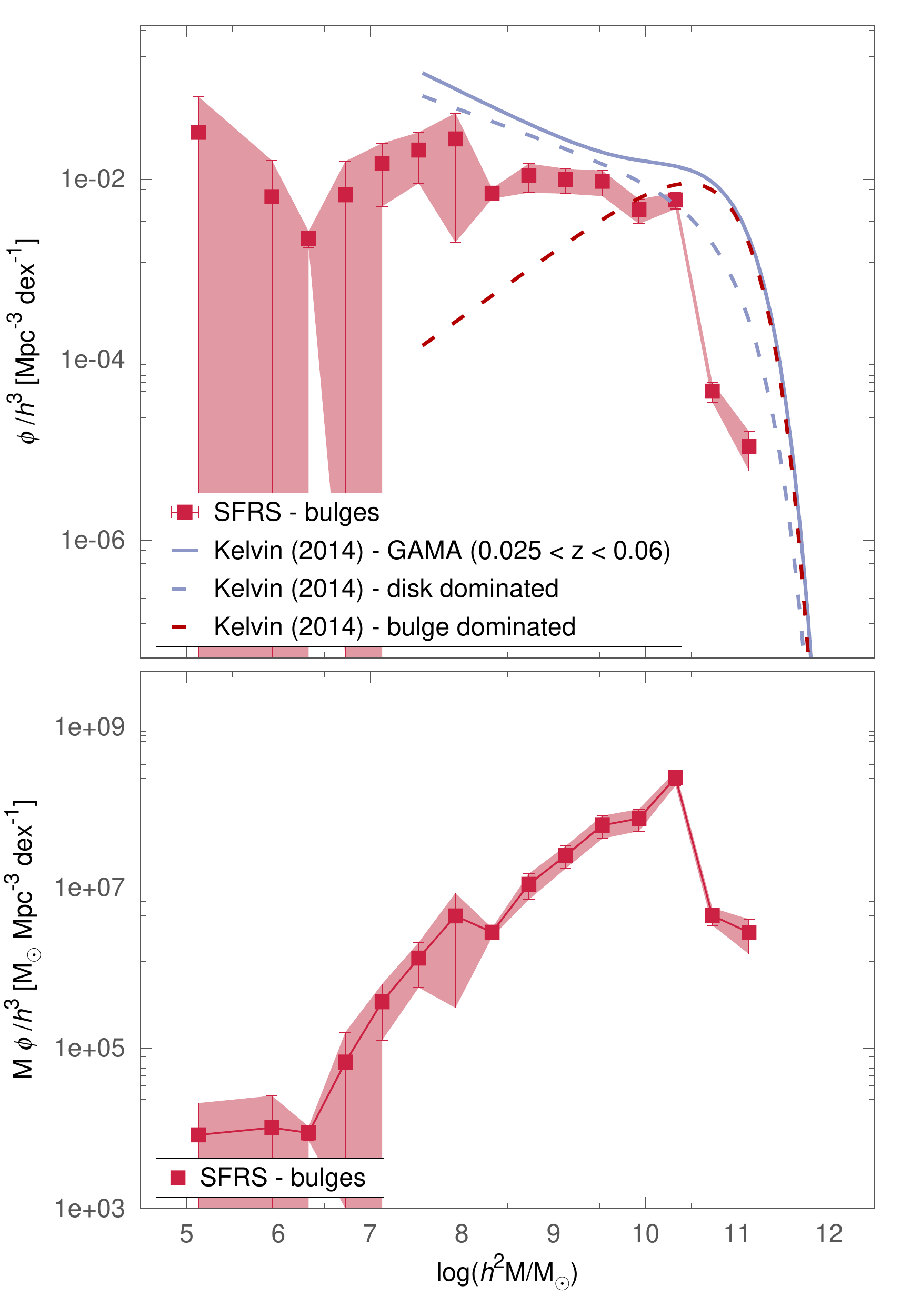}
 \caption[Stellar mass function and mass-density function for disks]{
  \emph{Top ---}
  Stellar mass function for the disk (\emph{left}) and bulge (\emph{right})
  sub-components, binned in 0.4~dex bins, with shaded bands showing the 1-$\sigma$
  uncertainties.
  As a comparison we overplot the mass function for the GAMA sample of
  nearby galaxies by \citet[][solid curve]{kelvin}, which is further subdivided
  into their MFs for the disk-dominated (blue dashed line) and bulge-dominated
  (red dashed line) objects.  
  \emph{Bottom ---} Mass-density function for the disk (\emph{left}) and bulge
  (\emph{right}) sub-components defined by Equation \ref{equation:MDF}.
  \label{figure:MF_disks_bulges}
 }
 
 \vspace{0.5cm}
 
\begin{tabular}{cc}
\begin{minipage}{0.48\linewidth}
  \begin{tabular*}{\textwidth}{c @{\extracolsep{\fill}} ccc}
  \hline
  \multicolumn{4}{c}{\textsc{Mass Function: Disks}} \\
  \hline
  \hline
  \addlinespace 
  M                                  & $dN$      & $\phi$                                    & $\delta\phi$                             \\
  ~[log($h^{2}$ M/M$_{\odot}$)] &             & [$dN$~$h^{3}$ Mpc$^{-3}$ dex$^{-1}$] & [$dN$~$h^{3}$ Mpc$^{-3}$ dex$^{-1}$] \\
  \tiny{(1)}                        & \tiny{(2)} & \tiny{(3)}                                  & \tiny{(4)}                                   \\

  \hline
  \addlinespace 
  7.93 & 4 & 4.4e$-$02 & 5.0e$-$02 \\
  8.33 & 2 & 9.0e$-$03 & 1.1e$-$02 \\
  8.73 & 10 & 4.8e$-$02 & 3.1e$-$02 \\
  9.13 & 21 & 3.1e$-$02 & 1.4e$-$02 \\
  9.53 & 32 & 2.0e$-$02 & 3.6e$-$03 \\
  9.93 & 40 & 1.8e$-$02 & 3.9e$-$03 \\
  10.33 & 81 & 1.4e$-$02 & 1.6e$-$03 \\
  10.73 & 97 & 6.4e$-$03 & 1.2e$-$03 \\
  11.13 & 35 & 2.0e$-$04 & 2.7e$-$05 \\
  11.53 & 4 & 1.6e$-$06 & 1.4e$-$06 \\
  \hline
  \addlinespace 
  \multicolumn{4}{c}{$\rho_{M,disks}$ = 3.35 $\pm$ 1.82 $\times$ 10$^{8}$ [M$_{\odot}$ $h^{-1}$ Mpc$^{-3}$]}\\
  \addlinespace 
  \hline
 \end{tabular*}

 \captionof{table}{
  Values for the disk MF (Figure \ref{figure:MF_disks_bulges})
  and mass density $\rho_{M,disks}$.
  \label{table:MF_disks}
 }
 \
 $^{(1)}$ Mass bin central value;
 $^{(2)}$ Number of SFRS sources in the bin;
 $^{(3)}$ Source density;
 $^{(4)}$ Poisson uncertainty on source density.

\end{minipage}
 & \begin{minipage}{0.48\linewidth}
  \begin{tabular*}{\textwidth}{c @{\extracolsep{\fill}} ccc}
  \hline
  \multicolumn{4}{c}{\textsc{Mass Function: Bulges}} \\
  \hline
  \hline
  \addlinespace 
  M                                  & $dN$      & $\phi$                                    & $\delta\phi$                             \\
  ~[log($h^{2}$ M/M$_{\odot}$)] &             & [$dN$~$h^{3}$ Mpc$^{-3}$ dex$^{-1}$] & [$dN$~$h^{3}$ Mpc$^{-3}$ dex$^{-1}$] \\
  \tiny{(1)}                        & \tiny{(2)} & \tiny{(3)}                                  & \tiny{(4)}                                  \\

  \hline
  \addlinespace 
  5.13 & 1 & 3.3e$-$02 & 4.9e$-$02 \\
  5.93 & 1 & 6.4e$-$03 & 9.7e$-$03 \\
  6.33 & 2 & 2.2e$-$03 & 4.4e$-$04 \\
  6.73 & 2 & 6.7e$-$03 & 9.1e$-$03 \\
  7.13 & 9 & 1.5e$-$02 & 1.0e$-$02 \\
  7.53 & 10 & 2.1e$-$02 & 1.2e$-$02 \\
  7.93 & 15 & 2.8e$-$02 & 2.6e$-$02 \\
  8.33 & 14 & 7.0e$-$03 & 9.0e$-$04 \\
  8.73 & 19 & 1.1e$-$02 & 3.9e$-$03 \\
  9.13 & 31 & 1.0e$-$02 & 3.1e$-$03 \\
  9.53 & 59 & 9.5e$-$03 & 3.0e$-$03 \\
  9.93 & 69 & 4.6e$-$03 & 1.4e$-$03 \\
  10.33 & 50 & 5.9e$-$03 & 1.2e$-$03 \\
  10.73 & 12 & 4.5e$-$05 & 1.1e$-$05 \\
  11.13 & 6 & 1.1e$-$05 & 5.1e$-$06 \\
  \hline
  \addlinespace 
  \multicolumn{4}{c}{$\rho_{M,bulges}$ = 7.49 $\pm$ 4.09 $\times$ 10$^{7}$ [M$_{\odot}$ $h^{-1}$ Mpc$^{-3}$]}\\
  \addlinespace 
  \hline
 \end{tabular*}

 \captionof{table}{
  Values for the bulge MF (Figure \ref{figure:MF_disks_bulges})
  and mass density $\rho_{M,bulges}$.
  \label{table:MF_bulges}
 }
 \
 $^{(1)}$ Mass bin central value;
 $^{(2)}$ Number of SFRS sources in the bin;
 $^{(3)}$ Source density;
 $^{(4)}$ Poisson uncertainty on source density.

\end{minipage}
 \\
\end{tabular}

\end{figure*}

 We built the mass function MF, i.e., the number of sources per mass bin per
 Mpc$^{3}$, by simply summing the SFRS sources within each mass bin, after balancing 
 by their weight and maximum volume, as calculated in the previous sections.
 Formally:
 
 \vspace{-0.3cm}
 \begin{eqnarray}
  &\phi({\Delta}M_{j}) = \sum \frac{w^{i}}{V_{MAX,eff}^{i}}& \\[0.5em]
  &\forall{i}~|~M_{j} - {\Delta}M_{j} < M_{i} < M_{j} - {\Delta}M_{j}~~,& \nonumber
  \label{equation:MF} 
 \end{eqnarray}
 
 \noindent
 where $\phi({\Delta}M_{j})$ is the mass function corresponding to the mass bin
 $M_{j} \pm 0.5~{\Delta}M_{j}$, the weight $w^{i}$ is defined as
 discussed in $\S$\ref{Correction for sky coverage },
 and $V_{MAX,eff}^{i}$ is simply the $V^{bin}_{MAX,eff}$ value for the
 60$\mu$m luminosity bin in which galaxy $i$ resides
 (see Equations \ref{equation:dN_dV} and \ref{equation:V_max_eff}). 
 Figure \ref{figure:MF} and the top-left and top-right panels of Figure
 \ref{figure:MF_disks_bulges} show the total, disk and bulge mass functions,
 respectively.
 The uncertainties were calculated by propagating the Poisson noise of the
 source counts in each bin, as in the $K_{s}$-band luminosity function.
 In Appendix \ref{Mass Function uncertainties assessment} we
 present the results of a Monte-Carlo simulation that we performed in
 order to verify that the Poissonian uncertainties we adopt are reliable.
 The top panel of Figure \ref{figure:MF} shows a Schechter fit to the
 total MF of the form:

 \vspace{-0.3cm}
 \begin{eqnarray}
  \begin{aligned}
   \phi(M)dM = \ln(10) (10^{M}) {{\phi*} \over M^{*}} \left(10^{M \over M^{*}}\right)^{\alpha} \exp\left(-10^{M \over M^{*}}\right)~~,
  \end{aligned}
  \label{equation:Schechter_MF}
 \end{eqnarray}

 \noindent
 whose best-fit parameters are reported in Table \ref{table:Schecter_fit}.

 Using the morphological information derived from the 2D surface brightness fit
 ($\S$\ref{Best-fit model selection}), complemented by the concentration
 index decomposition ($\S$\ref{Further decomposition of mixed components}), we
 could measure the separate masses of the disk and bulge components of the
 sample galaxies.
 The $M/L$ ratios for each component were calculated using the \cite{bell:2003}
 prescription
 (Appendix \ref{Zero-points, luminosities, and stellar mass calculation};
 Equation \ref{ML}).
 We calculated the stellar mass density for the whole sample ($\rho_{M}$), and for
 the disks ($\rho_{M,disks}$) and bulges ($\rho_{M,bulges}$) sub-distributions,
 using the same formalism as in Equation \ref{equation:j}, obtaining
 $\rho_{M}$        = 4.61 $\pm$ 2.40 $\times$ 10$^{8}$ M$_{\odot}$ $h^{-1}$ Mpc$^{-3}$,
 $\rho_{M,disks}$  = 3.35 $\pm$ 1.82 $\times$ 10$^{8}$ M$_{\odot}$ $h^{-1}$ Mpc$^{-3}$,
 and
 $\rho_{M,bulges}$ = 7.49 $\pm$ 4.09 $\times$ 10$^{7}$ M$_{\odot}$ $h^{-1}$ Mpc$^{-3}$.

\begin{figure*}
  \begin{center}
   \includegraphics[width=0.8\textwidth,angle=0]{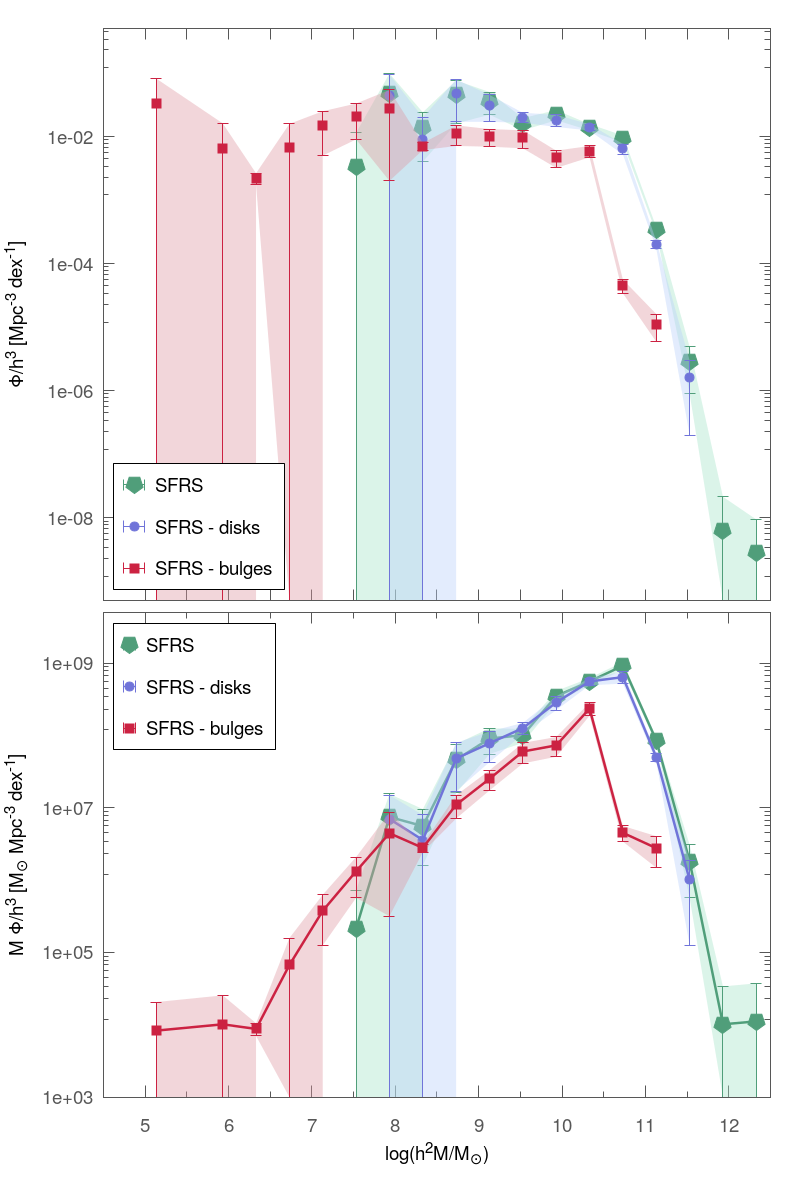}

   \begin{minipage}{0.8\textwidth}
    \caption[Comparison of total, disk, and bulge mass and mass-density functions]{
     Comparison of total (green diamonds), disk (blue dots), and bulge (red squares)
     mass (\emph{top}; Equation \ref{equation:MF}) and mass-density (\emph{bottom};
     Equation \ref{equation:MDF}) functions, binned in 0.4~dex bins.
     The 1-$\sigma$ confidence range shown by the shaded regions were calculated
     by propagating the Poisson uncertainty over the source counts in each bin.
     \label{figure:MF_comparison}
    }
   \end{minipage} 
  \end{center}
 \end{figure*}

 A direct comparison between the disk and bulge mass functions is reproduced in the
 top panel of Figure \ref{figure:MF_comparison}.
 Notice that --- in this representation --- the sum of the two curves does
 \emph{not} yield the total mass function value in the corresponding bin because
 each galaxy is accounted for in both the disk and bulge mass functions (except
 for pure disks/bulges).
 In order to better visualize the contribution to the stellar mass
 budget per unit volume due to each stellar component, we define the
 ``mass-density function'', i.e., the mass function multiplied by the central
 value of the mass bin:

 \vspace{-0.3cm}
 \begin{eqnarray}
  \phi(M) M \Delta{M} = \int^{M+\Delta{M}}_{M-\Delta{M}} \phi(M) M \delta{M}~~.
  \label{equation:MDF} 
 \end{eqnarray}

 \noindent
 The bottom panels of Figure \ref{figure:MF_disks_bulges} show the mass-density
 functions for disks and bulges, respectively, while
 in the bottom panel of Figure \ref{figure:MF_comparison} the two are compared
 against the total mass-density function.
 In this representation, the \emph{integral} mass densities of the disks and bulges
 \emph{do} sum up to the total mass\footnote{
  With the only caveat that the SFRS mass density for the \sampleNoBestFits{}
  unsuccessful fits (see $\S$\ref{Best-fit model selection})
  was integrated using 2MASS data, while the same was not possible for the single
  disks or bulges
 }.

 As a final remark, we stress that the MFs presented in this work refer
 to \emph{star-forming galaxies}, and to their bulges and disks.
 Passive galaxies will be and additional contribution to the MF.
 The underlying caveats of our work are the following.
  
 \begin{cutenumerate}[label=\arabic*)~]
  \itemsep0.5em  

  \newcounter{caveat}

  \stepcounter{caveat}
  \item[$\bullet$~] \emph{Assumption on volume correction ---}
   For the sub-component mass functions we adopted the volume correction
   calculated for the total MF \citep[e.g.,][]{benson}.
   This implies that if a bulge/disk is present, then we detected it.

  \stepcounter{caveat}
  \item[$\bullet$~] \emph{Caveat on completeness correction ---}
   The SFRS objects are selected from the \PSCz{} catalogue
   by binning it over
   the 3D parameter space defined by SFR, sSFR and dust temperature, and then
   selecting the brightest objects in each
   bin. 
   The representativeness of a SFRS target is then given by its weight in that
   3D bin ($\S$\ref{Correction for sky coverage } and $\S$\ref{Calculation of V_MAX}).
   Because the selection of the objects in each bin is based on flux, and not on
   luminosity (which could potentially be correlated with morphology), it does
   \emph{not} bias the morphological demographics with respect to the parent sample.
   Therefore, the SFRS weights applied for the total MF also apply to
   the bulge and disk MFs.
   
 \end{cutenumerate}

\section[Discussion]{Discussion}
\label{Discussion}
 
\subsection[Comparisons against other samples]{Comparisons against other samples}
\label{Comparisons against other samples}

 \noindent
 Several factors can hamper the comparison between LFs (or MFs) derived in
 different studies.
 These include factors related to the methodology (e.g., selection functions
 and completeness corrections, photometric systems, or even the analytical method
 used to build the LF/MF), or physical effects such as sample selection, redshift
 range, estimation of mass-to-light ratios, etc.
 With these limitations in mind, we present in this section a comparison
 of our functions against a selection of relevant studies addressing galaxies in
 the Local Universe.
 The cosmologies have been transformed to the one we assumed in $\S$\ref{Introduction}
 and the magnitudes converted to the Vega system.
 The bottom panel of Figure \ref{figure:LF_K} (Figure \ref{figure:MF}) show the
 all the comparative LFs (MFs) discussed below.
 
 \medskip 
 \noindent
 \emph{Luminosity Function} ---
 First, we consider $K$-band LFs based on samples directly selected in
 the $K$-band.
 These are:
 the 2MASS galaxies of \citet[][4\,192 galaxies, of which 2\,192 are late-type objects]{kochanek}
 and \citet[][1\,613 bright galaxies]{devereux},
 and the collection of 40\,111 galaxies from the UKIRT Infrared Deep Sky Survey (UKIDSS)
 of \citet[][]{smith}.
 The first two studies have been chosen for a direct comparison against our
 2MASS magnitudes, the last because it represents an extension of 2MASS
 towards fainter magnitudes \citep[\mbox{$K$ $\sim$ 18.3~mag};][]{hewett}.
 We additionally consider the LF derived by \cite{driver:GAMA} for the Galaxy And Mass
 Assembly (GAMA) sample: this refers to a slightly higher redshift than SFRS
 (\mbox{0.013 $< z <$ 0.1}) than that of SFRS (\mbox{$z \lesssim$ 0.05}), however
 it potentially encompasses all galaxy types in the nearby Universe and hence it
 offers a valuable reference for the overall galaxy population.
 Finally, we compare our LF against the one derived by \cite{xu}
 for a sample of 161 galaxies selected in the far IR (FIR), similarly to SFRS
 \footnote{
  For the sake of clarity, notice that \cite{xu} present the separate LFs of a
  UV-selected and of a FIR-selected galaxy sample.
  In our comparison we are solely referring to the latter sample.
 }.

 The most evident trend in \ref{figure:LF_K} is the progressive disappearance of
 low-luminosity systems as source-selection wavelength increases.
 The SFRS sample was selected at 60~$\mu$m to represent star-forming galaxies, and
 passive galaxies are therefore under-represented.
 Despite that, the SFRS agrees best with the $K$-band selected samples.
 In particular, our LF is a nearly exact match with the UKIDSS LF, and our LF
 exceeds the LF of the \cite{kochanek} late-type subsample, as it should
 (despite the selection wavelength, the SFRS does include some early-type galaxies).
 Relative to the GAMA visible-light survey \cite{driver:GAMA}, our sample shows a
 deficiency of dwarf systems, presumably passive dwarf ellipticals.
 However, in the range $-24 \la M_K \la -19$~mag, our results agree with those of GAMA,
 implying that star-forming galaxies constitute nearly all the galaxies in that
 magnitude range.
 One possible caveat is that our sample is limited to a closer redshift range than
 GAMA or UKIDSS, and local density enhancements can yield overestimates up to a
 factor of two\citep[e.g.][]{karachentsev}.
 In comparison to the SFRS and all the other surveys, the FIR LF of \cite{xu} gives
 systematically smaller density counts for luminosities smaller than $L^*$ ($M_K\ga -23$). 
 However, it shall be noted that their data have large uncertainties due to the low-number
 statistics (see their Figure 7) which can partially reconcile their LF with ours.
 If the bias was real, it could be attributed  to the significantly different morphological
 census: that of  \cite{xu} is composed of $\sim$39\% interacting systems or peculiar galaxies,
 while SFRS contains a much lower fraction of such objects. 
 We can rule out an inaccurate completeness correction as the source of the
 difference, because both our study (see Figure \ref{figure:LF_60mu}) and that of
 \citet[][]{xu} correctly reproduce the 60$\mu$m LF of the parent sample\footnote{
  The completeness for the \cite{xu} sample has been studied by \cite{iglesias}.
 }.
 We argue instead that the difference arises from the construction of the LF:
 \cite{xu} built their $K$-band LF using the conditional probability to
 find a FIR-selected galaxy of a given $K$ magnitude within the combined
 \mbox{UV + 60$\mu$m} luminosity bin (see their Section 3.3 and their Equation 1
 and 2), while we directly calculate the $K$-band number counts.
 Alternatively, we also consider that SFRS may be biased against metal-poor
 dwarf galaxies, which present low IR emission but strong UV luminosities.

 \medskip 
 \noindent
 \emph{Mass Function} ---
 As a first comparison we consider the mass function by \cite{cole},
 which is based on a NIR-selected sample (2MASS).
 Secondly, we consider the work by \cite{bell:2003}, whose prescriptions
 we used to estimate the $M/L$ ratios and ultimately the SFRS stellar masses
 (Equation \ref{ML}).
 Their study regards a magnitude-limited, optically-selected sample of SDSS
 galaxies with available 2MASS counterparts.
 Finally --- as for the LF --- we selected the GAMA survey as a benchmark to evaluate
 the selection bias in our IR-selected sample with respect to the general galaxy
 population.
 \cite{baldry:GAMA_MF} presented a GAMA MF restricted to a redshift
 range (\mbox{$z < 0.06$}) very close to that of SFRS (\mbox{$z \lesssim$ 0.05}).
 \cite{kelvin} studied the GAMA MF on a slightly narrower redshift range
 (\mbox{0.025 $< z <$ 0.06}), but offered a decomposition of their MF into
 disk- and bulge-dominated systems.

 Our total mass function exactly traces the data points of \cite{cole} throughout
 the extent of their mass sampling.
 Our MF is also largely consistent with the curve of \cite{bell:2003} at the low mass
 range (\mbox{log($h^2$~M/M$_{\odot}$) $\sim$ 7.5--10.5}).
 Similarly to the LF, the low-end ``upturn'' displayed by the GAMA mass functions
 is not observed in the SFRS MF.
 Additionally, the comparison against both the GAMA and the \cite{bell:2003} curves
 highlights a lack of objects at masses larger than M$^{*}$ (this feature is less
 evident in the LF). 
 These are likely to be passive galaxies, which are missing from the SFRS.
  
 \medskip 
 \noindent
 \emph{Lack of upturn at the low-end} --- 
 As mentioned above, one of the most noticeable features of both our
 LF and MF is that they fall below (but consistent within the uncertainties) the
 studies based on visible-light selection \citep{baldry}.
 Most probably this happens because low luminosity/mass galaxies lack the
 necessary dust amount to efficiently reprocess the radiation from young stellar
 populations into IR emission \citep[e.g.,][]{dale,calzetti:2010}.
 In fact, dwarf galaxies tend to have lower metallicities and hence higher
 gas-to-dust ratios than more massive systems.
 It follows that these objects are under-represented in our sample, which is
 primarily selected based on 60$\mu$m luminosity
 ($\S$\ref{The Star Formation Reference Survey (SFRS)}).
 Interestingly, the upturn is undetected in $K$-band studies as well
 (see the UKIDSS LF or the \citealt{bell:2003} MF).
 \cite{baldry} argued that those surveys do not properly probe down to 
 their declared surface brightness limit: a similar issue might in principle
 affect our data.
 However, \cite{driver:GAMA} argued that the upturn in their GAMA LF is due to
 the combination of the LF of elliptical galaxies --- forming a plateau below
 $M^{*}$ --- plus the tail of LF of the star-forming sub-population
 (see their Figure 13).
 This is consistent with the lack of upturn in our functions, because in the SFRS
 sample ETGs are rare by construction.

 \medskip 
 \noindent
 \emph{Behaviour at the high-end} ---
 The SFRS MF, and to a lesser degree the SFRS LF, display a clear deficiency of high
 mass/luminosity objects with respect to the general galaxy population of GAMA.
 The morphologically-separated curves by \cite{kelvin} show that the lack of objects
 above M$^{*}$ is likely the result of the different sample selection: the
 SFRS consists of star-forming galaxies, and
 therefore it is biased against passive bulge-dominated objects which populate
 the high-mass end.
 In the last 1--2 data points corresponding to the most luminous (massive) LF
 (MF) bins, 
 our data show a deficiency with respect to the general trend, although
 such difference is not significant given the large uncertainties.
 
 \cite{driver:GAMA} observed a similar feature in their data and argued that this is
 related to a sub-population of high-\Sersic{} index ETGs.
 The average \Sersic{} index we obtained by the single \Sersic{} fit
 ($\S$\ref{Fit procedure}) to the 10 most luminous SFRS galaxies
 is only $\sim$4.5, far from being in the highest \Sersic{} index tier of our
 sample (see Figure \ref{figure:Sersic_n}).
 Moreover, as shown by Figure \ref{figure:MF_comparison}, disks can significantly
 contribute to the high-end of the SFRS MF, hence we exclude any involvement of
 high-\Sersic{} index ETGs in our case.
 
 \medskip
 \noindent
 \emph{Morphological type and colour} ---
 Remarkably, the LF of late-type galaxies of \citet[][Figure \ref{figure:LF_K}]{kochanek}
 does not completely follow the SFRS LF throughout the entire magnitude range:
 there is agreement only at the low-end.
 Similarly, neither the bulge- nor the disk-dominated curve of
 \citet[][Figure \ref{figure:MF}]{kelvin} agrees with the SFRS MF over the
 entire mass range.
 This shall not be surprising since star-forming galaxies are not expected to be
 exclusively disk-dominated (or budge-dominated, for what matters).
 Instead, they contain a varying mixture of both bulge and disk components.
 In other words, star-forming galaxies cannot be sharply identified by their
 morphological appearance; on the contrary, they populate all mass
 ranges and all Hubble types. 
 This is consistent with the trends in the \cite{kelvin} MFs, where the
 bulge-dominated MF agrees well with the high-end of the SFRS MF, while the
 disk-dominated MF of \cite{kelvin} agrees with the low-mass end of our MF.

\subsection[Mass repartition in bulges and disks]{Mass repartition in bulges and disks}
\label{Mass repartition in bulges and disks}
 
 \noindent
 The discussion in $\S$\ref{Comparisons against other samples} lead to the conclusion
 that an arbitrary morphological separation into late- and early-types or disk- and
 bulge-dominated galaxies does not uniquely characterize the population of
 star-forming galaxies: star formation takes place in a wide range of 
 galaxies' morphological components \citep[e.g.,][]{kennicutt:optically_thick_limit}.
 
 In this context, it is more meaningful to study the distribution of mass into
 disks and bulges.
 Our mass-density functions (bottom panel of Figure \ref{figure:MF_comparison}) show
 that for the range 10$^{8}$--10$^{10}$~M$_{\odot}$ the contribution
 of disks and bulges to the mass density in the Local Universe is closely comparable.
 Moreover, both distributions have their maximum close to the global MF maximum
 (M$^{*}$; i.e., \mbox{log($h^{2}$ M/M$_{\odot}$) $\sim$ 10.41}, see Table 
 \ref{table:Schecter_fit}), where they also display similar peak values (within 
 a factor of $\sim$3 of each other).
 By broadly associating bulge mass with an old stellar population and disk mass
 with recently formed stars, the similarity of the peak locations and amplitudes
 implies that the \emph{average} 
 star-forming galaxy assembled as much mass in the distant past as it did in its
 latest evolution.
 The galaxy mass functions are a key metric for testing and calibrating cosmological
 galaxy evolution models \citep[e.g.,][]{crain:2015}.
 The evolution of galaxy morphology (e.g.\ fraction of disk and bulge dominated
 galaxies) is another such metric \citep[e.g.,][]{perry:2009}.
 The presented bulge and disk mass functions provide yet an other metric
 for testing and calibrating galaxy evolution simulations.  
 
 A more detailed inspection of the mass-density functions reveals that bulges alone
 populate the bottom range of the distribution
 (\mbox{log($h^{2}$ M/M$_{\odot}$) $\lesssim$ 7.5}),
 while disks largely monopolize the high-mass end, indicating that
 star-forming galaxies host most of their mass in their disks.
 Adopting the previous assumption that disk (bulge) mass is associated with recent (past)
 stellar formation, the ratio of the disk to bulge density functions can serve as a
 proxy of the ``timescale'' of star formation as a function of mass.
 For example, following on the consideration that in massive objects the
 disk density largely dominates, the high-mass end is where most recent star
 formation happened.
 Starting from their  morphologically decomposed MFs and assuming an average $B/T$,
 \cite{kelvin} estimated that 50\% of the stellar mass in the Local Universe resides
 in bulge structures.
 With our bulge/disk decomposition we show that this conclusion does not hold,
 at least for star-forming galaxies.
 Instead, our integrated stellar densities (Tables \ref{table:MF_disks} and
 \ref{table:MF_bulges}) indicate a ratio 4:1 in favour of disks.

 The fact that star formation activity happens where most of the mass is
 concentrated (i.e., in disks) further confirms that mass is the primary
 parameter characterizing galaxy evolution.
 This is in agreement with the galaxy main sequence \citep[e.g.,][]{elbaz} as
 well as with the sub-galactic main sequence \citep[e.g.,][]{maragkoudakis:main_sequence},
 where we find that more massive systems (either galaxies or individual star-forming
 regions) also host higher SFR.
 To further explore this, in a forthcoming SFRS paper we will additionally
 present the specific SFR (sSFR) and the bivariate (mass \emph{vs.} sSFR)
 volume-weighted functions, the latter being the most accurate representation
 of the star-forming main sequence.

\section[Summary and Conclusions]{Summary and Conclusions}
\label{Summary and Conclusions}

\noindent
We obtained archival $K_{s}$-band 2MASS data ($\S$\ref{K-band data}) for the galaxies
composing the sample of the Star Formation Reference Survey (SFRS), with the ultimate
intent of studying the stellar mass function in star-forming galaxies, and the
distribution of stellar mass in their disk and bulge sub-components.

We designed a sophisticated bulge/disk decomposition routine based on the
software \GALFIT{} for the 2D fit of the surface brightness of galaxies
($\S$\ref{The 2D fit of SFRS galaxies}).
The \emph{best-fit} model was automatically selected primarily based on the
``excess variance'' \citep{vaughan}, providing a better statistic than the
commonly adopted $F$-test.
The parametric components of the selected model were attributed a
physical meaning (i.e., bulge, disk, or AGN) through a decisional tree which
accounts for unresolved bulges ($\S$\ref{Separation of disk/bulge components}).
The galaxy luminosities were converted to stellar masses using the \cite{bell:2003}
prescriptions and using archival SDSS-DR12 ($u - r$) colors as proxies for the global
stellar population.
For the disk/bulge sub-components we instead determined the separate colors by
applying our decomposition to the SDSS $u$ and $r$ images
($\S$\ref{Disk/bulge optical colors}), producing the bivariate color distribution,
which provides the joint probability of a galaxy to host a disk \emph{and} a bulge with
a given color combination (Figure \ref{figure:colors}).
We built the SFRS luminosity function (Figure \ref{figure:LF_K}) using the $1/V_{MAX}$
method, after correcting the source density by selection biases
($\S$\ref{Correction for sky coverage }), and
correcting the volume by the effects of lost survey area ($\S$\ref{Calculation of V_MAX}).
Similarly, we produced the mass function for the SFRS galaxies (Figure \ref{figure:MF})
and for their disk- and bulge- sub-components (Figure \ref{figure:MF_disks_bulges}),
which we also represented in terms of mass-density function
(Figures \ref{figure:MF_disks_bulges} and \ref{figure:MF_comparison}) to highlight
contribution of each component to the total mass density.

\medskip
\noindent
Our main conclusions can be summarised as follows.

\begin{cutenumerate}[label=---~]
 \itemsep0.5em  
 
  \item The SRFS $K_{s}$-band LF and MF agree with previous studies based
        on similarly selected samples.
        In particular, the LF resembles that of $K$-band selected samples
        and is broadly compatible with that of FIR-selected studies.
        Likewise, the SFRS MF traces remarkably well that of 2MASS- and
        IR-selected samples ($\S$\ref{Comparisons against other samples}).

  \item The star-forming galaxy number density is lower than that of the general
        galaxy population --- as probed by the GAMA and $K$-band surveys --- at
        both ends of the LF and MF.
        We interpret these features in terms of the different sample selection
        ($\S$\ref{Comparisons against other samples}).
        At the faint/low-mass end, dwarf galaxies are under-represented in the SFRS
        due to the lack of dust content able to produce significant emission
        in the 60$\mu$m luminosity band, which is the primary selection
        band for the SFRS sample.
        At the bright/high-mass end, ``dead'' early-type systems do not
        contribute to the population of star-forming galaxies, and hence the
        LF (MF) is intrinsically lower.
 
 \item Star-forming galaxies cannot be identified with specific galaxy
       sub-samples identified via a morphological separation (e.g., late-type or
       disk-dominated systems) or via their color
       ($\S$\ref{Comparisons against other samples}).

 \item Despite the \emph{mean} star-forming galaxy roughly equally sharing its
       stellar mass between its disk and bulge components, 
       most of the mass of currently star-forming objects is stored in their
       disks rather than bulges (4:1; $\S$\ref{Mass repartition in bulges and disks}).
       This information, combined with the fact that disks dominate the high-mass
       end of the MF, has two implications: 1) that most recent star formation
       happened in massive systems, and 2) that mass is the primary parameter in
       the study of star forming objects.

\end{cutenumerate}

\noindent
The last result confirms the general scenario depicted by the main sequence of
star-forming galaxies \citep[e.g.,][]{elbaz}.
However, our mass-density functions now constitute an accurate benchmark for
models addressing the mass assembly in bulges/disks in that they quantify for the first
time the relative density of stellar mass stored in the sub-components of star-forming
objects (which --- as our MF shows --- compose the vast majority of local galaxies).
This can in turn be used to investigate the evolution of star formation
activity.

\section*{Acknowledgments}
\label{Acknowledgments}

We wish to thank T.~Bitsakis, and J.~Fritz
for the useful assistance along the development of the
analysis techniques.
PB and AZ acknowledge support by the EU IRG grant 224878.
Space Astrophysics at the University of Crete was supported by EU FP7-REGPOT grant
206469 (ASTROSPACE).
This project has received funding from the European Union's Horizon 2020 research
and  innovation programme under the Marie Sklodowska-Curie RISE action, grant
agreement No 691164 (ASTROSTAT).
The research leading to these results has received funding from the
European Research Council under the European Union's Seventh Framework
Programme (FP/2007-2013) / ERC Grant Agreement n. 617001.
This publication makes use of data products from the Two Micron All Sky Survey,
which is a joint project of the University of Massachusetts and the Infrared
Processing and Analysis Center/California Institute of Technology, funded by the
National Aeronautics and Space Administration and the National Science Foundation.
This research has made use of the VizieR catalogue access tool, CDS, Strasbourg,
France.

\section*{Data Availability}
\label{Data Availability}

\noindent
The data underlying this article are available in the article and in its
online supplementary material.
The photometric measurements from the 2MASS-PSC and -XSC catalogues were derived
via Vizier: \url{https://vizier.u-strasbg.fr/viz-bin/VizieR}
The 2MASS datasets were derived from sources in the public domain:
\url{https://irsa.ipac.caltech.edu/applications/2MASS/IM}.
The SDSS datasets were derived from sources in the public domain:
\url{http://skyserver.sdss.org/dr12/en/home.aspx}.




\appendix
\onecolumn

\section{Pre-fitting procedures}
\label{Pre-fitting procedures}

\subsection[Setting up GALFIT input]{Setting up GALFIT input}
\label{Setting up GALFIT input}

\noindent
Every $\chi^{2}$ minimization-based algorithm, such as \GALFIT{}, requires an
initial parameter set from which to start mapping the $\chi^{2}$ space determined
by the model parameters.
Appropriate choice of these initial values speeds-up the convergence to the the
\emph{best-fit} parameters and avoids local minima.
In order to determine initial photometric and morphological parameters for the target
galaxies (and any other source which is fitted simultaneously; see
$\S$ \ref{Fit procedure}), we used \textsc{SExtractor} v2.8.6
\citep[][]{sextractor}\footnote{
 \url{www.astromatic.net/software/sextractor}
}.
This package performs source detection and deblending, and provides basic photometry
for each source after estimating and subtracting the local background.
A summary of the quantities of interest for this preliminary analysis, and the 
relevant \textsc{SExtractor} \emph{keys}, are reported in
Table \ref{table:SExtractor parameters} and described in further detail in
Appendix \ref{SExtractor measurements}.

\medskip
\noindent
We tuned the \textsc{SExtractor} configuration file, and especially the detection,
analysis, and deblending thresholds, by extensively testing the software on the
2MASS images.
For the whole SFRS data set we chose a single \textsc{SExtractor} setup, which
represented the optimal compromise between the maximum number of detected sources
and the limit at which the sub-components of the SFRS targets (e.g., arms,
star-associations, etc.) start being deblended.
For just $\sim$10 problematic objects (contaminated by nearby stars or in
close interacting pairs), we had to customize the \textsc{SExtractor} parameters in
order to perform accurate photometry.
One additional required setup parameter, used as an initial value for the
\textsc{SExtractor} neural network which computes the stellarity index\footnote{
 An estimate of the difference between the radial profile of an object and that of a
 point-like source.
}, is an estimate
of the seeing (or, equivalently, the stellar FWHM).
We adopted the average FWHM of the stars within an 2MASS Atlas image and checked
it was stable for different observation dates.
In any case, this value was only used to perform the initial separation between
point-like and extended sources.
A deeper analysis of the PSF was anyway performed at the \GALFIT{} fitting stage
(see Appendix \ref{PSF creation}).

  \begin{center}
   \includegraphics[width=0.8\textwidth,angle=0]{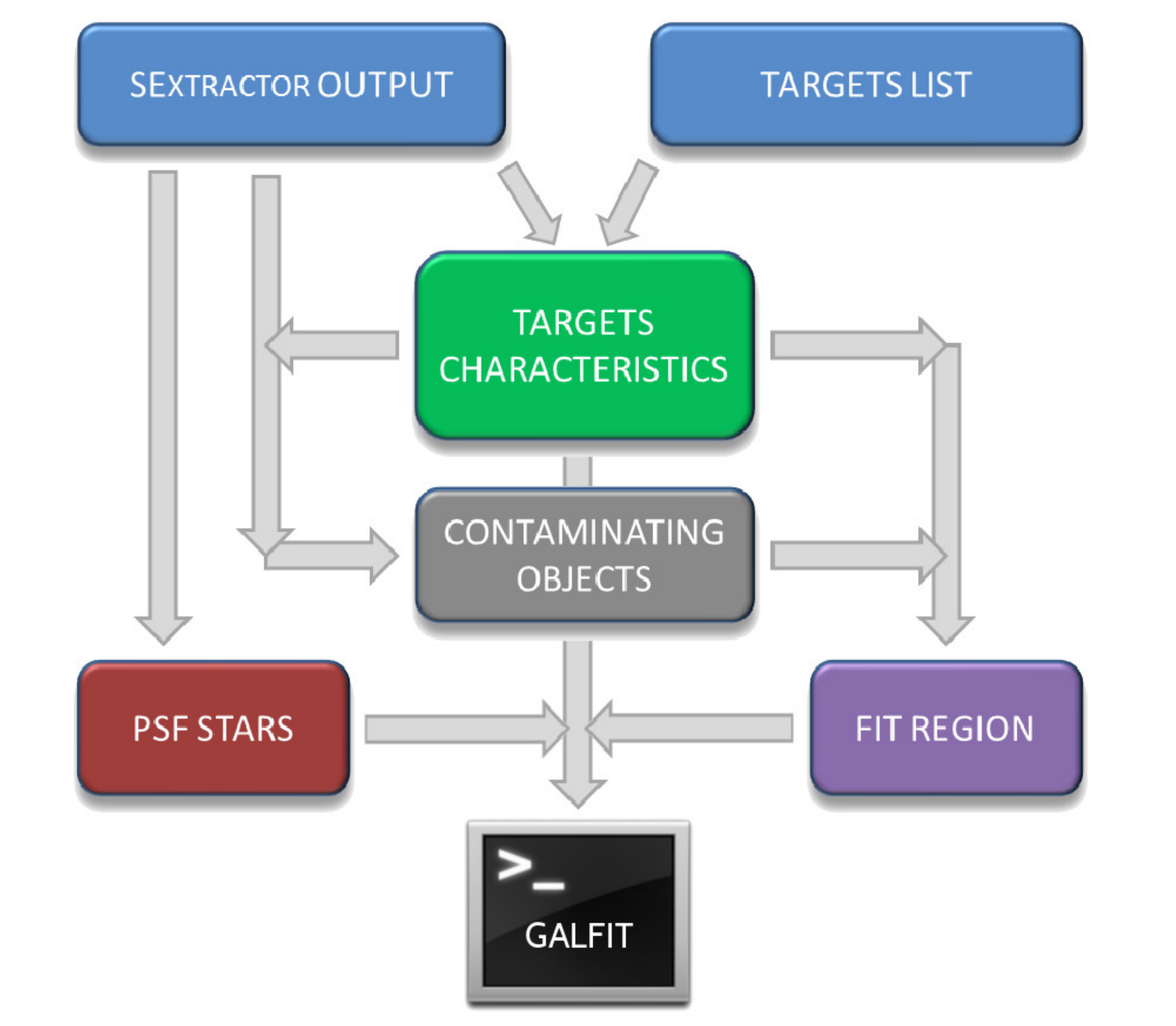}
   \captionof{figure}{
    Flowchart representing the preparation of the \GALFIT{} input.
    The \textsc{SExtractor} results were cross-matched against the input target list
    to associate each SFRS target to the corresponding Atlas image and to the
    relevant attributes produced by \textsc{SExtractor}.
    Contaminating objects (overlapping/blended) were identified based on their
    projected separation from the SFRS target.
    The best-quality stars were chosen as candidates to generate the model PSF
    required by the fit algorithm and to calculate the flux calibration.
    A fit region was cropped around the objects (or PSF stars) to be fit,
    based on their spatial extent, in order to speed up the fit calculations.
    \label{figure:GALFIT_input}
   }
  \end{center}

\medskip
\noindent
The\textsc{SExtractor} results were processed as illustrated in Figure
\ref{figure:GALFIT_input} in order to accomplish the following preparatory tasks.

\begin{cutenumerate}[label=$\blacktriangleright$~]
 \itemsep0.5em  

 \item The SFRS targets were associated to their corresponding image files through a
       coordinate cross-match.
       In fact, given a list of targets and a list of images, our pipeline finds
       the correct association by comparing the WCS of the target with the FOV of
       the image.
        
 \item All the objects not associated with a SFRS target (i.e., overlapping/blended
       objects) were marked to be fit together with the target in order to remove
       their contribution.
       A source was labelled as ``contaminating'' if it laid within a circle
       centered on the SFRS target and of radius:

       \vspace{-0.3cm}
       \begin{eqnarray}
         R  =  5 \times (R^{eff}_{obj} + R^{eff}_{target})~~,
         \label{r_contaminating} 
       \end{eqnarray}
       
       \noindent
       where R$^{eff}_{obj}$ and R$^{eff}_{target}$ are the effective radius of
       the object and the target, respectively.
       The multiplicative constant was chosen to account for the wings of the
       radial light distribution of the most extended sources.
       Point-like and extended contaminating sources were distinguished based
       on their \textsc{SExtractor} stellarity index (threshold set at
       CLASS\_STAR = 0.3), and they were
       fit with different models (see $\S$\ref{Fit procedure}).
     
 \item Some of the stars were marked as candidates to be used for the generation
       of the model PSF required by the fit algorithm and to calculate the
       photometric zero-point of the image (``PSF stars''; see
       $\S$\ref{Zero-points, luminosities, and stellar mass calculation}).
       We imposed several criteria for a star to be considered eligible:
       (\emph{a}) the star could not be a contaminating object;
       (\emph{b}) the star could not be close to the image borders in order to
                  have sufficient area to estimate the local background;
       (\emph{c}) stellarity index > 0.8;
       (\emph{d}) 0.7 $\times$ FWHM$_{PSF}$ $<$ FWHM$_{star}$ $<$ 1.5 $\times$
                  FWHM$_{PSF}$,
                  where FWHM$_{PSF}$ is the input estimate for the PSF
                  FWHM (determined as discussed in \ref{PSF creation});
       (\emph{e}) axis ratio $b/a$ $>$  0.8;
       (\emph{f}) the star had to be isolated, i.e., it could not be blended
                  with other objects according to \textsc{SExtractor}.

 \item A ``fit region'' was cropped around the group of objects to be fit in order
       to speed up the fitting calculations.
       The size of this area (based on the effective radii of the objects) was
       designed to be as small as possible, while still guaranteeing complete
       coverage of the extended wings of the sources light profiles and a sufficient 
       sampling of the background.
       Similarly, a fit region was cropped around each PSF star.
       Figure \ref{figure:masks} (first panel on the left) shows an example of the
       source effective radii and fit regions for the SFRS targets NGC~4435
       and NGC~4438.

 \end{cutenumerate}

\medskip
\noindent
Although we heavily relied on \textsc{SExtractor} for the initial photometry
of the sources, we preferred not to use its background maps, which
are significantly contaminated by the light of actual sources\footnote{
 See e.g.,: \url{http://users.obs.carnegiescience.edu/peng/work/galfit/TOP10.html}
}.
This implies a systematic bias in the estimation of the background noise, which
inevitably leads to erroneous weighting of the data values
(see $\S$\ref{Fit procedure}) and ultimately biases or yields additional
uncertainties on the fit parameters.
Therefore, in order to achieve a more robust estimate of the background and
its noise, we developed a technique which heavily masks all the sources before evaluating
the sky statistics.
To do so, we took advantage of the \textsc{SExtractor} ``segmentation'' maps:
these are FITS images in which every pixel deemed part of a source is masked out,
essentially creating a map of the pixels belonging to detected objects.
We created a \textsc{Perl} script\footnote{
 Publicly accessible from:
 \url{https://paolobonfini.wordpress.com/2016/05/04/mask-borders-of-a-fits-image}
}
which expands isometrically the border of each object in the segmentation image,
leaving ``unmasked'' only the pixels associated with a ``secure'' background.
In this analysis, we roughly tripled the number of masked pixels with respect to the
\textsc{SExtractor} segmentation maps.
By applying this conservative masking on the data image, we measured the fiducial
background mean and root-mean-square (RMS).
The background level so measured was held fixed along the fitting process in order to
avoid known degeneracies with model parameters such as, e.g., the \Sersic{} index and
effective radius \citep[e.g.,][]{graham:1996}.

\medskip
\noindent
A final remark regards the setup of the saturation level (used by \textsc{Sextractor}
to reject bad pixels) and the camera gain (used by both \textsc{Sextractor} and
\GALFIT{} for noise statistics) for each image.
In fact, those quantities are exposure-dependent, as they depend on the actual 2MASS
telescope that performed the observation (either Mt. Hopkins, Arizona or
Cerro Tololo, Chile).
The 2MASS cameras were modified during the course of the observing
operations, which started in 1997 and were completed in 2001, and the SFRS
sources have been observed with either cameras.
We obtained the proper gain value using the time evolution of the 2MASS $K_{s}$-band gains,
available online\footnote{
 Available at: \url{www.ipac.caltech.edu/2mass/releases/allsky/doc/sec6_8a.html}
}.
As for the saturation levels, we used the median values of the 2MASS
Saturation Threshold Maps\footnote{
 Available at: \url{www.ipac.caltech.edu/2mass/releases/allsky/doc/sec4_2.html}
}.

\subsection{SExtractor measurements}
\label{SExtractor measurements}

\noindent
In order to automate the choice of initial photometric and morphological parameters
for the target galaxies (and any other source which is fitted simultaneously), we
used \textsc{SExtractor} v2.8.6 \citep[][]{sextractor}.
The \textsc{SExtractor} \emph{keys}\footnote{
 For a detailed description of the \textsc{SExtractor} \emph{keys}, please refer to the
 \href{http://www.astromatic.net/software/sextractor}{\textsc{SExtractor} manual},
 or to the ``Source Extractor for Dummies'' \citep{sextractor_for_dummies}
} corresponding to each measurement are listed
in Table \ref{table:SExtractor parameters}.

\medskip
\noindent
\emph{Source equatorial coordinates --- Used to automatically identify which, among
      the input images, contains a given target}.
      \textsc{SExtractor} uses the header WCS information to provide
      the equatorial coordinates of each object: these are the coordinates used for
      the ``blind search'' (Appendix \ref{Setting up GALFIT input}).

\medskip
\noindent
\emph{Pixel position of source within the image --- Used to locate a given
     target within a 2MASS Atlas image and define a fit area around the target
     (Appendix \ref{Setting up GALFIT input})}.
     \textsc{SExtractor} determines the pixel position of a source by
     evaluating its centroid within the isophotal boundary calculated
     according to the limiting $S/N$ desired by the user.
     It also offers an equivalent measurement performed within a circular
     Gaussian ``window'' (whose FWHM is equivalent to the half-light radius)
     intended to avoid the limitations which affect
     the isophotal boundaries (i.e., changes in the detection threshold,
     and azimuthal irregularities).
     We opted to use the ``windowed'' centroids, as suggested by
     \cite{sextractor}, due to their higher reliability.

 \begin{wraptable}{r}{5.5cm} 
 \centering
  \begin{tabular}{lc}
   \hline
   \multicolumn{2}{c}{\textsc{SExtractor: Keys of Interest}} \\
   \hline
   \hline
   \addlinespace 
   \multicolumn{1}{c}{Quantity}   & \textsc{SExtractor} Key \\
   \hline
   \addlinespace 
   R.A.                &  ALPHA\_J2000 \\
   Dec                 &  DELTA\_J2000 \\
   X pixel position    &  XWIN\_IMAGE  \\
   Y pixel position    &  YWIN\_IMAGE  \\
   Flux                &  FLUX\_ISOCOR \\
   Source size         &  FLUX\_RADIUS \\
   P.A.                &  THETA        \\
   Axis ratio          &  ELONGATION   \\
   Object type         &  CLASS\_STAR  \\
   Data quality        &  FLAGS        \\
   \hline
  \end{tabular}
  \caption[List of \textsc{SExtractor} outputs used to derive first-guess fit parameters]{
   List of \textsc{SExtractor} outputs used to derive first-guess fit parameters.
   \label{table:SExtractor parameters}
  }
 \vspace{-0.5cm}
\end{wraptable} 

\medskip
\noindent
\emph{Source Flux --- Used to calculate fist-guess object magnitude
     (Appendix \ref{Setting up GALFIT input})}.
     \textsc{SExtractor} offers flux measurements for a variety of apertures.
     We used for the isophotal-corrected aperture which
     is a circularization of the isophotal boundary (which is  discontinuous
     because its profile follows a pattern defined by the pixel borders) based
     on the assumption of a symmetric Gaussian profile for the object.
     This aperture is intended to recover the flux lost in the wings of
     the object profile. Although an elliptical aperture would better estimate
     galaxian fluxes, the isophotal-corrected aperture applies particularly
     well to stars, which compose the vast majority of objects we had to fit
     with our pipeline.
     In any case, the flux measured in this way for the objects of interest is
     only the initial parameter in the \GALFIT{} analysis, which will
     provide the ultimate photometry.
 
\medskip
\noindent
\emph{Source size --- Used to define the size of fit area around the target
     (Appendix \ref{Setting up GALFIT input})}.
     We set the PHOT\_FLUXFRAC \emph{key} to 0.5 in order to
     obtain the effective radius R$_{e}$ of the target.
 
\medskip
\noindent
\emph{Position Angle (P.A.) and axis ratio --- Used as initial orientation
        and elongation of the fit model}.
        
\medskip
\noindent
\emph{Object type --- Used to separate stars from extended objects in order
     to decide the relevant fit model (Appendix \ref{Setting up GALFIT input})
     and to produce the model PSF (Appendix \ref{PSF creation})}. 
     To each cluster of pixels above the desired ``analysis threshold'',
     \textsc{SExtractor} attributes  a ``stellarity-index'', whose value varies
     between 1 (secure point-like source) and 0 (secure extended source).

\medskip
\noindent
\emph{Data quality --- Used to select the best candidate stars to be combined
     in the production of the model PSF (Appendix \ref{PSF creation})}.
     \textsc{SExtractor} flags any object affected by issues such as blending or
     data corruption (e.g., saturated pixels within the aperture, memory
     overflow, etc.).
     Stars presenting any of these flaws were excluded from the pipeline creating
     the model PSF.
	
  \begin{center}
  \vspace{0.5cm}
   \makebox[\linewidth]{
    \fbox{\includegraphics[width=0.23\textwidth,angle=0]{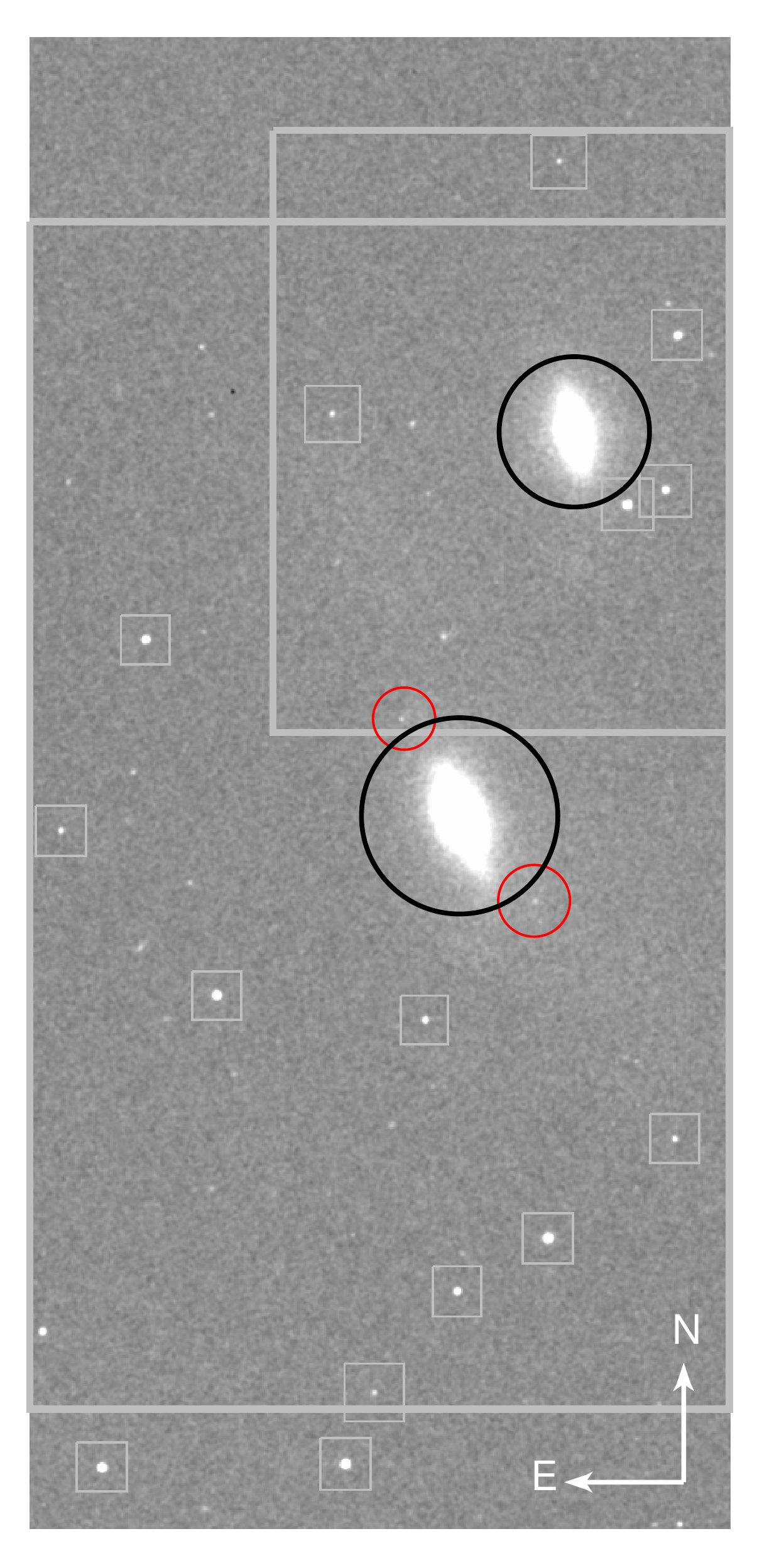}}
    \fbox{\includegraphics[width=0.23\textwidth,angle=0]{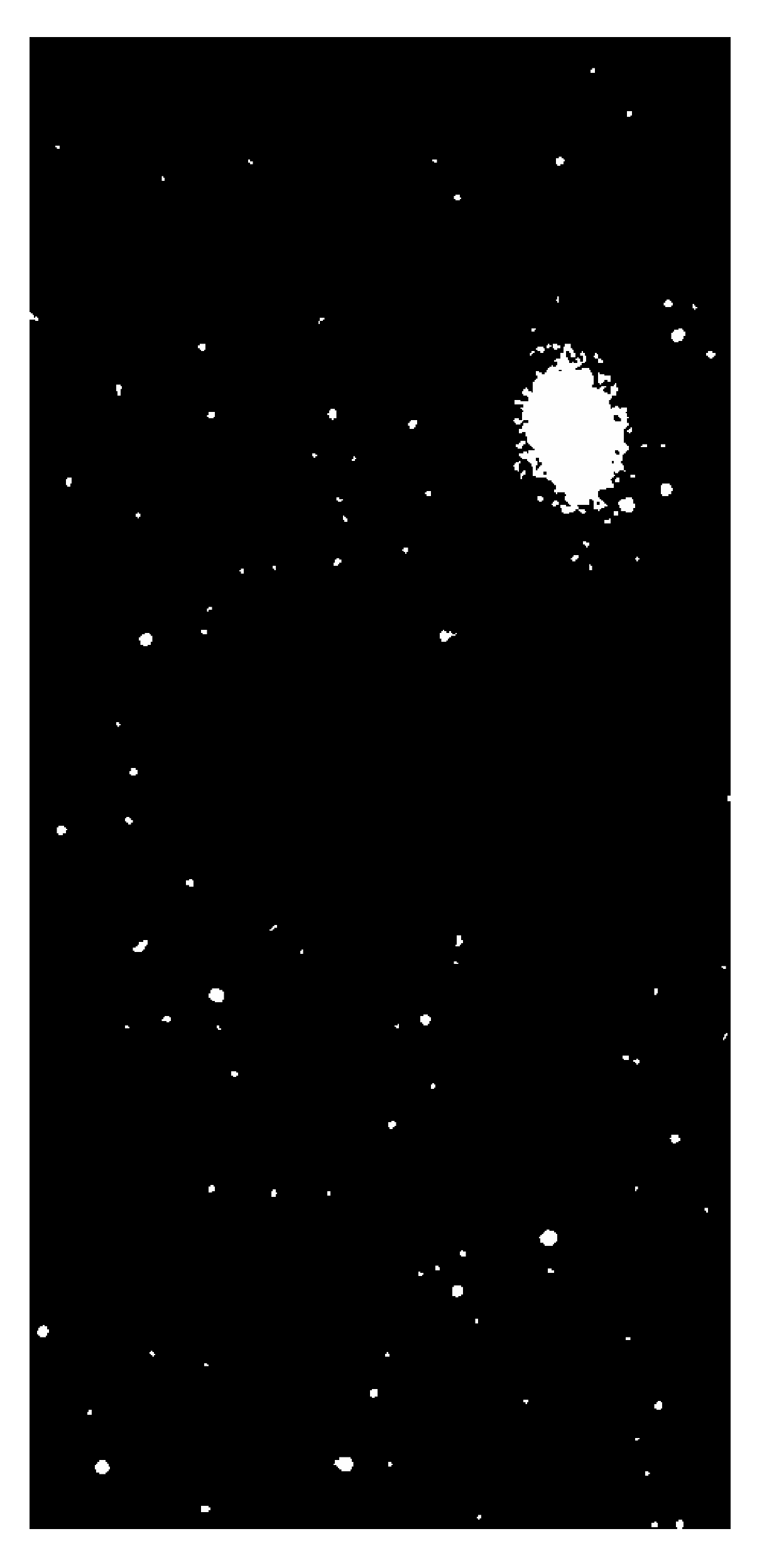}}
    \fbox{\includegraphics[width=0.23\textwidth,angle=0]{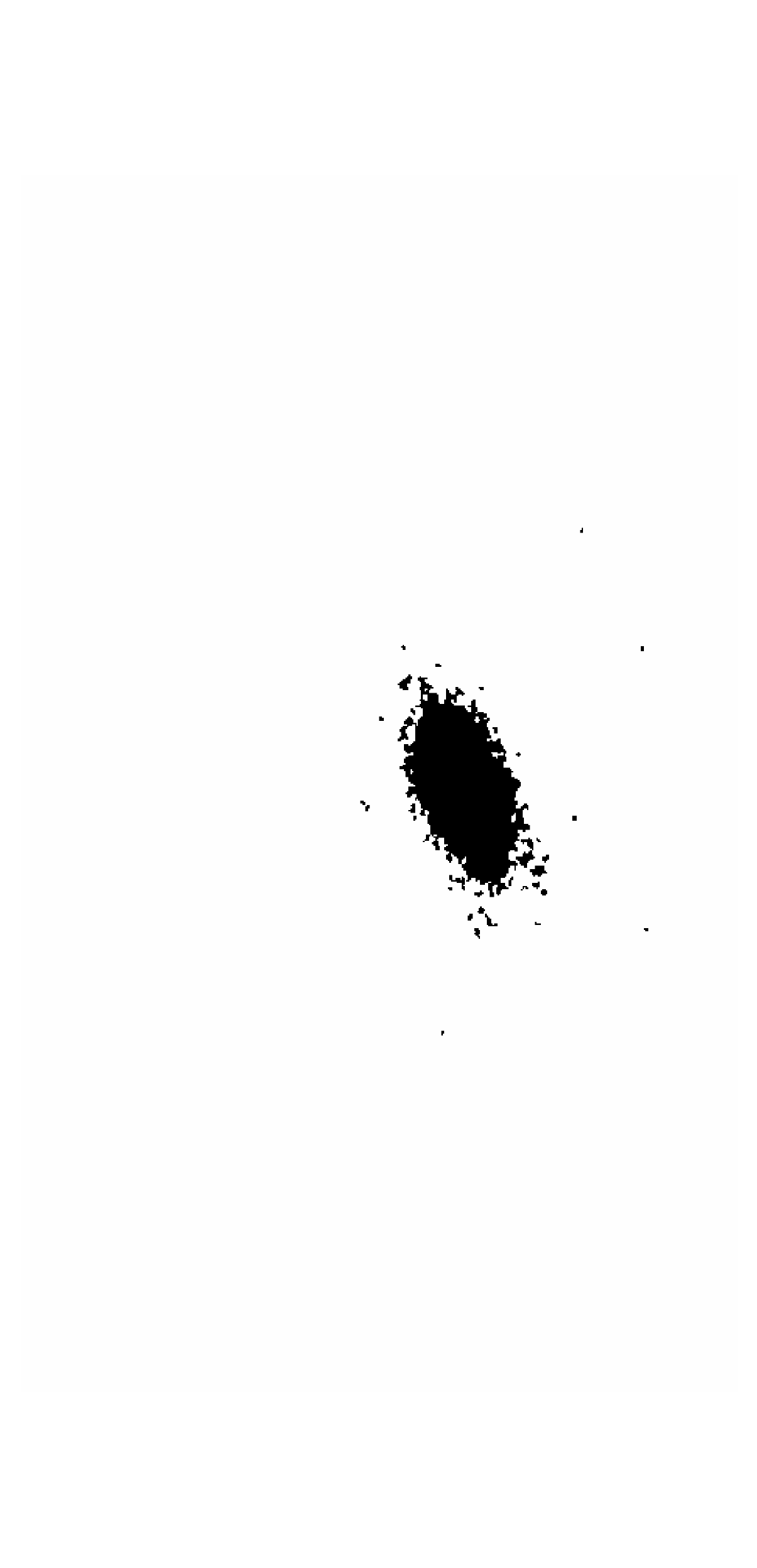}}
    \fbox{\includegraphics[width=0.23\textwidth,angle=0]{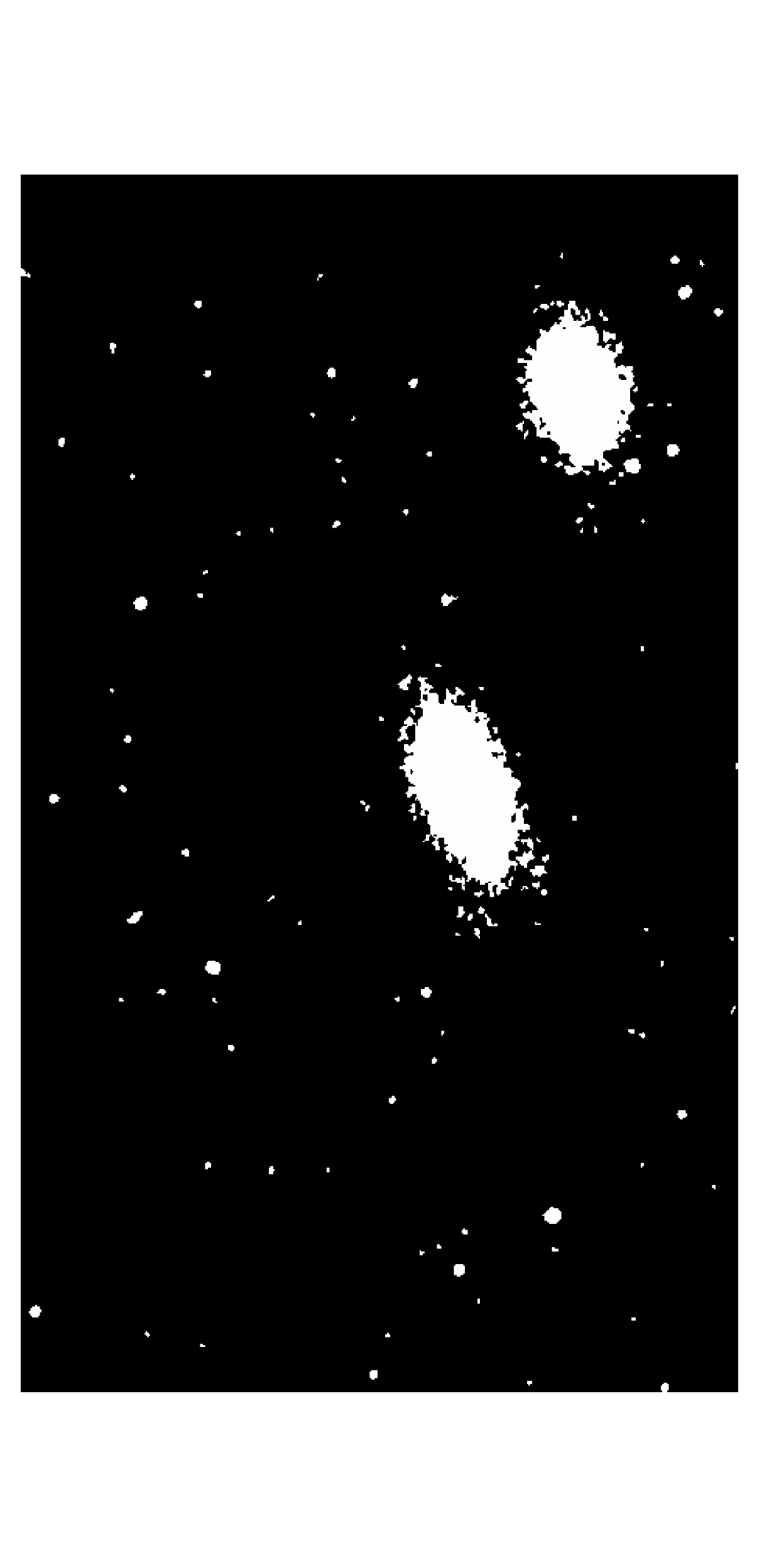}}
   }
   \captionof{figure}{
    Results of the masking procedure for the SFRS target NGC~4438.
    \emph{Left ---}
    On a 2MASS $K$-band Atlas image we overplot the effective radii of the
    detected sources (black circles) and the fit regions (squares) as determined
    by the procedure described in \ref{Setting up GALFIT input}.
    This specific Atlas image contains two SFRS targets: NGC~4438 (center) and
    NGC~4435 (N-W), which are identified and processed simultaneously by the
    pipeline.
    Circles are centered on the source coordinates and have a radius equal to 5
    times the sources' effective radii as measured by \textsc{SExtractor}.
    Black circles mark the target of interest, while red circles mark contaminating
    objects.
    The fit of each target was performed only within the wide grey box
    enclosing the object.
    The smaller boxes identify the PSF stars and their fit region.
    \emph{Center-left ---}
    Mask image for NGC~4438: all targets but NGC~4438 and its contaminating
    objects are masked.
    \emph{Center-right ---} and \emph{Right ---} panels represent respectively
    the ``negative mask''  and the ``negative background mask'' for NGC~4438, which
    we used to select the best-model (see $\S$\ref{Best-fit model selection}).
    The negative masks are smaller because they are defined only within the
    limits of the fitting region for a given object.
    In all masks, a white pixel defines a ``masked'' location, while a black
    pixel corresponds to a valid data point.
    \label{figure:masks}
   }
  \end{center}

\subsection[Mask creation]{Mask creation}
\label{Mask creation}

\GALFIT{} allows the usage of pixel masks in order to ``hide'' the objects whose flux
will not be accounted for in the fit: all the masked pixels are simply ignored.
We created mask images by exploiting the \textsc{SExtractor} segmentation maps,
in which clusters of connected pixels attributed to a specific detected object
are grouped in patches.
We modified these segmentation maps in order to mask the undesired sources
within the fit regions.

\medskip
\noindent
At this stage, we also produced two ``negative'' masks for the objects to be fit
and for the background to mark the areas over which we evaluated the fit
residuals used for the selection of the \emph{best-fit} model
(Section \ref{Best-fit model selection}).
An example of the masks used for the SFRS target NGC~4438 is shown in
Figure \ref{figure:masks}.

\subsection[PSF creation]{PSF creation}
\label{PSF creation}

\begin{wrapfigure}{r}{0.48\textwidth} 
 \vspace{-0.5cm}
 \includegraphics[align=c,width=0.48\textwidth,angle=0]{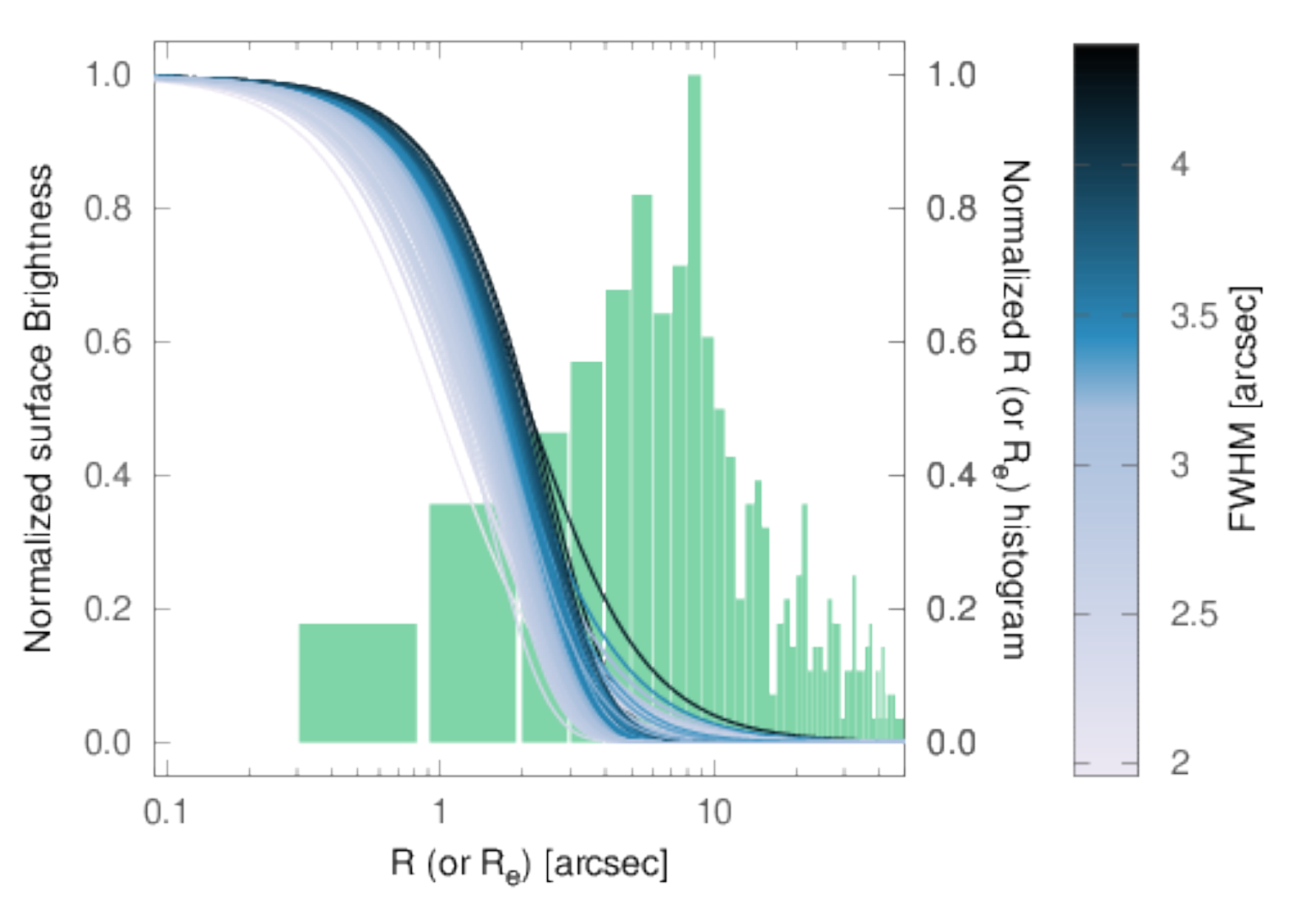}
 \caption[Comparison of the 2MASS seeing against the SFRS target sizes]{
  Comparison of the 2MASS seeing against the SFRS target sizes.
  The solid curves represent the Moffat profile fits to the
  surface brightness of 1\,000 PSF stars randomly selected among the 2MASS
  Atlas images.
  The curves are normalized to a unit peak intensity and are
  color-coded by FWHM, with darker curves showing more extended profiles.
  The green histogram shows the distribution of effective radii
  ($R_e$) of the  SFRS galaxies, obtained through the fit of a single
  \Sersic{} model (as described in $\S$\ref{The 2D fit of SFRS galaxies});
  the normalization is arbitrary.
  \label{figure:PSFs} 
 }
\end{wrapfigure}

\textsc{Galfit} adopts a PSF-fitting procedure which performs a convolution of the
target model with the image PSF before evaluating the statistics of each fit
iteration.
Therefore, an accurate reproduction of the PSF profile is the most critical (and
difficult) step in the fitting process. An ideal PSF is required to have high
$S/N$, account for geometric distortions (e.g., ellipticity), and of course be at
least Nyquist sampled.

\medskip
\noindent
In general, the 2MASS optical system guarantees a PSF FWHM $<$ 2\arcsec~across the
whole 512\arcsec~ field of view of the camera.
Gradients along the CCD are further minimized (averaged) by the scanning technique
\citep{beichman} and subsequent image
combination (Section \ref{The 2MASS data for the SFRS targets}).
Although the PSF along a 2MASS scan is fairly constant in standard observational
conditions \citep[][Figure 5]{jarrett}, it may vary
between exposures due to thermal and atmospheric condition differences.
For this reason, we had to define a PSF model for each observation independently.

\medskip
\noindent
Using as a model PSF a background-subtracted star from the native image is the
simplest approach, but it does not always satisfy the required characteristics.
Moreover, \GALFIT{} is extremely sensitive to the PSF shape.
PSFs derived from individual stars led to failed fits for the vast majority of
the objects.
On the other hand, a single \emph{theoretical} PSF profile (e.g., Moffat) used
in all data would not account for peculiar features such as
diffraction rings and spikes, nor be representative of the PSF across the field
or at different times.
For these reasons, we decided to pursue a hybrid approach by modelling the
best-quality stars in the field (``PSF stars''; see \ref{Setting up GALFIT input})
and then combining these model PSFs to produce the final PSF for each image.
First,  we used \GALFIT{} to fit all the PSF stars assuming a Moffat function
\citep{moffat}, widely used to describe star profiles, plus a background component.
Then, each background-subtracted model was re-projected onto a canvas
30 times the PSF FWHM, wide enough to include the PSF wings (while keeping the
computational time within a reasonable limit).
Smaller PSF canvas sizes resulted in evident truncations in the final target models.
The pixel scale of these PSF images was set to match that of the native
Atlas images in order to avoid the complications related to oversampling;
no additional resolution is needed as the 2MASS PSF is already Nyquist sampled,
its FWHM covering approximately 3.5~pixels.
Finally, the normalized PSF images associated to the models characterized by the best
\emph{$\chi^{2}$} were median combined \citep[using Swarp;][]{swarp} to create the
adopted PSF image.
This way we generated for each image a model PSF that is representative of the
PSF shapes present across the field.
We additionally compared our PSFs against the PSFs generated via
\textsc{Psf Extractor} \citep{psfex} as a combination of snap images of real stars
in the field.
Our PSFs --- being constructed from a combination of ``smooth''
Moffat models --- were marginally superior in that they were \emph{not} presenting
any artefact (due e.g., to bad pixels) which would ultimately reflect in the galaxy
model.

\medskip
\noindent
Figure \ref{figure:PSFs} shows the Moffat fits to the radial profiles of
a sample of 1\,000 randomly selected PSF stars.
In the same figure, the FWHM of the stars is compared with the distribution of
effective radii of the SFRS galaxies (as derived from the single \Sersic{} fits;
see $\S$\ref{Fit procedure}) to show that only a tiny minority of the targets
are partially resolved.
For a few fields, due to a lack of candidate PSF stars, the image PSF was
substituted by one adopted from to another image, chosen to be the closest
in time, with the obvious constraint that it should have been acquired by the
same 2MASS telescope and camera.

\newpage

\section[Zero-points, luminosities, and stellar mass calculation]{Zero-points, luminosities, and stellar mass calculation}
\label{Zero-points, luminosities, and stellar mass calculation}

\begin{wrapfigure}{r}{0.48\textwidth}
 \includegraphics[align=c,width=0.48\textwidth,angle=0]{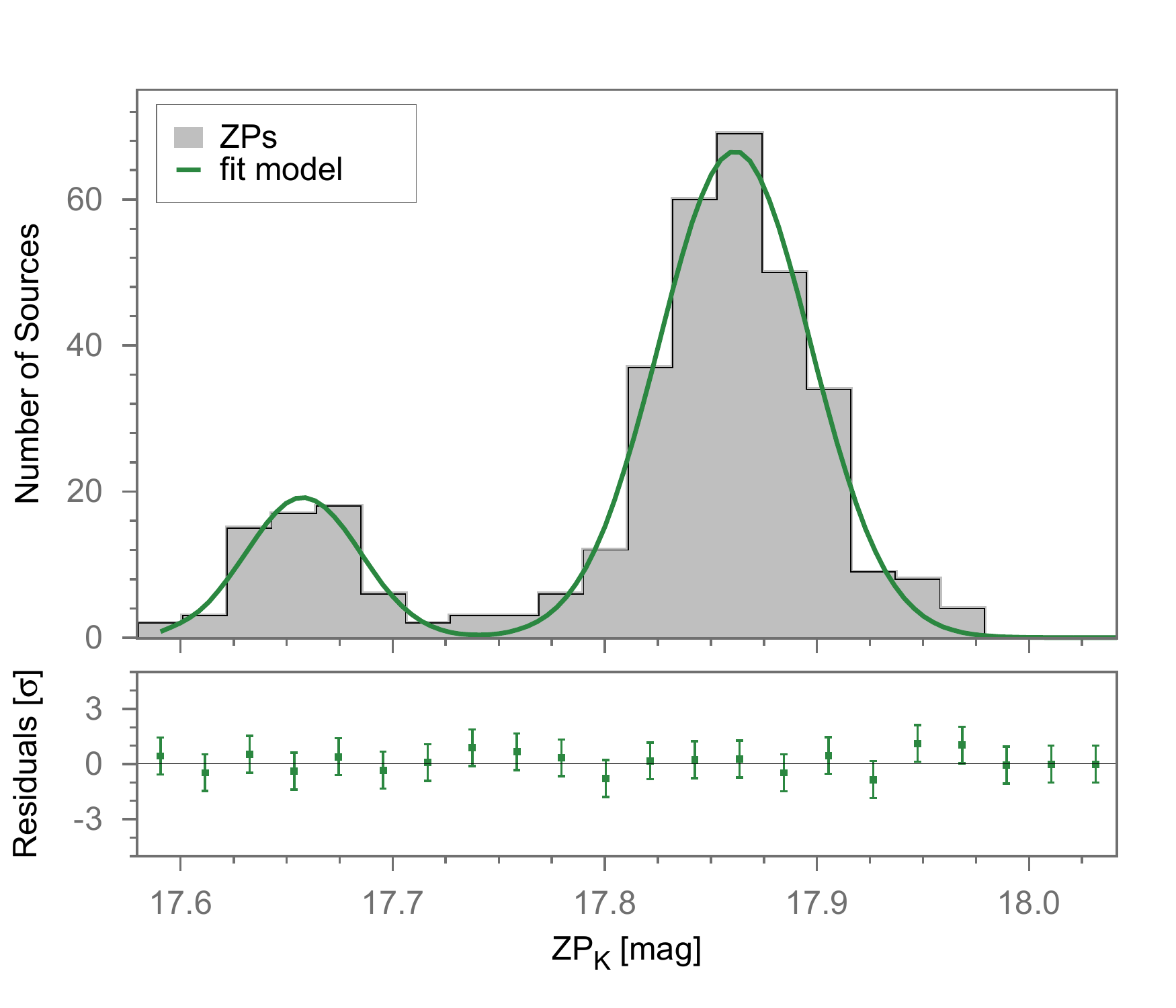}
 \caption[Distribution of the zero-points for the 2MASS Atlas images containing the SFRS targets]{
  Distribution of the zero-points (binned in 0.02~mag bins) for the 2MASS
  Atlas images containing the SFRS targets, over which we plot the results of a
  double Gaussian fit (green line).
  The bottom panel shows the fit residuals in units of standard deviation.
  \label{figure:K_zp_fit} 
 }
\end{wrapfigure} 

The zero-point was derived by comparing the magnitudes resulting from the fit of
the PSF stars in each image against the corresponding source flux densities
published in the 2MASS-PSC \citep{2MASS-PSC}.
We cross-correlated our PSF stars against the 2MASS-PSC sources encompassed by the
field of an Atlas image, using a search radius of 1\arcsec~(which is $\sim$2$\sigma$
the centroid accuracy of the 2MASS catalog), selecting the matches with the minimum
projected separation.
Zero-points were separately calculated for each PSF star, using the following relation:

\vspace{-0.3cm}
\begin{eqnarray}
 ZP^{\bigstar}        &=& mag^{2MASS} - mag^{GALFIT}                                 \\[0.5em]
 \delta ZP^{\bigstar} &=& \sqrt{(\delta mag^{2MASS})^2 + (\delta mag^{GALFIT})^2}~~, \nonumber
 \label{equation:ZP_star} 
\end{eqnarray}

\noindent
where $mag^{2MASS}$ is the star flux in magnitude units in the 2MASS-PSC,
$mag^{GALFIT}$ is the measured, uncalibrated source intensity for each source
(instrumental magnitude), $\delta{ZP}^{\bigstar}$ is the uncertainty on the
zero-point estimated for the star, and $\delta mag$ are the uncertainties
of the 2MASS and \GALFIT{} magnitudes.

The zero-point of a field ($ZP$) was simply estimated as the weighted average
of the zero-points $ZP^{\bigstar}$ of the stars within the image.
The calculation of the uncertainty of the $ZP$ (i.e., $\delta{ZP}$) required instead
a more sophisticated approach.
Although ideally one would calculate $\delta{ZP}$ as the standard
deviation of the $ZP^{\bigstar}$ of the stars in each field, the majority
of the images offered fewer than 5 stars suitable to calibrate the zero-point
(see selection criteria in Appendix \ref{Setting up GALFIT input}),
insufficient for this type of analysis.
In order to evaluate $\delta{ZP}$, we instead studied the distribution of the
zero-points of the whole 2MASS images sample (presented in Figure \ref{table:K_zp_fit}).
The distribution is bimodal, with the peaks corresponding to the zero-points
for the two 2MASS instruments (Mt. Hopkins, Arizona and Cerro Tololo/CTIO, Chile).
Assuming that the variation of the 2MASS zero-points around the mean values is due
only to photometric uncertainties, we fitted the distribution with two superimposed
Gaussian components using the \emph{Sherpa} package v4.2.1 (which is part of the
{\scriptsize CIAO} tool suite v4.2).
The uncertainties in the height of each bin were calculated using the \cite{gehrels}
approximation (Equation \ref{equation:Gehrels}).

The parameters of the \emph{best-fit} model (overplotted in Figure \ref{figure:K_zp_fit})
are reported in Table \ref{table:K_zp_fit}, where the error bars refer to the 1-$\sigma$
(68.3\%) confidence bounds from the $\Delta\chi^{2}$ projection for 1 interesting
parameter.
The FWHMs of the two components were used as a statistical estimate
of the uncertainty of the zero-points: in particular, we attributed a $\delta{ZP}$
equivalent to the FWHM of the model \mbox{``Gaussian 1''}, and \mbox{``Gaussian 2''}
(see Table \ref{table:K_zp_fit}) to all the images observed with the CTIO, and
Mt.~Hopkins instrument, respectively.
Similarly, for the images without a sufficient number of stars to confidently
estimate the $ZP$, we used the mean fit value of the gaussian corresponding to
the relevant instrument.

\begin{wraptable}{r}{9.5cm}
 \vspace*{0.5cm}
 \centering
 \begin{tabular*}{0.48\textwidth}{l @{\extracolsep{\fill}} ccc}
  \hline
  \multicolumn{4}{c}{\textsc{Fit Results for the Zero-point Distribution}} \\
  \hline
  \hline
  \addlinespace 
  \multicolumn{1}{c}{Component}  & Amplitude                      & Peak                           & FWHM                           \\
                                 &                                & [mag]                          & [mag]                          \\
  \hline
  \addlinespace 
  Gaussian 1 & 19.25$_{-4.09}^{+4.23}$ & 17.66$_{-0.01}^{+0.01}$ & 0.06$_{-0.01}^{+0.01}$ \\
  \addlinespace 
  Gaussian 2 & 66.61$_{-6.00}^{+6.12}$ & 17.86$_{-0.00}^{+0.00}$ & 0.08$_{-0.01}^{+0.01}$ \\
  \addlinespace 
  \hline
  \addlinespace 
  \multicolumn{4}{l}{$\chi_{\nu}^{2}$ = 0.38, 18 d.o.f. ($P_{(Q-value)}$ = 0.99\%)} \\
  \addlinespace 
  \hline
 \end{tabular*}
 \caption{
  Fit results for the double Gaussian fit of the zero-point distribution \mbox{(Figure \ref{figure:K_zp_fit})}.
  \label{table:K_zp_fit} 
 }
 \vspace*{0.3cm}
\end{wraptable}

Finally, the integrated magnitudes for each target, as well as the magnitudes for
each model sub-component were converted to fluxes (erg sec$^{-1}$ cm$^{-2}$) using the
flux-to-magnitude conversion:

\vspace{-0.3cm}
\begin{eqnarray}
 F_{K} = \nu_{K} \times (f_{0,K} \times 10^{-0.4 \cdotp M_{K}}) \times 10^{-23}~~,
\end{eqnarray}

\noindent
where $f_{0,K}$ is the zero-magnitude zero point conversion value (666.7~Jy)\footnote{
 Available at: \url{http://www.ipac.caltech.edu/2mass/releases/allsky/faq.html\#jansky}
}, $\nu_{K}$ (1.4 $\times$ 10$^{14}$~Hz) is the central wavelength of
the $K_{s}$-band filter, $M_{K}$ is the object (model sub-component) magnitude, and the
term 10$^{-23}$ is the conversion factor from Jy to  erg s$^{-1}$ cm$^{-2}$ Hz$^{-1}$.
The $K$-band K-correction was performed following the prescriptions by
\cite{K-correction}, where we used the 2MASS $J - K$ color
\citep[][their Table 9]{ashby} as a proxy for the galaxy spectrum.
We derived the luminosities of the SFRS objects by converting their fluxes
using the distances tabulated in \cite{ashby}.

Based on the mass-to-light ratio ($M/L$) for a given stellar population, one
can convert the $K_{s}$-band luminosity of a galaxy to its stellar mass.
However, this conversion factor is a function of the age of the population,
mainly through the non-linear dependence of the luminosity of individual stars
on their mass ($L$ $\propto$ M$^{3.5}$), and its metallicity.
For example, the $M/L$ changes by more than 10\% when the assumed stellar age
increases by 1~Gyr, and by more than 20\% when the metallicity [Fe/H] increases
by 0.5~dex.
These figures are based on the evolutionary models by
\cite{worthey}\footnote{
 Through the ``Worthey model interpolation engine'' applet
 \url{http://astro.wsu.edu/worthey/dial/dial_a_model.html}
},
where we assumed as central
values a solar metallicity and a light-weighted age of $\sim$2~Gyr
\citep[appropriate for our star-forming galaxies; e.g.,][]{peng:2015} and adopted
the default Salpeter initial mass function prescriptions.

As a first-order proxy of the age/metallicity of the stellar population, one can use
its colour \citep[e.g.,][]{bruzual:2003}.
In our analysis, we estimated the stellar mass of the galaxies (and of their
sub-components) using the prescriptions of \cite{bell:2003}, which provides
mass-to-light ratios as a function of colour for several 2MASS and SDSS passbands.
In particular, we used the formulation:

\vspace{-0.3cm}
\begin{eqnarray}
 {\rm M \over \rm M_{\odot}} = 10^{-0.273 + (0.091 \times (u-r))} \times {\rm L_{K} \over \rm L_{\odot}}~~,
 \label{ML}
\end{eqnarray}

\noindent
where $u$ and $r$ are the corresponding SDSS model magnitudes
(see $\S$\ref{Disk/bulge optical colors}), $L_{K}$ is the target
luminosity in $K_{s}$-band, and M$_{\odot}$ and $L_{\odot}$ are the solar mass and
luminosity, respectively (M$_{\odot}$ = 1.989 $\times$ 10$^{33}$~g,
$L_{\odot}$ = 4.97 $\times$ 10$^{32}$~erg s$^{-1}$).
The ($u$-$r$) SDSS colours for the SFRS galaxies (see Table \ref{table:SFRS_2MASS_SDSS})
are based on their integrated SDSS photometry, therefore they reflect the average
colour of the stellar content.
For the calculation of the separate masses of the disk/bulge sub-components we
adopted the disk/bulge colors we derived via 2D fitting of the SDSS-DR12 imaging, as
discussed in $\S$\ref{Disk/bulge optical colors}.

\section{Completeness test}
\label{Completeness test}

\noindent
We verify the accuracy of the completeness correction evaluated in
$\S$\ref{Correction for sky coverage } by applying it to the construction of
the 60$\mu$m LF for the SFRS sample, which we produced following the same methodology
as described in $\S$\ref{Luminosity and Mass Functions}.
Given our definition of completeness, the 60$\mu$m SFRS LF is effectively the LF
of the \PSCz{} galaxies, which we compared against the \PSCz{} LF derived by
\cite{takeuchi}.
The two functions are shown in Figure \ref{figure:LF_60mu} along with the
reference LF of 2\,818 IRAS selected galaxies by \cite{saunders:LF}.
The remarkable agreement between the curves confirms the reliability of our
completeness correction.

\begin{center}
  \includegraphics[align=c,width=0.85\textwidth,angle=0]{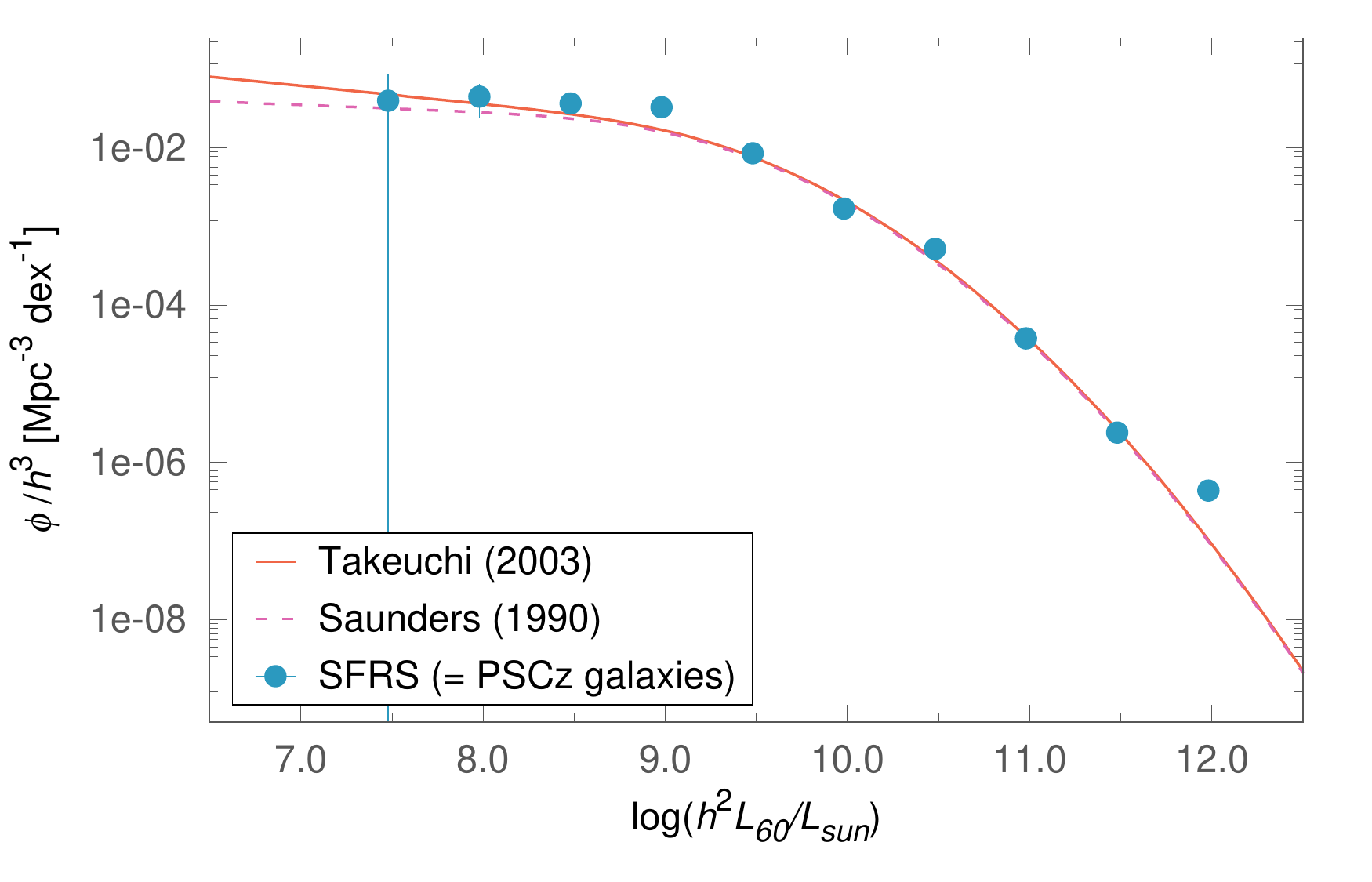}
  \captionof{figure}{
   Comparison with published 60$\mu$m \PSCz{} LFs.
   The 60$\mu$m LFs built from the SFRS sample by applying he completeness correction
   presented in $\S$\ref{Correction for sky coverage } effectively corresponds to the
   LF of the galaxies within the \PSCz{} catalogue (blue data points).
   The SFRS data shall be compared against the corresponding \PSCz{} LF by
   \citet[][solid curve]{takeuchi}.
   As a reference, the dashed curve shows the LF for the IRAS-selected galaxies
   of~\protect\cite{saunders:LF}.
   \label{figure:LF_60mu} 
  } 
\end{center}

\section{Mass Function uncertainties assessment}
\label{Mass Function uncertainties assessment}

\begin{figure*} 
 \begin{center}
  \includegraphics[align=c,width=0.85\textwidth,angle=0]{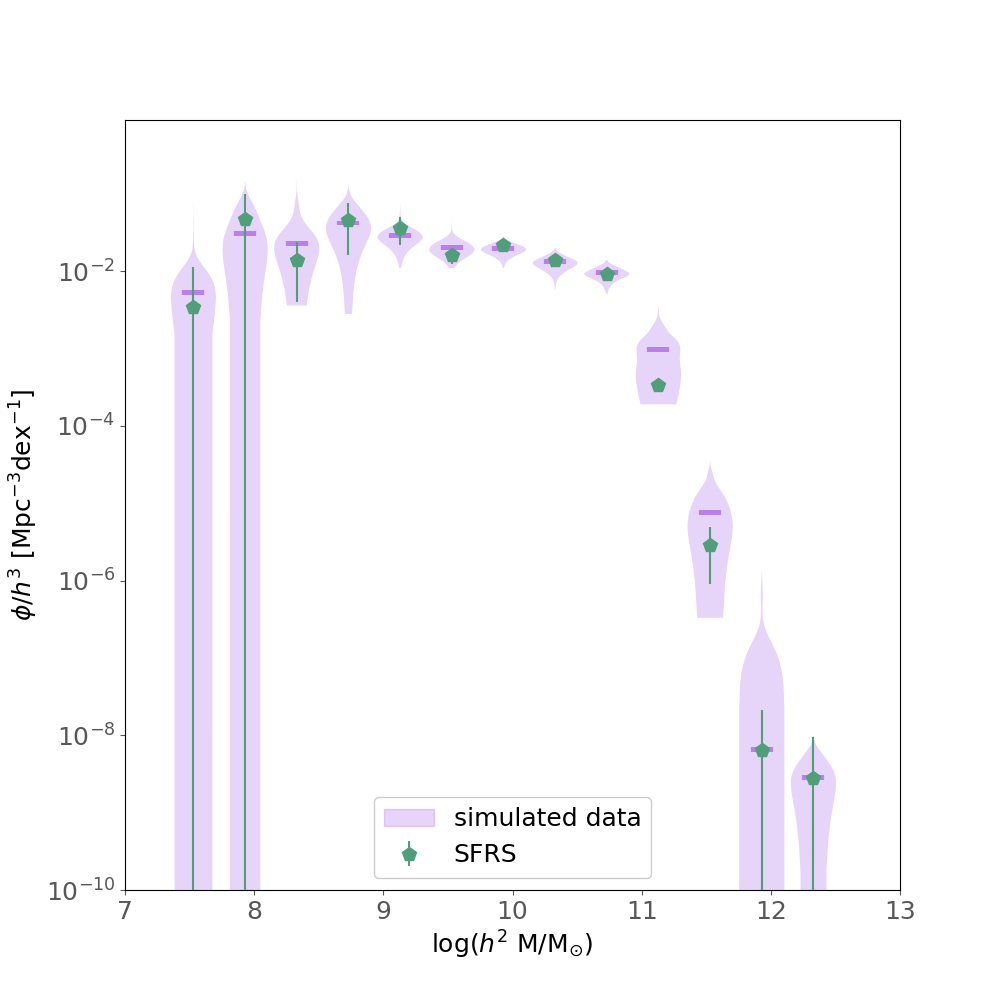}
  
   \begin{minipage}{0.8\textwidth}
    \caption{
     MF derived from the Monte Carlo simulation described in
     Appendix~\ref{Mass Function uncertainties assessment}.
     The green diamond points and error bars represent the calculated
     SFRS MF as shown in Figure~\ref{figure:MF}.
     The purple violins show the density distribution of the simulated values
     (estimated through a Gaussian kernel), while the thick horizontal bars
     indicate the median of the simulated values within each bin.
    \label{figure:simulated_MF} 
   } 
  \end{minipage}

 \end{center}
\end{figure*} 

\noindent
The Poissonian uncertainties assumed in our MF ($\S$\ref{Luminosity and Mass Functions})
represent a lower limit on the uncertainties.
In addition, there are the following factors that may contribute to the uncertainties
of the derived MF.
\begin{cutenumerate}[label=\alph{*})]
 \itemsep0.5em  
 \item \emph{Uncertainty in selecting the SFRS galaxies from the \PSCz{} catalogue.}
 As a reminder, the SFRS galaxies were selected by first binning the \PSCz{}
 in the three-dimensional space of IRAS 60~$\mu$m
 luminosity and the two flux density ratios $F_{100}/F_{60}$ and $F_{60}/K_{s}$
 and then selecting up to 10 galaxies from the \PSCz{} galaxies in each 3D bin.
 The weight assigned to each SFRS galaxy is
 the number of \PSCz{} galaxies it represents, i.e., number of
 \PSCz{} galaxies in the bin divided by the number selected for the SFRS\null.
 (Refer to $\S$\ref{The Star Formation Reference Survey (SFRS)} and 
 \citealt{ashby} for more details on the sample creation, including
 the bin definitions and how the numbers for each bin were chosen.)
 We therefore want to address the question:
 "if we had randomly chosen a different set of SFRS galaxies, how
 much different 
 would the number counts in each mass bin and resulting MF be?".
 \item \emph{Uncertainties in the binning.}
 The SFRS bin boundaries could have been chosen differently, and the
 observational data used to assign \PSCz{} galaxies to bins have
 finite uncertainties.  We therefore want to address what would have
 happened with slightly different binning.
 \item \emph{Uncertainty in the mass measurement.}
 Errors in the mass measurements could shift SFRS galaxies to a
 different mass bin.
\end{cutenumerate}
\noindent
To assess how the above uncertainties propagate to the final MF, we
performed a Monte-Carlo simulation.  For each trial, we varied the
galaxy number counts in the mass bins, 
the galaxy weights, and the galaxy masses and then calculated a new MF\null.
Corresponding to the items above,
\begin{cutenumerate}[label=\alph{*})]
 \itemsep0.5em  
 \item the number of galaxies in each trial bin was varied from the actual
   number according to a Poisson distribution. 
 \item the weights of the galaxies in each bin were varied according
   to a Poisson distribution of the weights, and then
       the trial total bin weight was equally re-distributed over the
       galaxies in the trial bin.  The sum of the SFRS bin weights
   is by construction the  number of \PSCz{}
       galaxies in the bin, and therefore  a Poisson distribution
       should represent the uncertainties in counting \PSCz{} galaxies.
       \noindent
       In this step we also ensured that the sum of all weights
       was within $\pm$1 of the number of galaxies in the \PSCz{},
       rejecting any trial that did not satisfy this
       condition.
 \item To account for the uncertainty in the mass measurement, the mass
       of each galaxy was sampled from a Gaussian distribution
       centered on the actual mass 
       and with standard deviation equal to the mass uncertainty.
\end{cutenumerate}
The Monte Carlo results of $\sim$1\,000 accepted trials are shown in
Figure~\ref{figure:simulated_MF}. The range among different trials
should represent the uncertainties mentioned above. As can be seen,
the ranges for most bins are comparable to the Poisson uncertainties
with little  systematic offsets.  This gives confidence in the
derived MF and its uncertainties.
This strongly suggests that our derivation of the MF in
$\S$\ref{SFRS stellar mass functions} is robust
against these additional uncertainty factors.

\newpage

\begin{landscape}
 \small 
 \begin{longtable}{llccccccc}
  \caption[SDSS and 2MASS magnitudes for the SFRS sample]{
    List of SDSS (Data Release 12) $u$, $r$, and 2MASS $J$, $H$, $K$ magnitudes for the SFRS sample.
    \\
    \emph{--- The full version of this table is available online ---}
  }\\
  \hline
  \multicolumn{9}{c}{\textsc{SFRS --- 2MASS and SDSS Photometry}} \\
  \hline
  \addlinespace 
  \hline
  \addlinespace 
  \multicolumn{1}{c}{SFRS}       & \multicolumn{1}{c}{Target}	  & \multicolumn{3}{c}{\hrulefill~~SDSS~~\hrulefill} & \multicolumn{4}{c}{\hrulefill~~2MASS~~\hrulefill}             \\
  \multicolumn{1}{c}{ID}         & \multicolumn{1}{c}{Name}	  & ID	       & $u^{\dagger}$ & $r^{\dagger}$       & ID	  & $J^{\ddagger}$ & $H^{\ddagger}$ & $K^{\ddagger}$ \\ 
  \multicolumn{1}{c}{\tiny{(1)}} & \multicolumn{1}{c}{\tiny{(2)}} & \tiny{(3)} & \tiny{(4)}    & \tiny{(5)}          & \tiny{(6)} & \tiny{(7)}     & \tiny{(8)}     & \tiny{(9)}     \\
  \hline
  \addlinespace 
  \endfirsthead

  \multicolumn{8}{@{}l}{Table \ref{table:SFRS_2MASS_SDSS} - \emph{continued}}\\
  \hline
  \multicolumn{1}{c}{SFRS}       & \multicolumn{1}{c}{Target}	  & \multicolumn{3}{c}{\hrulefill~~SDSS~~\hrulefill} & \multicolumn{4}{c}{\hrulefill~~2MASS~~\hrulefill}             \\
  \multicolumn{1}{c}{ID}         & \multicolumn{1}{c}{Name}	  & ID	       & $u^{\dagger}$ & $r^{\dagger}$       & ID	  & $J^{\ddagger}$ & $H^{\ddagger}$ & $K^{\ddagger}$ \\ 
  \multicolumn{1}{c}{\tiny{(1)}} & \multicolumn{1}{c}{\tiny{(2)}} & \tiny{(3)} & \tiny{(4)}    & \tiny{(5)}          & \tiny{(6)} & \tiny{(7)}     & \tiny{(8)}     & \tiny{(9)}     \\
  \hline
  \addlinespace 
  \endhead
1   & IC486                          & SDSSJ080020.98+263648.8  & 15.856(0.006) & 13.377(0.002) & XSC---08002097+2636483  & 11.514(0.020) & 10.826(0.021) & 10.496(0.028) \\
2   & IC2217                         & SDSSJ080049.73+273001.5  & 15.269(0.005) & 13.838(0.002) & XSC---08004972+2730010  & 11.834(0.024) & 11.186(0.027) & 10.898(0.040) \\
3   & NGC2500                        & SDSSJ080153.17+504413.5  & 14.104(0.005) & 12.263(0.002) & XSC---08015322+5044135  & 10.790(0.021) & 10.160(0.030) &  9.276(0.065) \\
4   & NGC2512                        & SDSSJ080307.84+232330.5  & 15.081(0.005) & 12.865(0.002) & XSC---08030785+2323308  & 10.859(0.020) & 10.117(0.026) &  9.820(0.028) \\
5   & MCG6-18-009                    & SDSSJ080328.94+332744.5  & 15.502(0.006) & 13.642(0.002) & XSC---08032897+3327444  & 11.803(0.034) & 11.135(0.034) & 10.763(0.040) \\
6   & MK1212                         & SDSSJ080705.51+270733.7  & 16.365(0.007) & 14.588(0.002) & XSC---08070551+2707336  & 12.594(0.026) & 11.995(0.032) & 11.589(0.045) \\
7   & IRAS08072+1847                 & SDSSJ081006.99+183818.0  & 18.271(0.020) & 15.590(0.003) & XSC---08100697+1838176  & 13.270(0.038) & 12.684(0.059) & 12.189(0.056) \\
8   & NGC2532                        & SDSSJ081015.17+335723.9  & 15.549(0.009) & 12.983(0.002) & XSC---08101519+3357233  & 10.643(0.022) & 10.004(0.029) &  9.615(0.036) \\
9   & UGC4261                        & SDSSJ081056.21+364941.3  & 15.771(0.006) & 14.984(0.003) & XSC---08105615+3649435  & 12.887(0.037) & 12.291(0.056) & 12.027(0.067) \\
10  & NGC2535                        & SDSSJ081113.48+251224.8  & 15.227(0.008) & 12.980(0.002) & XSC---08111348+2512249  & 10.993(0.027) & 10.341(0.029) & 10.120(0.043) \\
11  & NGC2543                        & SDSSJ081257.92+361516.6  & 15.083(0.007) & 12.514(0.002) & XSC---08125795+3615162  & 10.532(0.023) &  9.745(0.026) &  9.431(0.030) \\
12  & NGC2537                        & SDSSJ081314.69+455922.0  & 13.880(0.004) & 12.318(0.002) & XSC---08131464+4559232  &  9.949(0.022) &  9.366(0.025) &  9.130(0.034) \\
13  & IC2233                         & SDSSJ081358.77+454441.9  & 20.296(0.052) & 22.550(0.250) & XSC---08135890+4544317  & 11.463(0.032) & 10.950(0.042) & 10.754(0.062) \\
14  & IC2239                         & SDSSJ081406.79+235158.9  & 15.931(0.007) & 13.527(0.002) & XSC---08140678+2351588  & 11.363(0.021) & 10.687(0.025) & 10.339(0.028) \\
15  & UGC4286                        & SDSSJ081416.47+182626.3  & 16.045(0.011) & 13.570(0.002) & XSC---08141648+1826258  & 11.897(0.025) & 11.175(0.033) & 10.868(0.038) \\
16  & UGC4306                        & SDSSJ081736.78+352644.3  & 16.221(0.011) & 13.585(0.002) & XSC---08173676+3526455  & 11.449(0.023) & 10.660(0.024) & 10.333(0.030) \\
17  & NGC2552                        & SDSSJ081919.17+500017.6  & 22.230(0.295) & 20.584(0.047) & XSC---08192055+5000351  & 14.524(0.088) & 13.859(0.122) & 13.607(0.148) \\
18  & UGC4383                        & SDSSJ082334.21+212051.5  & 15.614(0.006) & 13.892(0.002) & XSC---08233424+2120514  & 12.244(0.037) & 11.575(0.040) & 11.368(0.047) \\
19  & IRAS08234+1054                 & SDSSJ082607.85+104451.0  & 21.853(0.272) & 18.768(0.014) & XSC---08260790+1044514  & 13.927(0.094) & 13.053(0.086) & 12.778(0.167) \\
20  & IRAS08269+1514                 & SDSSJ082945.19+150439.3  & 17.804(0.031) & 15.671(0.005) & XSC---08294519+1504394  & 13.671(0.075) & 12.934(0.089) & 12.373(0.090) \\
21  & NGC2604                        & SDSSJ083323.13+293219.7  & 14.412(0.004) & 13.363(0.002) & XSC---08332305+2932190  & 12.041(0.037) & 11.422(0.050) & 11.044(0.060) \\
22  & NGC2608                        & SDSSJ083517.33+282824.3  & 15.016(0.007) & 12.838(0.002) & XSC---08351736+2828246  & 10.299(0.019) &  9.611(0.023) &  9.326(0.030) \\
23  & MK92                           & SDSSJ083539.96+462928.1  & 16.065(0.005) & 14.632(0.002) & XSC---08353993+4629279  & 12.493(0.071) & 11.797(0.066) & 11.468(0.053) \\
24  & NGC2623                        & SDSSJ083824.01+254516.2  & 16.134(0.008) & 13.802(0.002) & XSC---08382409+2545167  & 11.572(0.020) & 10.808(0.025) & 10.427(0.027) \\
25  & CGCG120-018                    & SDSSJ083950.75+230836.1  & 16.946(0.014) & 14.502(0.002) & XSC---08395080+2308357  & 12.582(0.032) & 11.803(0.038) & 11.488(0.051) \\
26  & NGC2644                        & SDSSJ084131.84+045851.3  & 14.925(0.005) & 13.184(0.002) & XSC---08413189+0458488  & 11.232(0.034) & 10.646(0.043) & 10.359(0.061) \\
27  & UGC4572                        & SDSSJ084537.85+365604.7  & 15.566(0.005) & 13.276(0.002) & XSC---08453783+3656047  & 11.546(0.024) & 10.896(0.023) & 10.649(0.039) \\
28  & UGC4653                        & SDSSJ085354.62+350844.0  & 16.746(0.013) & 14.370(0.002) & XSC---08535462+3508439  & 11.960(0.104) & 11.150(0.040) & 10.654(0.033) \\
29  & IRAS08512+2727                 & SDSSJ085416.77+271559.5  & 16.640(0.008) & 14.837(0.002) & XSC---08541681+2715592  & 12.952(0.043) & 12.392(0.069) & 11.924(0.067) \\
30  & OJ287                          & SDSSJ085448.87+200630.6  & 16.573(0.007) & 15.421(0.004) & XSC---08544889+2006307  & 13.503(0.039) & 12.567(0.039) & 11.768(0.032) \\
  \dots & \dots & \dots & \dots & \dots & \dots & \dots & \dots & \dots \\
  \dots & \dots & \dots & \dots & \dots & \dots & \dots & \dots & \dots \\
  \dots & \dots & \dots & \dots & \dots & \dots & \dots & \dots & \dots \\
  \addlinespace 
  \hline
  \addlinespace 

  \label{table:SFRS_2MASS_SDSS}
 \end{longtable}
 
 \vspace{-1cm} 
 
 \begin{center}
  \begin{minipage}{0.85\linewidth}
   \noindent
   $^{(1)}$ SFRS target ID;
   $^{(2)}$ Galaxy name;
   $^{(3)}$ SDSS-DR12 ID;
   $^{(4)}$ SDSS $u$ magnitude (SDSS $ugriz$ filter set);
   $^{(5)}$ SDSS $r$ magnitude (SDSS $ugriz$ filter set);
   $^{(6)}$ 2MASS Extended Source Catalogue (XSC) or Point Source Catalogue (PSC) ID;
   $^{(7)}$ 2MASS $J$-band magnitude;
   $^{(8)}$ 2MASS $H$-band magnitude;
   $^{(9)}$ 2MASS $K$-band magnitude.
   \\
   $\dagger$: SDSS modelMag;
   $\ddagger$: 2MASS-XSC values refer to the ``extrapolated magnitudes''.
   ($J.ext$, $H.ext$, and $K.ext$; see $\S$\ref{The 2MASS Extended Source Catalogue photometry}).

  \end{minipage}
 \end{center}
\end{landscape}

\clearpage

 
\begin{landscape}
 {
  \definecolor{LightGray}{rgb}{0.92,0.92,0.92}
  \definecolor{White}{rgb}{1.0,1.0,1.0}
  \footnotesize
  \renewcommand{\tabcolsep}{0.3em}
  \renewcommand{\arraystretch}{0.95}
  \begin{longtable}[!h]{ll @{\hskip 0.35cm} ccccccccccccccccc}
   \caption[\GALFIT{} model results for the SFRS sample]{{
     \normalsize \GALFIT{} model results for the SFRS sample.
     \\
     \emph{--- The full version of this table is available online ---}
   }}\\
   \hline
   \addlinespace 
   \multicolumn{19}{c}{{\normalsize \textsc{SFRS Targets --- Galfit Model Results}}}\\
   \hline
   \addlinespace 
   \hline
   \addlinespace 
   \multicolumn{1}{c}{Target}      & \multicolumn{1}{c}{Selected Model} & Model       & Comp.       & S\'{e}rsic  & $R$       & Comp. $m_{K}$ & K-corr      & Comp. $L_{K}$  & ($u-r$)    & Comp. $M/L$                   & $M/L$                         & Comp. M          & Total M$^{\dagger}$ & Decomp.      & $B/T$      & $D$	       & Nuc.         & Weight       \\
   \multicolumn{1}{c}{Name}        & \multicolumn{1}{c}{}               & Comp.       & Nature      & $n$        & [\arcsec]  & [mag]           &             & [erg s$^{-1}$] & [mag]        & [$L_{\odot}$/M$_{\odot}$] & [$L_{\odot}$/M$_{\odot}$] & [M$_{\odot}$] & [M$_{\odot}$]       & Method       &              & [Mpc]             & Class.       &              \\
   \multicolumn{1}{c}{\tiny{(1)}} & \multicolumn{1}{c}{\tiny{(2)}}    & \tiny{(3)} & \tiny{(4)} & \tiny{(5)}  & \tiny{(6)} & \tiny{(7)}     & \tiny{(8)} & \tiny{(9)}      & \tiny{(10)} & \tiny{(11)}                    & \tiny{(12)}                    & \tiny{(13)}     & \tiny{(14)}	          & \tiny{(15)} & \tiny{(16)} & \tiny{(17)} & \tiny{(18)} & \tiny{(19)} \\
   \hline
   \addlinespace 
   \endfirsthead

   \multicolumn{16}{@{}l}{{\normalsize Table \ref{table:GALFIT_selected_models} --- \emph{continued}}}\\
   \hline
   \addlinespace 
   \multicolumn{1}{c}{Target}      & \multicolumn{1}{c}{Selected Model} & Model       & Comp.       & S\'{e}rsic  & $R$       & Comp. $m_{K}$ & K-corr      & Comp. $L_{K}$  & ($u-r$)    & Comp. $M/L$                   & $M/L$                         & Comp. M          & Total M$^{\dagger}$ & Decomp.      & $B/T$      & $D$	       & Nuc.         & Weight       \\
   \multicolumn{1}{c}{Name}        & \multicolumn{1}{c}{}               & Comp.       & Nature      & $n$        & [\arcsec]  & [mag]           &             & [erg s$^{-1}$] & [mag]        & [$L_{\odot}$/M$_{\odot}$] & [$L_{\odot}$/M$_{\odot}$] & [M$_{\odot}$] & [M$_{\odot}$]       & Method       &              & [Mpc]             & Class.       &              \\
   \multicolumn{1}{c}{\tiny{(1)}} & \multicolumn{1}{c}{\tiny{(2)}}    & \tiny{(3)} & \tiny{(4)} & \tiny{(5)}  & \tiny{(6)} & \tiny{(7)}     & \tiny{(8)} & \tiny{(9)}      & \tiny{(10)} & \tiny{(11)}                    & \tiny{(12)}                    & \tiny{(13)}     & \tiny{(14)}	          & \tiny{(15)} & \tiny{(16)} & \tiny{(17)} & \tiny{(18)} & \tiny{(19)} \\
   \hline
   \addlinespace 
   \endhead
\addlinespace 
\rowcolor{LightGray}
IC2217                 & Sersic + exDisk          & Sersic         & bulge & 4.0 & 8.1   & 11.56$\pm$0.12 & 0.969 & 1.48e+43 & 2.48  & 0.897 & 0.720 & 2.67e+10 & 4.51e+10 & \textsc{Galfit}   & 0.473$\pm$0.125 & 7.61e+07 & H\,{\sevensize II} & 13.2\\ 
\rowcolor{LightGray}
IC2217                 & Sersic + exDisk          & exDisk         & disk  & --- & 5.9   & 11.44$\pm$0.07 & 0.969 & 1.65e+43 & 1.58  & 0.743 & 0.720 & 2.47e+10 & 4.51e+10 & \textsc{Galfit}   & 0.473$\pm$0.125 & 7.61e+07 & H\,{\sevensize II} & 13.2\\ 
\addlinespace 
\rowcolor{White}
NGC2500                & Sersic + psfAgn          & Sersic         & disk  & 1.8 & 89.1  & 8.51 $\pm$0.04 & 0.980 & 9.63e+42 & 1.50  & 0.730 & 0.784 & 1.41e+10 & 1.52e+10 & \textsc{Galfit}   & 0.001$\pm$0.001 & 1.50e+07 & LINER & 4.2 \\ 
\rowcolor{White}
NGC2500                & Sersic + psfAgn          & psfAgn         & bulge & --- & ---   & 16.07$\pm$0.13 & 0.980 & 9.12e+39 & 4.64  & 1.410 & 0.784 & 2.59e+07 & 1.52e+10 & \textsc{Galfit}   & 0.001$\pm$0.001 & 1.50e+07 & LINER & 4.2 \\ 
\addlinespace 
\rowcolor{LightGray}
NGC2512                & Sersic + psfAgn          & Sersic         & disk  & 1.5 & 14.0  & 9.92 $\pm$0.04 & 0.962 & 5.51e+43 & 2.11  & 0.830 & 0.849 & 9.20e+10 & 1.03e+11 & \textsc{Galfit}   & 0.089$\pm$0.009 & 6.93e+07 & H\,{\sevensize II} & 7.4 \\ 
\rowcolor{LightGray}
NGC2512                & Sersic + psfAgn          & psfAgn         & bulge & --- & ---   & 12.45$\pm$0.04 & 0.962 & 5.36e+42 & 1.51  & 0.732 & 0.849 & 7.89e+09 & 1.03e+11 & \textsc{Galfit}   & 0.089$\pm$0.009 & 6.93e+07 & H\,{\sevensize II} & 7.4 \\ 
\addlinespace 
\rowcolor{White}
MCG6-18-009            & Sersic                   & Sersic         & mixed & 5.6 & 8.8   & 10.49$\pm$0.04 & 0.924 & 1.76e+44 & NA    & NA    & 0.788 & ---   & 2.79e+11 & $C\mapsto$~B/T    & 0.204$\pm$0.054 & 1.64e+08 & H\,{\sevensize II} & 1.5 \\ 
\addlinespace 
\rowcolor{LightGray}
MK1212                 & Sersic                   & Sersic         & mixed & 4.1 & 2.1   & 11.57$\pm$0.04 & 0.928 & 7.27e+43 & NA    & NA    & 0.774 & ---   & 1.13e+11 & $C\mapsto$~B/T    & 0.184$\pm$0.049 & 1.73e+08 & H\,{\sevensize II} & 7.4 \\ 
\addlinespace 
\rowcolor{White}
IRAS08072+1847         & Sersic + psfAgn          & Sersic         & disk  & 2.0 & 1.7   & 12.58$\pm$0.06 & 0.957 & 4.94e+42 & 2.59  & 0.918 & 0.935 & 9.12e+09 & 1.31e+10 & \textsc{Galfit}   & 0.290$\pm$0.077 & 7.08e+07 & SN    & 8.1 \\ 
\rowcolor{White}
IRAS08072+1847         & Sersic + psfAgn          & psfAgn         & bulge & --- & ---   & 13.55$\pm$0.10 & 0.957 & 2.02e+42 & 1.68  & 0.758 & 0.935 & 3.08e+09 & 1.31e+10 & \textsc{Galfit}   & 0.290$\pm$0.077 & 7.08e+07 & SN    & 8.1 \\ 
\addlinespace 
\rowcolor{LightGray}
NGC2532                & Sersic + psfAgn          & Sersic         & disk  & 1.2 & 18.8  & 9.85 $\pm$0.04 & 0.959 & 7.35e+43 & 2.04  & 0.818 & 0.913 & 1.21e+11 & 1.47e+11 & \textsc{Galfit}   & 0.077$\pm$0.009 & 7.76e+07 & H\,{\sevensize II} & 5.0 \\ 
\rowcolor{LightGray}
NGC2532                & Sersic + psfAgn          & psfAgn         & bulge & --- & ---   & 12.55$\pm$0.04 & 0.959 & 6.11e+42 & 2.06  & 0.821 & 0.913 & 1.01e+10 & 1.47e+11 & \textsc{Galfit}   & 0.077$\pm$0.009 & 7.76e+07 & H\,{\sevensize II} & 5.0 \\ 
\addlinespace 
\rowcolor{White}
NGC2535                & Sersic + psfAgn          & Sersic         & disk  & 1.2 & 23.8  & 10.06$\pm$0.04 & 0.979 & 3.90e+43 & 1.70  & 0.762 & 0.854 & 5.98e+10 & 7.14e+10 & \textsc{Galfit}   & 0.065$\pm$0.006 & 6.16e+07 & H\,{\sevensize II} & 5.5 \\ 
\rowcolor{White}
NGC2535                & Sersic + psfAgn          & psfAgn         & bulge & --- & ---   & 12.96$\pm$0.04 & 0.979 & 2.70e+42 & 2.36  & 0.875 & 0.854 & 4.75e+09 & 7.14e+10 & \textsc{Galfit}   & 0.065$\pm$0.006 & 6.16e+07 & H\,{\sevensize II} & 5.5 \\ 
\addlinespace 
\rowcolor{LightGray}
NGC2543                & Sersic + psfAgn          & Sersic         & disk  & 3.5 & 43.1  & 9.04 $\pm$0.04 & 0.982 & 1.82e+43 & 2.42  & 0.886 & 0.914 & 3.24e+10 & 3.51e+10 & \textsc{Galfit}   & 0.043$\pm$0.010 & 2.63e+07 & H\,{\sevensize II} & 12.1\\ 
\rowcolor{LightGray}
NGC2543                & Sersic + psfAgn          & psfAgn         & bulge & --- & ---   & 12.41$\pm$0.04 & 0.982 & 8.18e+41 & 2.08  & 0.825 & 0.914 & 1.36e+09 & 3.51e+10 & \textsc{Galfit}   & 0.043$\pm$0.010 & 2.63e+07 & H\,{\sevensize II} & 12.1\\ 
\addlinespace 
\rowcolor{White}
NGC2537                & Sersic                   & Sersic         & disk  & 0.3 & 22.5  & 9.24 $\pm$0.04 & 0.996 & 5.00e+42 & 2.01  & 0.813 & 0.740 & 8.18e+09 & 7.44e+09 & \textsc{Galfit}   & 0    $\pm$0     & 1.50e+07 & H\,{\sevensize II} & 3.2 \\ 
\addlinespace 
\rowcolor{LightGray}
IC2233                 & Sersic                   & Sersic         & disk  & 1.0 & 50.1  & 10.73$\pm$0.04 & 0.998 & 1.06e+42 & 2.01  & 0.813 & 0.333 & 1.73e+09 & 7.09e+08 & \textsc{Galfit}   & 0    $\pm$0     & 1.37e+07 & H\,{\sevensize II} & 1.0 \\ 
\addlinespace 
\rowcolor{White}
IC2239                 & Sersic + psfAgn          & Sersic         & disk  & 3.5 & 7.3   & 10.41$\pm$0.04 & 0.955 & 5.68e+43 & 2.37  & 0.876 & 0.883 & 1.00e+11 & 1.15e+11 & \textsc{Galfit}   & 0.123$\pm$0.041 & 8.85e+07 & H\,{\sevensize II} & 13.2\\ 
\rowcolor{White}
IC2239                 & Sersic + psfAgn          & psfAgn         & bulge & --- & ---   & 12.54$\pm$0.06 & 0.955 & 7.98e+42 & 2.44  & 0.889 & 0.883 & 1.43e+10 & 1.15e+11 & \textsc{Galfit}   & 0.123$\pm$0.041 & 8.85e+07 & H\,{\sevensize II} & 13.2\\ 
\addlinespace 
\rowcolor{LightGray}
UGC4286                & Sersic + exDisk          & Sersic         & bulge & 4.0 & 1.1   & 12.97$\pm$0.09 & 0.961 & 3.73e+42 & 4.09  & 1.257 & 0.896 & 9.43e+09 & 4.44e+10 & \textsc{Galfit}   & 0.151$\pm$0.136 & 7.35e+07 & H\,{\sevensize II} & 3.0 \\ 
\rowcolor{LightGray}
UGC4286                & Sersic + exDisk          & exDisk         & disk  & --- & 7.4   & 11.10$\pm$0.04 & 0.961 & 2.09e+43 & 1.89  & 0.792 & 0.896 & 3.33e+10 & 4.44e+10 & \textsc{Galfit}   & 0.151$\pm$0.136 & 7.35e+07 & H\,{\sevensize II} & 3.0 \\ 
\addlinespace 
\dots                  & \dots                    & \dots          & \dots & \dots & \dots & \dots$\pm$\dots & \dots & \dots    & \dots & \dots & \dots & \dots & \dots & \dots             & \dots$\pm$\dots & \dots    & \dots & \dots\\ 
\addlinespace 
\dots                  & \dots                    & \dots          & \dots & \dots & \dots & \dots$\pm$\dots & \dots & \dots    & \dots & \dots & \dots & \dots & \dots & \dots             & \dots$\pm$\dots & \dots    & \dots & \dots\\ 
\addlinespace 
\dots                  & \dots                    & \dots          & \dots & \dots & \dots & \dots$\pm$\dots & \dots & \dots    & \dots & \dots & \dots & \dots & \dots & \dots             & \dots$\pm$\dots & \dots    & \dots & \dots\\ 
\addlinespace 

   \hline
   \addlinespace 

   \label{table:GALFIT_selected_models}
  \end{longtable}

  \vspace{-0.8cm}   

  \noindent 
  $^{(1)}$  Galaxy name;
  $^{(2)}$  Selected model (see definition in $\S$\ref{Fit procedure});
  $^{(3)}$  Selected model component (see definition in $\S$\ref{Fit procedure});
  $^{(4)}$  Physical interpretation of component;
  $^{(5)}$  S\'{e}rsic index;
  $^{(6)}$  Effective radius (for S\'{e}rsic), or scale height (for exDisk);
  $^{(7)}$  Component $K$-band magnitude (Vega system);
  $^{(8)}$  K correction factor;
  $^{(9)}$  Component $K$-band luminosity;
  $^{(10)}$ Component color form our decomposition (the global galaxy color is reported in Table \ref{table:SFRS_2MASS_SDSS};
  $^{(11)}$ Component mass-to-light;
  $^{(12)}$ Galaxy mass-to-light ratio;
  $^{(13)}$ Component mass;
  $^{(14)}$ Total mass$^{\dagger}$ (omitting AGN light, if any);
  $^{(15)}$ Decomposition method: labelled as ``\GALFIT{}'' in case it was
              obtained through 2D fitting, or ``$C\mapsto$~B/T''
              in case it was derived using the concentration index
	      (see $\S$\ref{Further decomposition of mixed components});
  $^{(16)}$ Bulge-to-total ratio;
  $^{(17)}$ Distance from \cite{ashby};
  $^{(18)}$ Nuclear classification by \cite{maragkoudakis:2018};
  $^{(19)}$ Galaxy representativeness weight from \cite{ashby}.
              If multiplied by a constant factor of 6.7, a weight provides the
	      number of galaxies that the SFRS target represents in
	      the \PSCz{} catalogue (see \citealt{ashby} for details).
	      
  \smallskip
  \noindent$^{\dagger}${\textsc NOTE}:
   Because a different $M/L$ is used for bulge, disk, and whole galaxy, the
   masses of the sub-components do \emph{not} necessarily sum up to the total
   mass.\\
 }

\end{landscape}

\clearpage

\label{lastpage}

\end{document}